\documentclass[fleqn,usenatbib]{mnras}

\usepackage{newtxtext,newtxmath}
\usepackage{lscape}

\usepackage[T1]{fontenc}

\DeclareRobustCommand{\VAN}[3]{#2}
\let\VANthebibliography\thebibliography
\def\thebibliography{\DeclareRobustCommand{\VAN}[3]{##3}\VANthebibliography}

\usepackage{graphicx}	
\usepackage{amsmath}	

\def\cm2{cm$^2$ }
\def\se1{s$^{-1}$ }

\newcommand{\angstrom}{\textup{\AA}}

\def\aj{AJ}%
\def\araa{ARA\&A}%
\def\apj{ApJ}%
\def\apjl{ApJ}%
\def\apjs{ApJS}%
\def\apss{Ap\&SS}%
\def\aap{A\&A}%
\def\aaps{A\&AS}%
\def\mnras{MNRAS}%
\def\pasp{PASP}%
\def\rmxaa{RMXAA}%
\def\nat{Nature}%
\title[Helium  abundances in  Seyfert 2s]{Chemical abundances in  Seyfert galaxies-- IX. Helium abundance estimates}

\author[Dors, et al.]{
O. L. Dors$^{1}$\thanks{E-mail: olidors@univap.br}, M. Valerdi$^{2}$, P. Freitas-Lemes$^{1}$, A. C. Krabbe$^{1}$, R.~A. Riffel$^{3}$, E. B. Am\^ores$^{4}$,  R. Riffel$^{5}$, \newauthor{M. Armah$^{5}$, A. F. Monteiro$^{6}$, C. B. Oliveira$^{1}$}
\\
$^{1}$UNIVAP - Universidade do Vale do Para{\'i}ba. Av. Shishima Hifumi, 2911, CEP: 12244-000, S{\~ a}o Jos{\'e} dos Campos, SP, Brazil\\
$^{2}$Instituto Nacional de Astrof{\'i}sica, \'Optica y Electr\'onica (INAOE), Luis E. Erro No. 1, Sta. Ma. Tonantzintla, Puebla, C.P. 72840, M\'exico. \\
$^{3}$Departamento de F\'isica, Centro de Ci\^encias Naturais e Exatas, Universidade Federal de Santa Maria, 97105-900, Santa Maria, RS, Brazil\\
$^{4}$Departamento de F{\'i}sica, Universidade Estadual de Feira de Santana, Av. Transnordenstina, S/N, CEP 44036-900 Feira de Santana, BA, Brazil \\
$^{5}$ Departamento de Astronomia, Universidade Federal do Rio Grande do Sul, Av. Bento Gon\c calves 9500, Porto Alegre, RS, Brazil\\
$^{6}$ Instituto Federal do Maranh\~ao. Av. Newton Bello s/n, CEP: 65906-335, Imperatriz, MA, Brazil
}

\date{Accepted XXX. Received YYY; in original form ZZZ}

\pubyear{2022}

\begin{document}

\label{firstpage}
\pagerange{\pageref{firstpage}--\pageref{lastpage}}
\maketitle

\begin{abstract}
For the first time, the helium abundance relative to hydrogen
(He/H), which relied on direct measurements of the electron temperature,  has been derived in the narrow line regions (NLRs)  from a local sample of Seyfert 2 nuclei. In view of this, optical emission line intensities [$3000 \: < \lambda($\AA$) \: < \: 7000$] of 65 local Seyfert 2 nuclei ($z\: < \: 0.2$),
taken from Sloan Digital Sky Survey Data Release 15  and additional compilation from the literature, were considered.
 We used photoionization model grid to derive an Ionization Correction Factor (ICF) for the neutral helium. The application of this ICF indicates that the NLRs of Seyfert~2 present a neutral helium fraction of $\sim50$ per cent in relation to the total helium abundance.
We find that Seyfert~2 nuclei present helium abundance
ranging from 0.60 to 2.50 times the solar value, while
 $\sim 85$ per cent of the sample present over-solar abundance values. The derived (He/H)-(O/H) abundance relation from the Seyfert~2 is stepper than that of star-forming regions (SFs) and this difference could be due to excess of helium injected into the Interstellar Medium  by the winds of Wolf Rayet stars. 
 From a regression to zero metallicity, by using  Seyfert~2 estimates combined with 
  SFs estimates, we obtained  a primordial helium  mass fraction  $Y_{\rm p}=0.2441\pm0.0037$, a value in good agreement with the one  inferred from the
temperature fluctuations of the cosmic microwave background by the
Planck Collaboration, i.e.   $Y_{\rm p}^{Planck}=0.2471\pm0.0003$.  
 
\end{abstract}


\date{Released 2021 Apr 20}



\section{Introduction}
The helium abundance determination in the gas phase of  Active Galactic Nuclei (AGNs) and gaseous nebulae (\ion{H}{ii} regions, \ion{H}{ii} galaxies, Planetary Nebulae) is essential to the characterization  of the primordial stellar nucleosynthesis  after the Big Bang as well as in the study of the
Interstellar Medium (ISM) enrichment    
of galaxies along the Hubble time.
 
The first  
helium and heavy element determinations which relied on the direct measurements of the electron temperature 
($T_{\rm e}$-method\footnote{For a review of the
$T_{\rm e}$-method see \citet{2017PASP..129h2001P}
and \citet{2017PASP..129d3001P}.})
in AGNs seems to have been carried out by \citet{1975ApJ...197..535O} for the radio galaxy 3C\,405 (Cygnus\,A). These authors calculated the He abundance  using  the $\rm He^{+}$ and $\rm He^{2+}$  ionic abundances and derived a value of $y\sim 0.10$, where
\begin{equation}
    y=N(\mathrm{He})/N(\mathrm{H})
\end{equation}
the ratio of the total number densities of helium to hydrogen.
Numerical simulations by 
\citet{1974ApJ...191..309S}, which were carried out to reproduce observational data for the Seyfert galaxy 3C\,120, 
inferred  a value of  $y=0.09$ (see also \citealt{1978ApJ...223...56K, 1991PASP..103..888C}) for this object. Based on the 
results above it seems that no abnormal helium abundance has been found in the local Seyferts in comparison to the solar value,
$y_{\odot}=0.10$ \citep{2010Ap&SS.328..179G}, which is generally used
as standard reference  and scale factor (see \citealt{2017MNRAS.466.4403N}).
However, helium abundance determination in a sample
of AGNs is necessary to confirm this result.

 On the other hand, quasars seem to exhibit a different
and wider  range of $y$ values. For instance,
\citet{1971ApJ...163..235B}, by using 
helium and hydrogen emission-lines observed in the 
optical and ultraviolet and assuming $y=y^{+}+y^{2+}$,
estimated  $y$ in the range 0.003-0.2 for 14 quasars.  \citet{1975ApJ...201...26B}, assuming a similar approach by \citet{1971ApJ...163..235B}, derived the $y$ for a sample of 14 low-redshift quasars ($z\approx0.2$) and found  values  from $\sim0.1$ to $\sim0.3$ (see also,
\citealt{1971ApJ...167L..27W, 1973ApJ...181..627J}). The helium estimates by these authors were based on the $T_{\rm e}$-method,  i.e. a conventional and reliable method
\citep{2003A&A...399.1003P, 2017MNRAS.467.3759T}.  

Regarding gaseous nebulae, the first He abundance in this class of object was obtained by 
\citet{1945ApJ...102..239A} for 7 Galactic Planetary Nebulae by using optical emission lines measured by \citet{1942ApJ....95..356W} and, derived the mean abundance value of the helium relative to hydrogen as  $\sim 0.10$. Thereafter this pioneering  work, \citet{1957ApJ...125..328M} obtained the first helium abundance
estimation  from \ion{H}{ii} regions, in this case in the  Orion Nebula, deriving  values  in the range $\sim$0.09--$\sim$0.16, depending on the electron temperature assumed in the calculations. Subsequent  studies have been extended to
extragalactic objects and the first helium abundance determination was obtained by \citet{1959PASP...71..301J} for 30 Doradus 
(for other pioneering works see:
\citealt{1962PASP...74..219A, 1962ApJ...136..374M, 1965PASP...77...90M, 1965MNRAS.130..393F, 1968ApJ...151..491A, 1970ApJ...159..809P}).  Thanks to the  (relatively) recent spectroscopic surveys, as the Sloan Digital Sky Survey (SDSS, \citealt{2000AJ....120.1579Y}), it has become possible
to derive the helium abundance for thousands of star-forming regions
(SFs; \ion{H}{ii} regions and star-forming galaxies) with a wide range of metallicity ($Z$, e.g. \citealt{2007ApJ...662...15I, 2013A&A...558A..57I, 2021MNRAS.502.3045K}), i.e. $ 7.7 \: \la 12+\log[(N(\mathrm{O})/N(\mathrm{H})] \la \: 8.7$  or $ 0.10 \: \la (Z/{\rm Z_{\odot}}) \la \: 1.0$, assuming the solar oxygen value $ \log[(N(\mathrm{O})/N(\mathrm{H})]_{\odot}=-3.31$ \citep{2001ApJ...556L..63A}, thus allowing the estimation of the primordial
helium abundance.

The usual formalism   \citep{1974ApJ...193..327P, 1976ApJ...203..581P} establishes that the  helium mass fraction, defined by
\begin{equation}
    Y=\frac{m_{\rm He}}{m_{\rm gas}},
\end{equation} 
has a relation with the metallicity ($Z$), given by
\begin{equation}
\label{eq_1a}
Y=\frac{4y(1-Z)}{1+4y}.
\end{equation}
 The  $Z$  can be traced by the $N(\mathrm{O})/N(\mathrm{H})$\footnote{For simplicity, along the paper, this ratio is defined by O/H. } abundance in SFs (e.g. \citealt{2012MNRAS.422..215Y, 2019ARA&A..57..511K}) and in AGNs (e.g. \citealt{1998AJ....115..909S, 2021MNRAS.507..466D}) because the oxygen is the most abundant metal  in the Universe and emission lines 
 (e.g. [\ion{O}{ii}]$\lambda$3726, $\lambda3729$, [\ion{O}{iii}]$\lambda$5007) of its most abundant ions
 are ubiquitous measured in the rest-frame optical regime (e.g. \citealt{2016ApJ...827..126B}).  Along this work $Z$ and O/H are assumed  interchangeably.
 Regarding the $Z-\rm (O/H)$ relation,
 \citet{2007ApJ...666..636P} pointed out that 
 the oxygen by mass is in order of $(55\pm10)$ per cent of the $Z$ value. It is beyond the scope of the present work
 to discuss the $Z-\rm (O/H)$ relation. In any case,
 \citet{2007ApJ...666..636P} reported that
  the error in the $\mathrm{O}/Z$ ratio translates into
an error slightly smaller than 0.0001 in the  determination of the primordial $Y$ value. We assumed the fixed relation $Z=20\rm(O/H)$ proposed by \citet{1992MNRAS.255..325P} and used recently by 
\citet{2021MNRAS.508.1084K}. Thus, Equation~\ref{eq_1a} is given by 
\begin{equation}
\label{eq_1}
Y=\frac{4y[1-20({\rm O/H})]}{1+4y}.     
\end{equation}

  The primordial helium
abundance $Y_{p}$ is derived by extrapolating the $Y-(\rm O/H)$ relation to oxygen abundance (or $Z$) equal to zero (for a review on uncertainties in
$Y$ determinations see \citealt{2003ASPC..297...81P, 2007ApJ...666..636P, 2004ApJ...617...29O}).
Recently, \citet{2021MNRAS.502.3045K}, by using  spectroscopic data
of 100 SFs taken  from the SDSS \citep{2000AJ....120.1579Y}, derived the primordial
helium mass $Y_{p}=0.2462\pm0.0022$, which this result is in consonance with the value inferred from the temperature fluctuations of the cosmic microwave background by the \cite{planck2018}, i.e. $Y_{p}^{Planck}=0.2471\pm0.0003$ (see also \citealt{2014MNRAS.445..778I, 2015JCAP...07..011A, 2016RMxAA..52..419P, 2018NatAs...2..957C, 2019MNRAS.487.3221F, 2019ApJ...876...98V,2021MNRAS.505.3624V, 2020ApJ...896...77H}).

The   extrapolation of the $Y-(\rm O/H)$ relation to derive $Y_{p}$ is very dependent on the extreme oxygen abundance values, i.e. on the lowest and highest  values derived for  line emitter objects. Along decades, efforts have been
made to derive He and O abundances in Extremely metal-poor galaxies (XMPs), in order to obtain $Y$ values close to $Y_{p}$
(e.g. \citealt{1983ApJ...273...81K, 1991PASP..103..919S, 1994ApJ...431..172S, 1994ApJ...426..123G, 1997ApJ...483..788O, 1998ApJ...500..188I,
2000ApJ...541..688P, 1999ApJ...527..757I, 2007ApJ...662...15I, 2009A&A...503...61I, 2014MNRAS.445..778I, 2019MNRAS.482.3892A, 2021MNRAS.505.5460V, 2021arXiv210900178A}).
However, despite efforts have been done to achieve direct 
oxygen estimation in SFs in the high abundance regime [$\rm 12+\log(O/H)\: \ga \: 8.7$ or ($Z/\rm Z_{\odot}) \: \ga \: 1.0]$   (e.g., see \citealt{1994A&A...282L..37K, 2004ApJ...615..228B,
2004ApJ...607L..21G, 2002MNRAS.329..315C, 2007A&A...473..411L, 2013ApJ...765..140A, 2016MNRAS.458.1529B, 2020ApJ...893...96B}),
helium   abundance determinations in these objects are barely found in the literature.

The inclusion of objects with high metallicity, for which direct  electron temperatures   as well as $\rm He^{+}$ and $\rm He^{2+}$ abundances estimates are possible, will result in a significant improvement in the determination of the
$Y-\rm O/H$ relation and, consequently, in the estimation  of $Y_{p}$. In this context, AGNs are ideal objects because they have higher gas ionization degree  (e.g. \citealt{2014MNRAS.437.2376R, 2021MNRAS.505.4289P}) and higher $Z$ 
(e.g. \citealt{1998AJ....115..909S, 2006MNRAS.371.1559G,  2018ApJ...856...46R, 2021ApJ...910..139R, 2020MNRAS.492..468D}) in comparison with \ion{H}{ii} regions.  These physical features make it possible to precisely measure  the total helium abundance, since \ion{He}{ii} lines are stronger in AGN spectra than
those in SFs and, they allow to constraint  the $Y$
value in the high metallicity regime.

 Large spectroscopic surveys, such as Sloan Digital Sky Survey (SDSS, \citealt{2000AJ....120.1579Y}), CALIFA \citep{2012A&A...538A...8S}, MaNGA \citep{2015ApJ...798....7B} and CHAOS \citep{2015ApJ...806...16B} have made  thousands of emission lines from SFs and AGNs available and these data have revolutionized our understanding of chemical evolution of galaxies (see, for instance,  \citealt{2003ApJ...584..210G, 2003ApJ...587...55G, 2004MNRAS.351.1151B, 2006A&A...448..955I, 2007MNRAS.381..263A, 2008ApJ...681.1183K, 2012MNRAS.421.1624P, 2017MNRAS.469.2121S, 2018MNRAS.474.2039E,  2020A&A...634A.107Y, 2020ApJ...893...96B}, among others). 
 However, majority of the data from these surveys have been used to derive abundances  mainly in SFs, with AGNs still being understudied. In fact, most of the AGN studies have
 relied on large surveys just to address  the oxygen abundance  and/or physical properties
 (e.g., ionization degree, electron temperature, electron density) of  these objects (e.g. \citealt{2006MNRAS.371.1559G, 2008ApJ...685L.109Z, 2012MNRAS.427.1266V, 2012ApJ...756...51L, 2013MNRAS.430.2605Z,
 2014MNRAS.437.2376R, 2015ApJ...801...35C, 2016MNRAS.456.3354F, 2017ApJ...842...44K, 2020MNRAS.492.5675C, 2020MNRAS.492..468D, 2021MNRAS.507..466D, 2021arXiv210807812A}). The abundances of other elements
 beyond the oxygen are poorly known in AGNs, in particular, the local Seyfert~2, for which large amount of spectroscopic data are available in the literature. 
 
 \citet{2017MNRAS.468L.113D}, who built detailed photoionization models
 to reproduce optical narrow emission line ratio intensities of a 
 sample of 44 Seyfert~2 nuclei, presented the first quantitative estimations of the nitrogen abundance for this class of object (see also \citealt{2019MNRAS.489.2652P, 2020MNRAS.496.1262J, 2020MNRAS.496.2191F, 2021MNRAS.501.1370D}).
 Recently, \citet{2020MNRAS.496.3209D} proposed a new methodology of
 the $T_{\rm e}$-method for AGNs which produce reliable O/H abundances, slightly lower ($\approx0.2$ dex) than those derived from detailed photoionization models. \citet{2021MNRAS.tmp.2390A} and \citet{2021MNRAS.508.3023M}, motivated by this new methodology  and by the availability of spectroscopic data in the
 literature, developed Ionization Correction Factors (ICFs) as well as electron temperature relations based on photoionization model results and,  for the first time, derived the neon and argon abundances in a sample of Seyfert~2 nuclei, respectively. 
 
 As a further study, in the present work, we apply the $T_{\rm e}$-method to a combined spectroscopy data  taken from the SDSS
 and additional data from distinct authors 
 with the goal to derive the $y$ values in a sample of local Seyfert~2 galaxies. These estimates allow access to the helium abundance in a higher  metallicity regime than that in SFs and produce important constraints to  primordial helium abundance as well as to the studies of stellar nucleosyntesis.  The present study is organized as follows.   In Sections~\ref{obs} and \ref{secabund}
the  observational data and the methodology
used to estimate the helium and oxygen abundances are presented, respectively.
The results and discussion are presented in Sect.~\ref{res}.
Finally, the conclusion of the outcome is given in Sect.~\ref{conc}.

\section{Observational data}
\label{obs}

\subsection{Seyfert~2 nuclei}
\label{sampleagn}

In the present work we studied a subsample of the sample presented by
\citet{2020MNRAS.492..468D}. These authors used the measurements
of the  SDSS-DR7  made available
by MPA/JHU\footnote{\url{https://wwwmpa.mpa-garching.mpg.de/SDSS/DR7/}} group 
to obtain a sample of Seyfert~2 nuclei. Initially, by using the MPA/JHU data, \citet{2020MNRAS.492..468D}
 carried out  a cross-correlation between the galaxy identification provided by the SDSS-DR7   and by the NED/IPAC\footnote{\url{ned.ipac.caltech.edu}}
and obtained the classification of each object in Seyfert~1 and 2 nuclei.
After  applying the
criteria to separate SF and AGN objects proposed by \citet{2001ApJ...556..121K, 2006MNRAS.372..961K} and \citet{perez13}, which consider a
set of emission line intensity ratios in diagnostic diagrams \citep{1981PASP...93....5B}, 463 Seyfert 2 AGNs   ($z \: \la 0.4$) with reliable optical emission lines
in the optical range $3000 \: \la \: \lambda(\angstrom) \: \la \: 7500$ were selected.

The data  from the MPA/JHU group do not contain measurements for the  
\ion{He}{ii}$\lambda$4686\AA. Therefore, we downloaded from SDSS-DR15   
database\footnote{\url{https://dr15.sdss.org/optical/spectrum/search}} the
spectra of the 463 objects selected by \citet{2020MNRAS.492..468D} in order
to obtain a sub-sample of Seyfert~2 nuclei in which it is possible to determine
the parameters necessary to calculate the He and O abundances through the $T_{\rm e}$-method. Firstly, 
in each spectrum, we performed the extinction correction using the \citet{1989ApJ...345..245C} law assuming 
the parameterized  extinction coefficient
$R_{V}$ = 3.1, a standard value for the diffuse
interstellar medium. The Galactic interstellar extinction values provided by \citet{1998ApJ...500..525S} and calibrated by \citet{2011ApJ...737..103S} obtained by using the {\sc VOTool} for interstellar extinction called GALExtin\footnote{\url{http://www.galextin.org}}~\citep{2021MNRAS.508.1788A} were also assumed. We shifted the spectra for the 
rest frame wavelength using the redsfhits available for each object and binned them with bin size equal to 1\AA.

 \begin{figure*}
 \begin{center}
  \includegraphics[angle=-90.0,width=15cm]{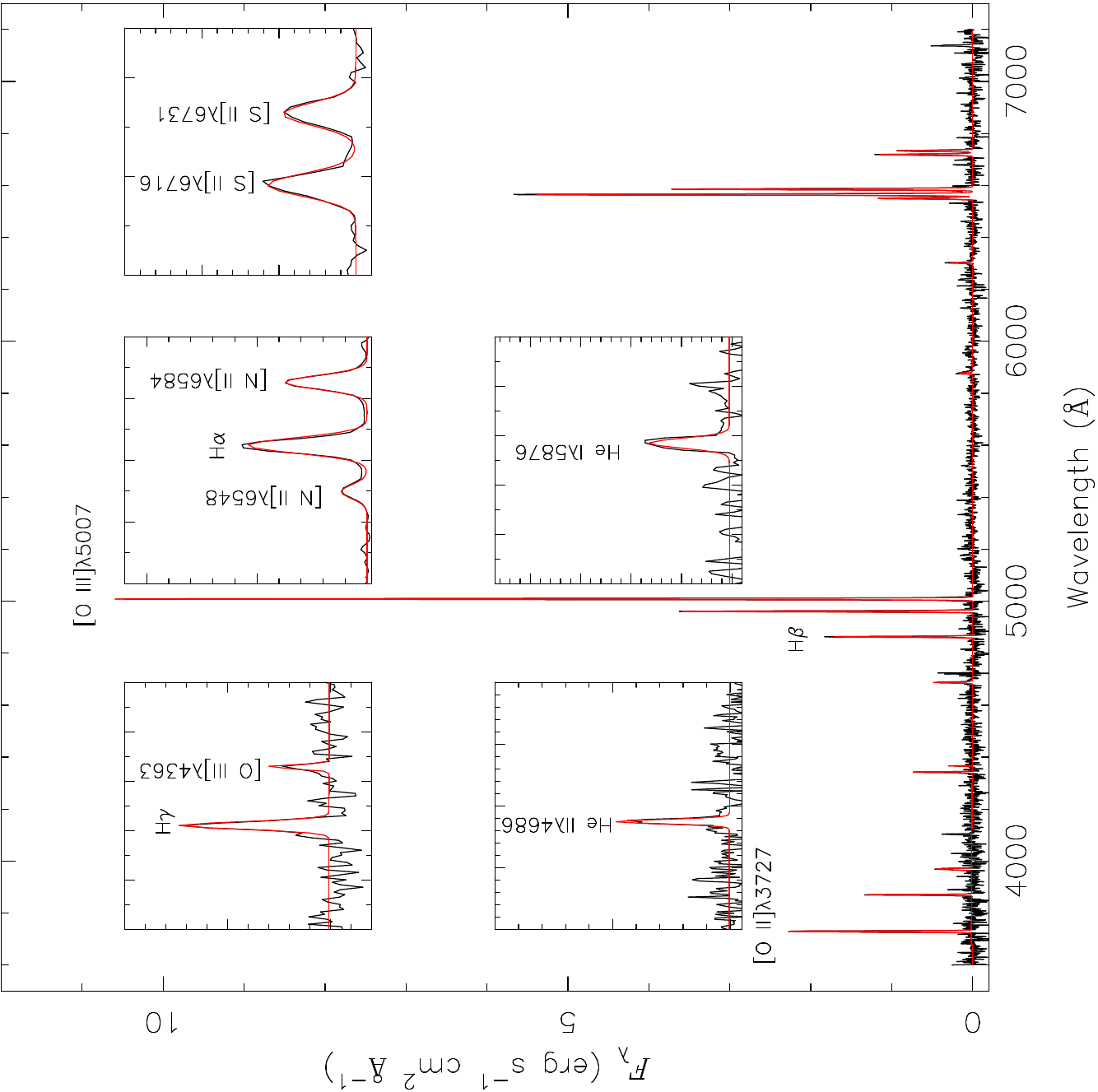}
 \end{center}
 \caption{Optical spectrum of one of the Seyfert~2 nucleus in our sample
 (see Sect.~\ref{sampleagn}) taken from SDSS DR15  and represented in black colour. The
 fitting to the emission-line profiles using the {\sc IFSCUBE code}
 \citep{daniel_ruschel} is represented in red colour. The measured emission lines and 
 corresponding wavelength are indicated. Boxes show a
 zoom in regions of some weak lines, as indicated.}
 \label{fig1}
 \end{figure*}
 
 The stellar population continuum was subtracted from  the  spectra to obtain the pure nebular spectra using  the stellar population synthesis {\scriptsize\,STARLIGHT} code
\citep{2005MNRAS.358..363C, 2006MNRAS.370..721M, 2007MNRAS.381..263A}. 
This code fits the observed spectrum of a galaxy using a combination of Simple Stellar Populations\,(SSPs), in different proportions and excluding the emission lines.
We used a  basis of  45 synthetic SSP spectra with three metallicities $Z$\,=\,0.004, 0.02  ($\mathrm{Z_{\odot}}$), and 0.05,
assuming 15 ages ranging from 1 Myr  to 13 Gyr, taken from the evolutionary synthesis models of \citet{2003MNRAS.344.1000B}.   
Prior to the fitting the synthetic spectra to the observational ones,
we convoluted the SSP stellar spectra  adopting  a 
Gaussian function to achieve the same spectral resolution of the observational data. A detailed description of the SSP spectra fitting to spectroscopic observational data is given by \citet{2011MNRAS.416...38K, 2017MNRAS.467...27K}.

After the pure nebular spectra for the sample were obtained by subtracting the contribution of the stellar component from the observed spectra, the emission-lines
were fitted using the publicly available  {\sc ifscube} package \citep{daniel_ruschel,2021MNRAS.507...74R}. Each emission line was fitted by a single Gaussian function and the kinematics (line width and centroid velocity) of lines from the same parent ion was kept tied. We also included in the fit a third order polynomial to account for any residual continuum emission.  The uncertainties on the fluxes were obtained by performing 100 iterations of Monte Carlo simulations of the emission-line fits as the standard deviation of the fluxes obtained from these simulations.  The fluxes were corrected for extinction following the procedure described by \citet{rogerio21}, using the $\rm (H\alpha/H\beta)$ line ratio to estimate the visual extinction, assuming the  theoretical value for the  $\rm (H\alpha/H\beta) = 2.86$ proposed  by \citet{hummer1987recombination} at a temperature of 10\,000 K and an electron density of 100 cm$^{- 3}$. A detailed analysis justifying the assumption for considering 
$\rm (H\alpha/H\beta) = 2.86$ for reddening correction in AGNs was presented by \citet{2021MNRAS.tmp.2390A}. We found values of the extinction coefficient $A_{v}$ ranging from 0.02 to 0.4 mag.

 It is clear hitherto that the most widely used method for estimating the dust content is based on the relative strengths of the lower Balmer H I lines i.e. H$\alpha$/H$\beta$. When using the gas emission lines as a dust tracer, the canonical assumption is that the gas emission comes from the same position as the emission from the ionization source, which is usually not the case, because the gas in a spiral galaxy for instance is contained in a plane. Generally, little is known about the nature and real distribution of dust in AGNs \citep[e.g.][and references therein]{2016MNRAS.461.4227H, 2016ApJ...832....8B,
2017MNRAS.467..226G}.  However, neglecting these uncertainties at first order and considering the fact that H$\alpha$ and H$\beta$ are the strongest recombination lines of hydrogen in the optical spectrum, the reddening effect can be written on the reliance of the ratio H$\alpha$/H$\beta$. 
Assigning a single extinction value is just a rough first approximation, therefore, irrespective of the dereddening method used, it is good practice to verify that the H$\alpha$/H$\beta$, H$\gamma$/H$\beta$ and H$\delta$/H$\beta$ emission line ratios have the expected values in comparison with the theoretical Case B ratios of 2.86 or 3.1, 0.468 and 0.259 \citep{halpern1982xray, halpern1983ionization,  hummer1987recombination, ost06}, respectively. On the other hand, H7 and H8 can not be considered in reddening correction because they are usually blended with other emission lines. Additionally, the signal-to-noise ratio for lines
such as H9 and H10 is often too low to allow for their detection with high precision in order to constrain the reddening and underlying stellar absorption. Hence, as part of the emission-line flux processing, the underlying continuum and absorption-line spectra were fitted and eliminated, removing the effects of stellar Balmer absorption on the H$\alpha$ and H$\beta$ \citep[e.g.][]{2006ApJS..164...81M}. Finally,  the amount of interstellar extinction obtained using H$\alpha$/H$\beta$ is relatively in agreement with the amount of extinction determined considering other line ratios, i.e.
 [\ion{S}{ii}] $\lambda4071$/[\ion{S}{ii}] $\lambda10320$ \citep{wampler71, shields75, 1978ApJ...223...56K} or \ion{He}{ii} $\lambda3203$/\ion{He}{ii} $\lambda4686$ \citep{1981ApJ...250...55S}. 

From the resulting sample, we selected only the objects that present the emission-lines
[\ion{O}{ii}]$\lambda3726,\lambda3729$ (hereafter [\ion{O}{ii}]$\lambda3727$),
[\ion{O}{iii}]$\lambda4363$, \ion{He}{ii}$\lambda4686$,
H$\beta$, [\ion{O}{iii}]$\lambda5007$,  \ion{He}{i}$\lambda5876$,
H$\alpha$, and [\ion{S}{ii}]$\lambda6716,\lambda6731$
 with a signal/noise ratio (S/N) higher than 2.0.
The final sample resulted in 9 objects with redshift  
$z \: \la \: 0.2$. In Figure~\ref{fig1}, an example of a pure Seyfert~2 nebular spectrum (in black) 
and the fitting (in red) produced by the {\sc ifscube} software \citep{daniel_ruschel} are
shown.

In addition to the SDSS data, we compiled from the  literature
fluxes of emission-lines from Seyfert~2 nuclei  obtained by different authors. We applied the same selection criteria used 
for the SDSS data to these additional selected objects. This sample consists of 94 Seyfert~2 nuclei whose emission-line intensities were reddening corrected by the authors from  which the data were taken. However, in the cases where
the reddening correction was not performed in the original works, the same procedure applied to the SDSS data was considered.
Since several measurements for the emission lines compiled from the literature do not have their uncertainties listed
in the original papers where the data were compiled,   we
adopted a typical error of $10$ per cent for strong emission-lines (e.g. [\ion{O}{iii}]$\lambda$5007) and $20$ per cent for weak emission lines (i.e. [\ion{O}{iii}]$\lambda$4363, \ion{He}{ii}$\lambda4686$ and \ion{He}{i}$\lambda5876$), as derived, for instance, by \citet{1994ApJ...435..171K}.
 
Finally, we applied  the criterion 
proposed by \citet{2001ApJ...556..121K}, 
\begin{equation} 
\label{eq1}
\rm log([O\:III]\lambda5007/H\beta) \: > \: \frac{0.61}{log([N\:II]\lambda6584/H\alpha)-0.47}+1.19
\end{equation}
to separate SF and AGN objects. Additionally,  the  
criterion proposed by \citet{2010MNRAS.403.1036C} to  separate AGN-like and  Low-ionization nuclear emission-line region (LINER) objects, given by
\begin{equation} 
\label{eq5}
\rm log([O\:III]\lambda5007/H\beta) \: > \: 0.47+log([N\:II]\lambda6584/H\alpha)\times1.10,
\end{equation}
was also considered. 

The final sample consists of 65 Seyfert~2 with redshift $z \: \la 0.2$ whose  reddening-correction line intensities (in relation to H$\beta=1.0$) are listed 
in Table~\ref{table1}  in the Appendix.
In Fig.~\ref{fig2}, a diagnostic diagram log([\ion{O}{iii}]$\lambda 5007$/H$\beta$) versus log([\ion{N}{ii}]$\lambda 6584$/H$\beta$), the observational data for the  objects as well as the curve representing the criteria above are shown.
It  can be seen that the objects of our sample
 cover a large range of ionization degree
and metallicity, hence a wide range of [\ion{O}{iii}]/H$\beta$ 
and [\ion{N}{ii}]/H$\alpha$ line ratio intensities are noted (e.g. \citealt{2006MNRAS.371.1559G, 2016MNRAS.456.3354F, 2020MNRAS.492.5675C}).

\begin{figure}
 \begin{center}
  \includegraphics[angle=-90.0,width=\columnwidth]{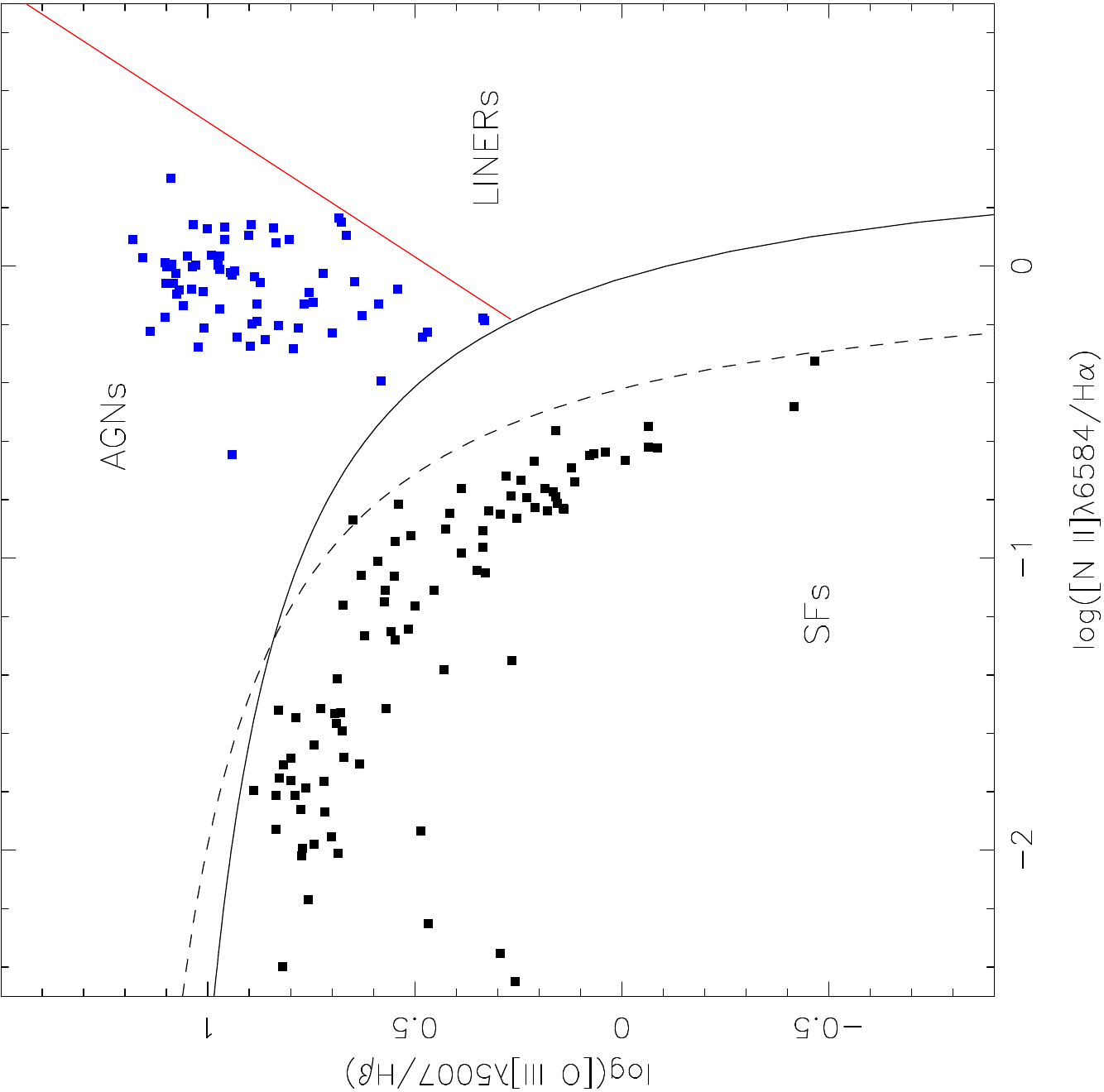}
 \end{center}
 \caption{Diagnostic diagram log([\ion{O}{iii}]$\lambda 5007$/H$\beta$]) versus log([\ion{N}{ii}]$\lambda 6584$/H$\alpha$]}). Blue points represent Seyfert~2 nuclei of our sample
 (see Sect.~\ref{sampleagn}) while  black points are for SFs (see Sect.~\ref{samplesf}).
 Solid and dashed curves represent  the criteria  
proposed by \citet{2001ApJ...556..121K} and \citet{2003MNRAS.346.1055K}, given by Equations~\ref{eq1}
and \ref{eq2a}, respectively, to separate SF and AGN objects. 
The red line represents the
criterion proposed by \citet{2010MNRAS.403.1036C} to  separate AGN-like and  Low-ionization nuclear emission-line region (LINER) objects, given by Eq.~\ref{eq5}.
\label{fig2}
 \end{figure}

 \begin{figure}
 \begin{center}
  \includegraphics[angle=-90.0,width=\columnwidth]{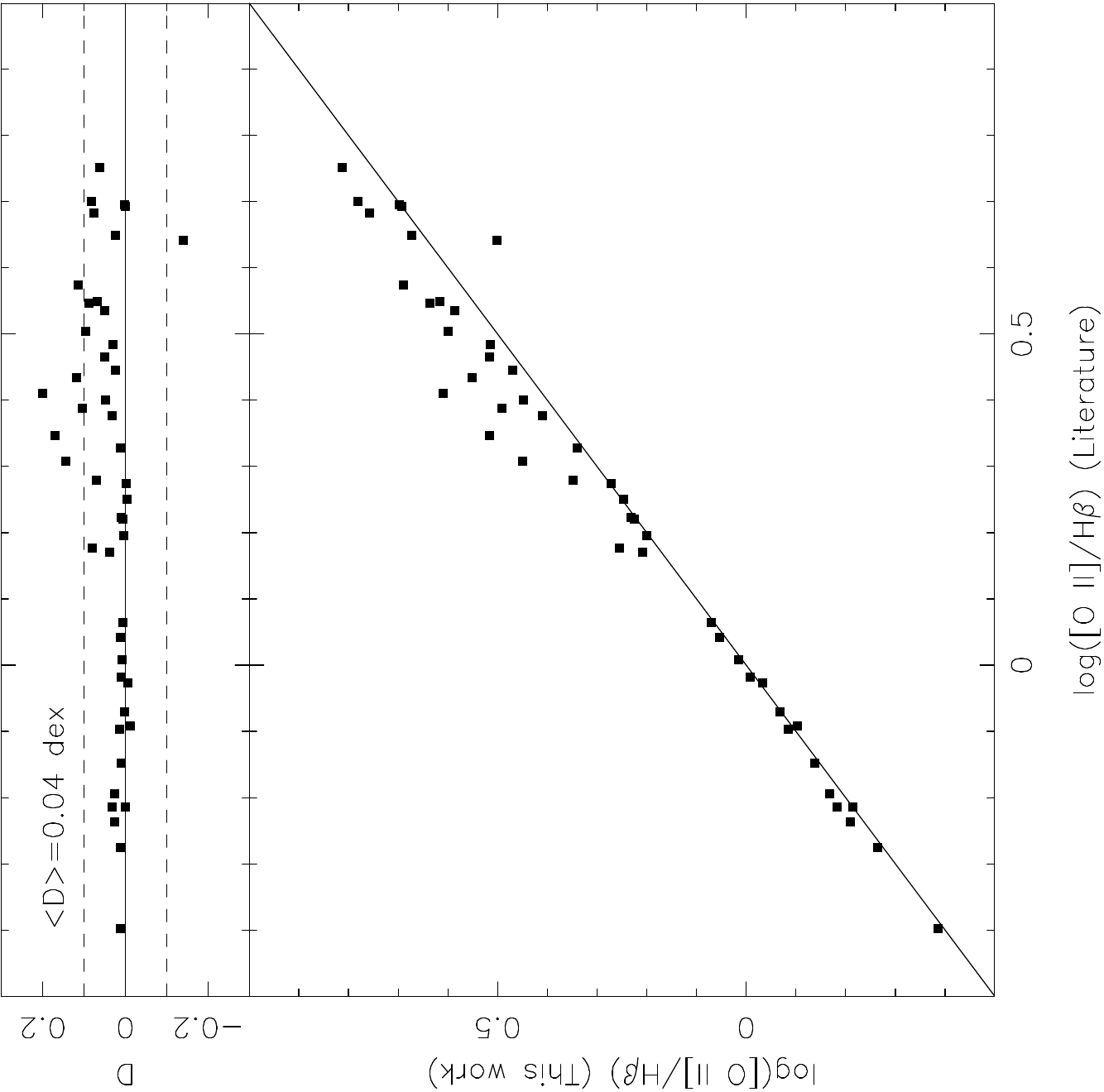}
 \end{center}
 \caption{Bottom panel: Comparison between the [\ion{O}{ii}]$\lambda$3727/H$\beta$ intensity line ratio, for part of the observational sample (AGNs and SFs), whose
 reddening correction were carried out following the methodology described
 in Sect.~\ref{obs} and performed by the authors from which the data were compiled.
 A one-to-one correlation is plotted as a solid line. Top panel: Difference (D=y-x)
 between the intensities. Solid line represent the null difference while dashed lines are the uncertainty of 0.1 dex in the line ratio measurements. The mean difference ($\rm <D>$) in the measurements is indicated.}
\label{figred}
 \end{figure}
 
\citet{2015MNRAS.453.4102D} and \citet{2017MNRAS.467.1507C} presented
a complete discussion on  the use of  heterogeneous sample
and its possible implications on abundance estimates.  Effects of aperture, electron density
variation along the AGN radius, X-Ray dominated regions, shock,
and electron temperature fluctuations in abundance determinations
have also been presented by \citet{2020MNRAS.492..468D, 2021MNRAS.501.1370D} and these are not repeated here.
Basically, the effects of these parameters on abundance
estimates produce uncertainties of $\sim0.1$ dex, i.e.
in order of those derived by applying the $T_{\rm e}$-method
(e.g. \citealt{2008ApJ...687..133I}) and
strong-line methods (e.g. \citealt{1998AJ....115..909S}).
Since it is not possible to estimate, for each AGN of our sample, the aforementioned uncertainty is not
considered in the resulting abundance values.

We present a comparison between the logarithm of the [\ion{O}{ii}]$\lambda$3727/H$\beta$ intensity line ratio   by adopting our
 extinction correction procedure  with that assumed by the authors from
 which the data were compiled. This comparison was possible only for
 46 objects (19 AGNs and 27 SFs) due to the fact that most of the original 
 works present only the reddening corrected intensity lines.
 In Fig.~\ref{figred}, bottom panel, this comparison is shown where there exist a good agreement between
them. In top panel of Fig.~\ref{figred}
 the difference (D=y-x) is presented, where one can see the uncertainty due the
 distinct approaches to extinction correction is in order of that produced
 by the error measurements ($\sim0.1$ dex, e.g. \citealt{2003ApJ...591..801K}), 
 with  the mean difference ($<\rm D>=0.04$ dex) being about null. Thus, distinct methodologies employed 
 for the extinction correction of  intensity line ratios  introduce a minimal uncertainty in our 
 abundance estimates.

\subsection{Star forming regions}
\label{samplesf}
We compiled  emission-line intensities
of SFs from the literature with the goal of comparing their estimates with our AGN results. We applied the same criteria above in the selection of SFs, i.e. we  selected only objects whose  emission-lines
 [\ion{O}{ii}]$\lambda3727$, [\ion{O}{iii}]$\lambda4363$, \ion{He}{ii}$\lambda4686$, H$\beta$, [\ion{O}{iii}]$\lambda5007$,  \ion{He}{i}$\lambda5876$, H$\alpha$, and [\ion{S}{ii}]$\lambda6716,\lambda6731$ were measured. The measurements of these lines make it possible to calculate the He and O abundances by using the $T_{\rm e}$-method following a similar methodology applied to the Seyfert~2 sample. Furthermore, we considered only  the objects which satisfy the empirical criterion proposed by \citet{2003MNRAS.346.1055K}
 
 \begin{equation} 
\label{eq2a}
\rm log([O\:III]\lambda5007/H\beta) \: < \: \frac{0.61}{log([N\:II]\lambda6584/H\alpha)-0.05}+1.3.
\end{equation}

 In Table~\ref{table1a}, the emission line intensities (in relation to H$\beta$=1) and the original works from which data  were taken  are listed. The
 data consist of reddening corrected emission lines of 85   \ion{H}{ii} regions and star-forming galaxies  with 
 redshift $z\: < \: 0.2$. In Fig.~\ref{fig2}, the
 SF sample is represented by black points, where we can see the very known sequence 
 (e.g. \citealt{1981PASP...93....5B, 2003MNRAS.346.1055K})
 formed by this object class.

\begin{figure}
 \begin{center}
  \includegraphics[angle=-90.0,width=\columnwidth]{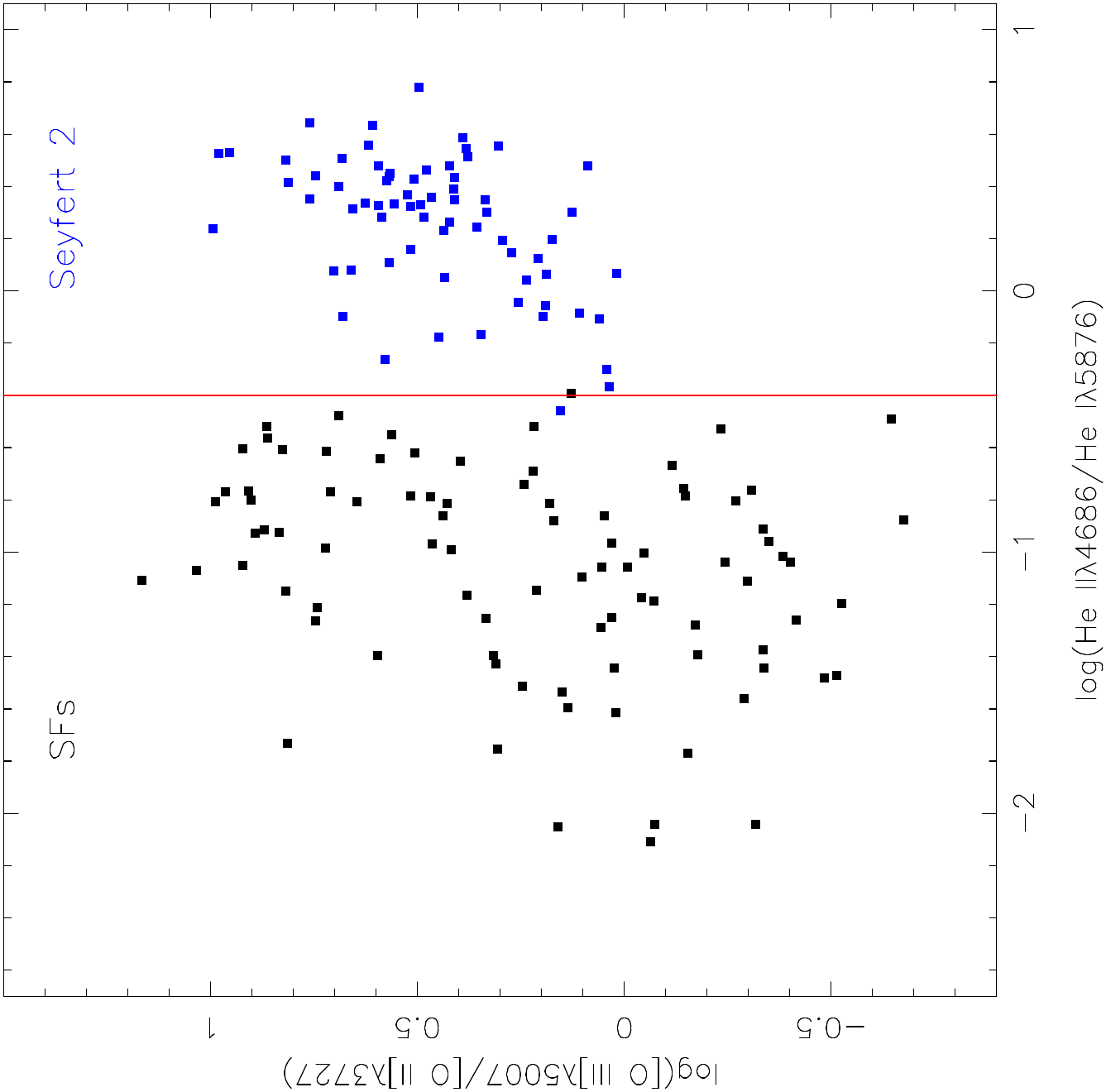}
 \end{center}
 \caption{Logarithm  of [\ion{O}{iii}]$\lambda$5007/[\ion{O}{ii}]$\lambda$3727
versus \ion{He}{ii}$\lambda$4686/\ion{He}{i}$\lambda$5876. Blue and black points represent Seyfert~2 nuclei (see Sect.~\ref{sampleagn}) and  SFs (see Sect.~\ref{samplesf}) of our samples, respectively, as indicated. Red line indicates the
separation criterion between Seyfert nuclei and SFs given by Eq.~\ref{eqsep}.}
 \label{fig3}
 \end{figure}

 In Fig.~\ref{fig3}, we plotted the logarithm of  [\ion{O}{iii}]$\lambda$5007/[\ion{O}{ii}]$\lambda$3727
versus \ion{He}{ii}$\lambda$4686/\ion{He}{i}$\lambda$5876 line ratios for the AGN sample (blue points) and for
the SF sample (black points).
Despite the scattering produced mainly for the SF data, it can be seen  the expected result that;   a clear correlation is derived (see also, for instance, \citealt{2000MNRAS.311..329D}) 
hence both line ratios are dependent on the ionization degree of the gas. Moreover, a distinction between Seyfert and SFs can be observed in Fig.~\ref{fig3} due to the high excitation of Seyfert in comparison with SFs. In Fig.~\ref{fig3}, we observe a clear 
separation criterion, where objects with
\begin{equation}
\label{eqsep}
\log \left(\frac{\ion{He}{ii}\lambda4686}{\ion{He}{i}\lambda5876} \right) \: \ga \: -0.4
\end{equation}
are classified as Seyfert
in this diagram, otherwise, as SF (see also \citealt{2022MNRAS.513.5134N}).
 
\section{Abundance determination}
\label{secabund}

We  calculated helium and oxygen abundances by using
the observational data  sample of Seyfert 2 nuclei 
and SFs described  in Sect.~\ref{obs}. Therefore,
the [\ion{O}{iii}]$(1.33\times\lambda5007)/\lambda4363$ and
[\ion{S}{ii}]$\lambda6716/\lambda6731$ line ratios were used to derive electron temperature and density, respectively,
thus obtaining the ionic abundances of these elements.
Thereafter, we used photoionization models built with the
\textsc{Cloudy} code  \citep{2013RMxAA..49..137F} to produce a correction for the presence of $\rm He^{0}$ in the calculation
of the total He. In what follows, a 
description of the methodology employed is presented for
both object classes.

\begin{table*} 
\centering 
\caption{Atomic dataset used for recombination and collisionally excited lines of selected  element ions.} 
\label{tatomic} 
\begin{tabular}{lccc} 
\hline 
& Transition probabilities &  \\ Ion & and energy levels & Collisional strengths &  Recombination coefficients \\ 
\hline 
He$^{+}$  & \citet{2012MNRAS.425L..28P} & --- & \citet{2012MNRAS.425L..28P} \\ 
He$^{2+}$ & \citet{1995MNRAS.272...41S} & ---  & \citet{1995MNRAS.272...41S} \\
O$^{+}$   & \citet{1996atpc.book.....W} & \citet{2009MNRAS.397..903K} & --- \\ 
O$^{2+}$  & \citet{2004ADNDT..87....1F}, \citet{2000MNRAS.312..813S} & \citet{2014MNRAS.441.3028S} & --- \\
\hline 
\end{tabular} 
\end{table*}

\subsection{Seyfert}
\subsubsection{Temperature and density \label{secTeNe}}

To determine the abundance of an element it is necessary to estimate a representative value for the electron temperature and density of the gas phase occupied by the ions of this element.

Since the elements of interest are the He and O, which are
in most part once and twice ionized, following \citet{2021MNRAS.505.3624V}, we adopted the two zone models, where the temperature for the high ionization zone, defined as $T_{\rm high}$ is derived from the relation between this parameter and the  [\ion{O}{iii}]($1.33 \times \lambda5007$)/$\lambda4363$ line ratio,
where we assumed, as usual, [\ion{O}{iii}]$(\lambda4959)/(\lambda5007)$  to be
0.33\footnote{For a detailed discussion on the value of
this ratio see \citet{2000MNRAS.312..813S}.} \citep{1985Msngr..39...15R}. To calculate $T_{\rm high}$, 
and the electron density ($N_{\rm e}$) from the $[\ion{S}{ii}]\lambda 6716/\lambda 6731$ line ratio,  
we used the version 1.1.13 of {\sc PyNeb} code \citep{Luridiana2015}, which permits an interactive procedure in the derivation of these parameters. We assumed the atomic data of recombination and collisionally excited lines
listed in Table~\ref{tatomic} for the temperature, density
and ionic abundance calculations.

For almost all the data in our sample it was not possible to obtain the direct determination of
$T_{\rm low}$  due to the lack of the   [\ion{N}{ii}]$\lambda$5755  and [\ion{O}{ii}]$\lambda$7319, $\lambda$7330 auroral line measurements. Thus, for consistence, we used the 
 theoretical relation between $T_{\rm low}$-$T_{\rm high}$ proposed by \citet{2020MNRAS.496.3209D}, obtained through a grid of photoionzation models simulating Narrow Line Regions (NLRs) of AGNs  built with the {\sc Cloudy} code  \citep{2013RMxAA..49..137F} by \citet{2020MNRAS.492.5675C}, given by 
 \begin{equation}
\label{t2t3new}
t_{\mathrm{low}}=({\rm a} \times t_{\mathrm{high}}^{3})+({\rm b} \times t_{\mathrm{high}}^{2})+({\rm c} \times t_{\mathrm{high}})+{\rm d},
\end{equation}
where $\rm a=0.17$, $\rm b=-1.07$, $\rm c=2.07$ and  $\rm d=-0.33$,
while $t_{\mathrm{low}}$ and $t_{\mathrm{high}}$ represent $T_{\rm low}$ and $T_{\rm high}$, respectively, in units of $10^{4}$ K. These temperatures
 predicted by the models correspond to the mean
temperature for $\rm O^{\rm +}$ ($T_{\rm low}$) and $\rm O^{2+}$
($T_{\rm high}$) over the nebular AGN radius times the
electron density.

In \citet{2021MNRAS.501L..54R}, the relation represented by the Eq.~\ref{t2t3new} was compared with electron temperature values derived from observational auroral emission lines for a sample
of 12 local AGNs, and a good agreement was found between them. However, as pointed out by these authors, some cautions must be taken into account in the use of Eq.~\ref{t2t3new} for AGNs with strong outflowing gas.
 
\subsubsection{Helium abundance \label{secHe}} 
 
The total helium abundance in relation to hydrogen
$y=N(\mathrm{He})/N(\mathrm{H})$ was considered to be
\begin{equation}
    y=y^{0}+y^{+}+y^{2+},
\end{equation}
 where $y^{0}= N(\mathrm{He^{0}})/N(\mathrm{H^{+}+H^{0}})$,  $y^{+}= N(\mathrm{He^{+}})/N(\mathrm{H^{+}})$
 and $y^{2+}= N(\mathrm{He^{2+}})/N(\mathrm{H^{+}})$. The ionic abundance ratio $y^{+}$ and $y^{2+}$
 were derived from the \ion{He}{i}$\lambda$5876/H$\beta$ and \ion{He}{ii}$\lambda$4696/H$\beta$ intensity
 line ratios, respectively. We used the two-zone ionization model (e.g. \citealt{2019ApJ...876...98V}), which $T_{\rm low}$ (from Eq.~\ref{t2t3new})
 and  $T_{\rm high}$ (from direct estimation) were assumed in the $y^{+}$ and $y^{2+}$ calculation, respectively. Recently, \citet{2021ApJ...922..170B} proposed a more
 detailed ionization model for \ion{H}{ii} regions, i.e. the 4-zones model, where the He$^{+}$ is
 located in the low, intermediate, and high zones, i.e. 
 in the zones occupied for the $\rm O^{+}$ and 
 $\rm O^{2+}$ ions. Thus, in principle,   $T_{\rm low}$ can be assumed 
 as a representative temperature value for the region occupied by the He$^{+}$.
 It is worth noting that
 this is not a conventional approach, therefore some previous studies assume $T_{\rm high}$ in the
 He$^{+}$ abundance calculations.
 Concerning He$^{2+}$, \citet{2021ApJ...922..170B} showed the it  is located mainly in the  very-high-ionization zone. To derive the temperature of the very-high-ionization zone  the measurement of the [\ion{Ne}{iii}]$(\lambda3342/\lambda3868)$ 
 ratio is  required, which
 is not available in our data sample. Therefore, we assumed
 $T_{\rm high}$ in the abundance calculation of this ion.
 Both ionic abundances 
 were computed using  the {\sc PyNeb} code
 \citep{Luridiana2015}. We consider
 the presence of $\rm He^{0}$ in the neutral and ionized  gas since this is
 expected in low ionization objects \citep{1986PASP...98.1061P, 1986ApJ...311...45D}.
 
 The fraction of $y^{0}$ in \ion{H}{ii} regions
 and Planetary Nebulae has been estimated to be
 in order of $\sim 3$ per cent of the total helium 
 (e.g. \citealt{1986PASP...98.1061P, 1992RMxAA..24..155P}) or even larger values could be predicted (e.g. \citealt{2000MNRAS.311..329D}). In fact, \citet{2022MNRAS.510.4436M} found that only 9/42 Galactic \ion{H}{ii} regions have negligible
 contributions of $y^{0}$ (see also \citealt{2014MNRAS.440..536D}).
 Since AGNs present  a large amount of  molecular gas (e.g., \citealt{2004A&A...425..457R, 2005ApJ...633..105D, 2005MNRAS.364.1041R, 2013MNRAS.428.2389M, 2013MNRAS.430.2002R, rogemar18, 2021MNRAS.504.3265R,rama19, 2019MNRAS.486.3228R,  jarvis20,alonso-herrero20}) and have similar ionization parameter ($U$) other than  SFs  \citep{2019MNRAS.489.2652P},  a 
 higher abundance of neutral and molecular gas are  expected in AGNs as  compared to \ion{H}{ii} regions, which favors the presence of larger contribution of $y^{0}$ in the
 total helium abundance.  In fact, radio observations  have shown the existence of a gas neutral reservoir
 in the central parts of galaxies containing 
 AGNs (e.g. \citealt{1982ApJ...259...55D, 1987AJ.....93....6H,  2007A&A...470..571B, 2008ApJS..177..103H, 2018ApJ...861...50B, 2019MNRAS.482.5694E, 2019A&A...623A..79C}) which can coexist with the ionized gas.
 \citet{2014A&A...567A.125G}, who used the Atacama Large Millimeter Array (ALMA) to map the emission of dense molecular gas in the Seyfert~2 NGC\,1068,
 showed an overlay of the CO(3–2) (a tracer of $\rm H_{2}$)  intensity contours on the Pa$\alpha$ emission obtained with the Hubble Space Telescope.  In addition, recently, \citet{2022MNRAS.513..807D} derived the O/H abundance in NLR of
 108 Seyfert nuclei and found that these values are lower (by 0.16 to 0.30 dex) than those inferred by the  radial gradients along their galaxy discs and  those from a matched control sample of no active galaxies.  This discrepancy could be due to the accretion of a  metal-poor gas to the AGN that feeds the nuclear supermassive black hole (SMBH), coming from the
 neutral reservoir.
 
  To estimate the abundance of $\rm y^{0}$, we used 
 a grid of photoionization models  similar to that of \citet{2020MNRAS.492.5675C}, built
   with the {\sc Cloudy} code version 17.00 \citep{2013RMxAA..49..137F}.  
 These models were also used to derive neon and argon ICFs by \citet{2021MNRAS.tmp.2390A} and \citet{2021MNRAS.508.3023M}, respectively,  considering a wide range of nebular parameters:
 \begin{itemize}
     \item Metallicity: ($Z/\rm Z_{\odot})=3.0, 2.0, 1.0, 0.75, 0.5$, and $0.2$.
     \item Electron density: $N_{\rm e}\: \rm (cm^{-3})=3000, 1500, 500, 100$.
     \item Ionization parameter ($U$): $\log U$  ranging from $-1.5$ to $-3.5$, with step of 0.5 dex.
     \item Spectra Electron Distribution (SED): The SED is parametrized by the  continuum between 2 keV and 2500\AA\ \citep{1979ApJ...234L...9T} and it is 
 described by a power law with a spectral index $\alpha_{ox}$=$-$0.8, $-$1.1 and $-1.4$.
 \end{itemize}
 
 The  outermost nebular radius
 was considered by \citet{2020MNRAS.492.5675C}  to be the one where the electron temperature reaches 4000 K (the default value of the 
 {\sc Cloudy} code), which produces  a region
 with hydrogen almost completely ionized. Since we are also interested in the neutral region, we rerun  the
 grid of models by \citet{2020MNRAS.492.5675C} but using the 
 version 17.02 of the {\sc Cloudy} code \citep{2013RMxAA..49..137F} and
 assuming the outermost radius where the electron temperature
 reaches 1000 K; as was done by \citet{2012MNRAS.422..252D}, who analysed the  dominant excitation mechanism of 
 [\ion{Fe}{ii}] and $\rm H_{2}$ emission
lines in AGNs. These new models allow us to consider a larger
part of the neutral gas instead of stopping the calculations
at outermost radius of the modelled nebula  where the temperature falls below  4000 K.
We added cosmic-ray background emission  as a second ionizing source with a value of  $\rm H_{2}$
ionization rate of $10^{-15}\: \rm  s^{-1}$, which is about the same rate found by
\citet{2003Natur.422..500M} for a Galactic line of sight.
The cosmic-ray background emission has little influence on 
the formation of emission lines located in the ionized gas and/or
in the ionization of elements with 
ionization potential higher than the hydrogen one.
 In the models the abundance of each ion was considered as the
average over the nebular radius times the electron density.
For a detailed description of the photoionization models 
see \citet{2020MNRAS.492.5675C}.
 
\begin{figure}
\begin{center}
\includegraphics[angle=-90.0,width=\columnwidth]{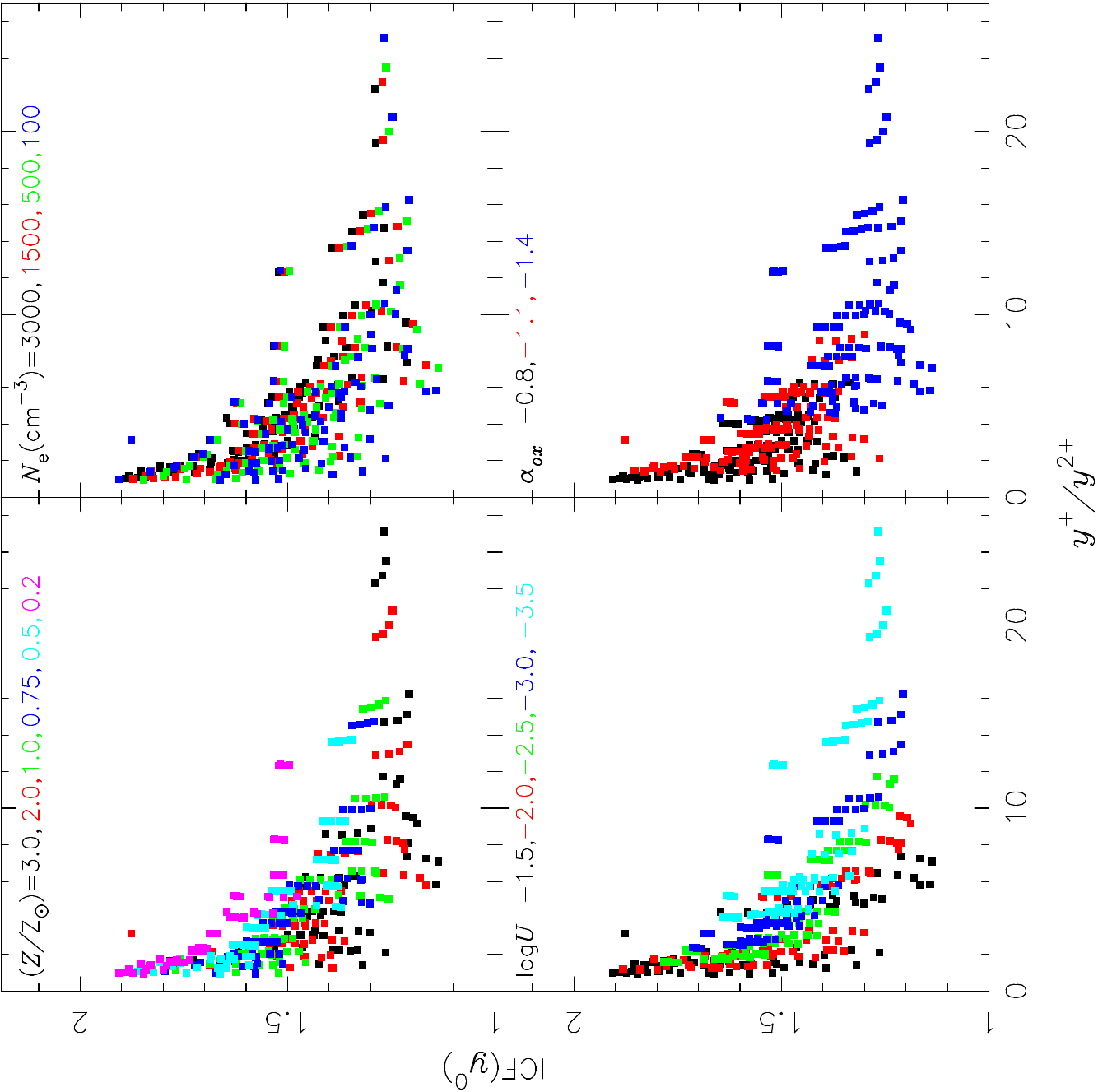}
\end{center}
\caption{Photoionization model  predictions (see Sect~\ref{secHe})
for $\mathrm{ ICF}(y^{0})$ versus $y^{+}/y^{2+}$. In each plot,
model results assuming distinct nebular parameters are shown by different colours, as indicated. $\mathrm{ICF}(y^{0})$ is defined 
as in Eq.~\ref{deficf}.}
\label{figicfa}
\end{figure}

Based on the model results, we assumed the ICF for the  $y^{0}$ to be
\begin{equation}
\label{deficf}
{\mathrm{ ICF}(y^{0})}= \frac{y}{y^{+}+y^{2+}}.
\end{equation}
In Fig.~\ref{figicfa}, the  photoionization model predictions
for $\mathrm{ICF}(y^{0})$ versus $\mathrm{ x}=y^{+}/y^{2+}$ are shown, where
the results are discriminated according to distinct nebular parameters, as indicated. Any dependence between the $\mathrm{ ICF}(y^{0})$-x relation and $N_{\rm e}$ and  $Z$ can be seen.  However, a clear
dependence of this relation with the ionization parameter $U$ and $\alpha_{ox}$ is noted, in the sense that a
 stepper  $\mathrm{ ICF}(y^{0})$-x relation is derived for 
 higher $U$ and $\alpha_{ox}$ values. The dependence of  this relation on these parameters is due to  the increase in the gas ionization degree, i.e. driven by $U$ and the hardness of the SED ($\alpha_{ox}$),  produces lower values of x.
 
 Estimations of  $\alpha_{ox}$ require observations of integrated nuclear flux at optical and UV wavelengths (e.g. \citealt{1987ApJ...313..596W, 1999ApJ...516..672H, 2011ApJ...726...20M, 2019MNRAS.482.2016Z, 2021MNRAS.505.1954Z}), and values of this parameter are often times unavailable for AGNs.
 Moreover, detailed photoionization models   by \citet{2017MNRAS.468L.113D}
 and Bayesian-like
approach by \citet{2019MNRAS.489.2652P}   have predicted  $\alpha_{ox}$ to be higher than $-1.2$ for Seyfert 2 nuclei. Conversely, observational estimations  of $\alpha_{ox}$  by \citet{2011ApJ...726...20M}  indicate that most AGNs have  $\alpha_{ox}$ in the order of  $-$1.4 and even lower values 
($\sim -2.0$) are derived for
these objects. This discrepancy is probably  due to an extra physical process is missing
in the models, probably shocks (e.g. \citealt{2019MNRAS.488.4487C}).   Therefore, as a result of few AGNs estimations of $\alpha_{ox}$ in comparison to those with emission-line measurements, and also due to the existence of the discrepancy between model predictions and observational estimations of $\alpha_{ox}$, this parameter was not directly taken into account in the expression for the helium ICF derived here. In any case, the gas ionization degree, traced by $U$, probably has a direct relation with $\alpha_{ox}$.

\begin{figure}
\begin{center}
\includegraphics[angle=0.0,width=\columnwidth]{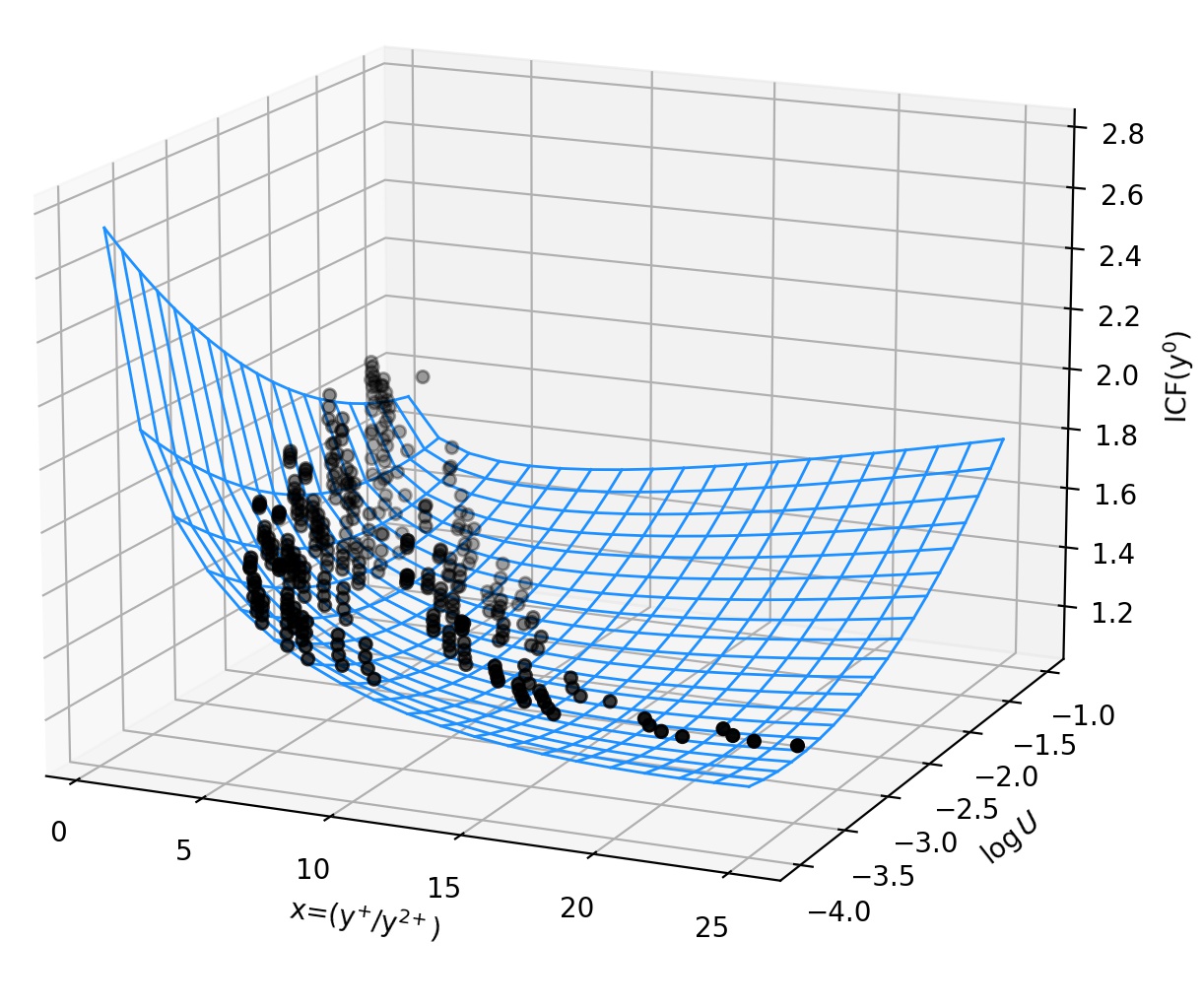}
\end{center}
\caption{Bi-parametric calibration among the 
$\mathrm{ ICF}(y^{0})$, $\mathrm{ x}= y^{+}/y^{2+}$ and
logarithm of the ionization parameter ($\log U$).
 Points represent the photoionization model results (see Sect~\ref{secHe}), 
 $\mathrm{ICF}(y^{0})$ is defined by Eq.~\ref{deficf}.
 The surface is given by Eq.~\ref{fitbi}.}
\label{figicfb}
\end{figure}

We derive a bi-parametric calibration 
$\mathrm{ICF}(y^{0})=\mathrm{f(x,} \log U)$ shown in 
Fig.~\ref{figicfb} and given by
\begin{equation}
\label{fitbi}
\begin{split}
\mathrm{ z =  [ (0.14\pm0.01)\times {\rm ln(x)} \times w ] + 
[(0.09\pm0.01)\times w{^2}] } \\
+ \mathrm{ [(0.03\pm0.01) \times \rm x] + [(0.18\pm0.07)\times \rm w] +  [(1.71\pm0.08)]},  
\end{split}
\end{equation}
where $\mathrm{ z=ICF}(y^{0})$, $\mathrm{x}= y^{+}/y^{2+}$
and $\mathrm{w}=\log U$.   The ionization parameter can be obtained using the semi-empirical calibration  proposed by \citet{2020MNRAS.492.5675C}
\begin{equation}
\label{eq2}
\log U=(0.57\pm 0.01 \: r^{2})+(1.38\pm 0.01 \: r) - (3.14\pm 0.01),
\end{equation}
where $r$=log([\ion{O}{iii}]$\lambda$5007/[\ion{O}{ii}]$\lambda$3727). This calibration was obtained through a comparison of observational optical narrow emission line ratios of a sample of local Seyfert 2 nuclei with those predicted by photoionization models. 
 An expression for deriving the helium ICF taking into account the ionization parameter but for \ion{H}{ii} regions was also derived by \citet{2002A&A...381..361S}, by using a grid of photoionization models.

\subsubsection{Oxygen abundance \label{secOx}}
 
The total oxygen abundance in relation to hydrogen
 was derived assuming
\begin{equation}
\label{eqt6}
\frac{N(\mathrm{O})}{N(\mathrm{H})}=\mathrm{ICF(O)}\: \times \: \left[\frac{N(\mathrm{O^{2+}})}{N(\mathrm{H^{+}})}+\frac{N(\mathrm{O^{+}})}{N(\mathrm{H^{+}})}\right], 
\end{equation}
where  ICF(O) is the Ionization Correction Factor for the oxygen which takes into account the contribution of  unobserved oxygen ions (e.g. $\rm O^{3+})$.
In fact, several studies  have found a strong intensity of the [\ion{O}{iv}]$\lambda$25.89$\mu$m
 emission line in AGN spectra (e.g. \citealt{2012ApJ...746..168D, 2016ApJS..226...19F}),
 indicating  a non-negligible fraction of the  $\rm O^{3+}$ abundance.

To calculate the ICF(O) the empirical expression proposed by \citet{1977RMxAA...2..181T}:
\begin{equation}
\label{icfox}
{\rm ICF(O)}=\frac{y^{+}+y^{2+}}{y^{+}},
\end{equation}
 used in SF (e.g. \citealt{2006A&A...448..955I, 2021MNRAS.505.3624V}) and
AGN (e.g. \citealt{2020MNRAS.496.2191F, 2020MNRAS.496.3209D}) abundances studies, was considered.
To calculate the ICF(O) for each object of the sample,  we consider  our helium abundance results derived using the {\sc PyNeb} code \citep{Luridiana2015}.

In Table~\ref{table2} presented in the Appendix, we show the  oxygen  (ionic and total)  and helium abundances. To estimate the abundance uncertainties, due to few objects present
observational error in the line measurements, for all objects of the AGN sample,  we use a Monte Carlo simulations. For each diagnostic line, we generate 1000 random values assuming a Gaussian distribution with a standard deviation equal to the associated uncertainty of the line intensity involved in the diagnostic.  

\subsection{Star-forming regions}
\label{sfsec}

To determine the He and O abundances  for the SFs,
we adopted a similar procedure considered for the Seyfert sample.
The ionic helium abundances were also computed using  the {\sc PyNeb} code \citep{Luridiana2015} which,  $y^{2+}$ and $\mathrm{O^{2+}}$ 
were derived assuming 
the $T_{\rm high}$  and 
$N_{\rm e}$, calculated for each object through the [\ion{O}{iii}]($1.33 \times \lambda5007$)/$\lambda4363$ and $[\ion{S}{ii}]\lambda 6716/\lambda 6731$ line ratios,  respectively. 

The $y^{+}$ and $\mathrm{O^{+}}$ were calculated assuming the $T_{\rm low}$  derived from the theoretical relation  
\begin{equation}
t_{\mathrm{low}}^{-1}\,=\,0.693\,t_{\mathrm{high}}^{-1}+0.281
\end{equation}
proposed by \citet{2003MNRAS.346..105P}. We assume $T_{\rm low} \equiv T_{\rm e}(\ion{N}{ii}) \approx T_{\rm e}(\ion{O}{ii})$.

The total oxygen abundance is obtained as described
previously:
\begin{equation}
    \mathrm{\frac{O}{H}=ICF(O)\times\frac{O^{+}}{H^{+}}+\frac{O^{2+}}{H^{+}}}, 
\end{equation}
where ICF(O) is same as Eq.~\ref{icfox}.

In the same way, we assume the expression
\begin{equation}
    y=\mathrm{ICF}(y^{0}) \times (y^{+}+y^{2+}),
\end{equation}
in the total helium abundance calculation. The $\mathrm{ICF}(y^{0})$  expression (Eq.~\ref{deficf}) proposed for 
Seyferts can not be applied to SFs due to 
the distinct ionization structure of these objects. Thus,
we build a grid of photoionization models simulating SFs
by using the \textsc{Cloudy} code. The nebular
parameters are similar to those adopted by \citet{2018MNRAS.479.2294D}
and they are summarised in what follows. 
\begin{itemize}
     \item Metallicity: $(Z/\mathrm{ Z_{\odot}})=0.03,$ 0.2, 0.5 and 1.0. This metallicity range includes values derived from the
     disk \ion{H}{ii} regions (e.g. \citealt{2020ApJ...893...96B}) and in XPMs (e.g. \citealt{1999ApJ...527..757I}).
     \item Electron density: $N_{\rm e}\: \rm (cm^{-3})= 100,$  500 and 1\,000, the same range
     derived by \citet{2013MNRAS.430.2605Z} for SFs
     and calculated from $[\ion{S}{ii}]\lambda6716/\lambda6731$ line ratio.
     \item Ionization parameter ($U$): $\log U$  ranging from $-1.5$ to $-3.5$, with step of 0.5 dex. Similar range of $\log U$ was
     derived by \citet{2011MNRAS.415.3616D} for local
     \ion{H}{ii} regions (see also \citealt{2022A&A...659A.112J, 2019MNRAS.483.1901Z, 2014MNRAS.441.2663P}).
     \item SED:  synthetic spectra of stellar
clusters formed by an instantaneous burst and with 
ages equal to 0.01, 1.0, 2.5, 4.0 and 5.0 Myr, built with the \textsc{Starburst99} \citep{1999ApJS..123....3L} were
considered as ionizing source. The value 0.01 Myr corresponds to the
lowest age considered in the \textsc{Starburst99}.
In the interval 2.5 -- 4.0 Myr  OB stars become Wolf-Rayet stars  producing an increase in the number of ionizing photons (e.g. \citealt{1995ApJS...96....9L, 2006ApJS..167..177D, 2016MNRAS.460.1739V}).
Nebulae older than about 5 Myr are  difficult
to observe because their original massive stars have cooled emitting little ionizing photons
(e.g., \citealt{1995A&AS..112...35G, 1996AJ....111.1252M}).
We  assumed the WM-basic stellar atmosphere models by \citet{2001A&A...375..161P}, and the GENEVA tracks with stellar rotation
\citep{2012ApJ...751...67L}.
 \end{itemize}

\begin{figure}
\begin{center}
\includegraphics[angle=-90.0,width=\columnwidth]{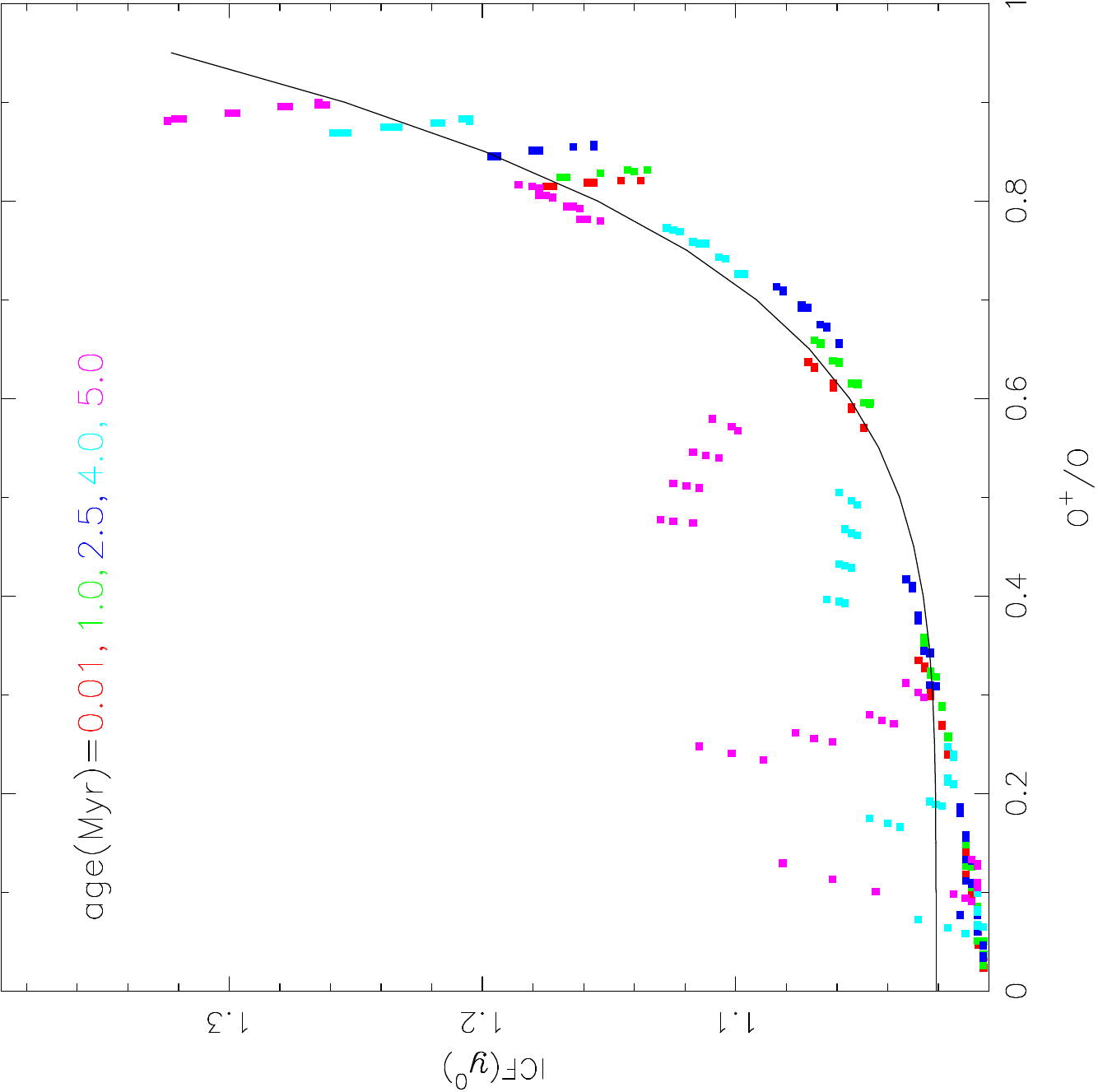}
\end{center}
\caption{Calibration between the 
$\mathrm{ ICF}(y^{0})$ and $(\rm O^{+}/O)$.
 Points represent  SF photoionization model results (see Sect~\ref{sfsec}) assuming different ages for the ionizing
 source, as indicated.  
 $\mathrm{ICF}(y^{0})$ and the curve are defined by Eq.~\ref{eqicfsf}.}
\label{icfsf}
\end{figure}

For these models the calculations stopped
at outermost nebular radius where the temperature falls below  4\,000 K. We note that the photoionization model results predict
a very low $\rm He^{2+}$ abundance in comparison
to those derived from our observational data sample.
This  indicates that real SEDs of the ionizing source of SFs are harder than the theoretical ones assumed here.
The  Binary Population and Spectral Synthesis code (\textsc{bpass}, \citealt{2018MNRAS.479...75S})
produces somewhat harder SEDs than the \textsc{Starburt99}
(e.g. \citealt{2019ApJ...878....2D}), however, the former might be more
appropriate for moderate/low metallicity SFs and less appropriate for high metallicity systems \citep{2022MNRAS.513.5134N}. Due to the problem pointed out above, we derive
a relation between the ICF($\rm He^{0}$) and the $\rm O^{+}/O$ abundance ratio instead of $y^{+}/y^{2+}$
as previously. In Fig.~\ref{icfsf}, the model results for
 the relation ICF($\rm He^{0}$)-($\rm O^{+}/O$) are shown, where models
assuming stellar clusters with distinct ages are indicated.
It can be seen  that, for $(\rm O^{+}/O) \: \la \: 0.6$, higher ICFs are obtained  for older models
(4 and 5 Myr). We did not find any dependence of the ICF($\rm He^{0}$)-($\rm O^{+}/O$) relation with other nebular parameters.
\citet{1981Ap&SS..80..267D} and
\citet{1986A&A...156..111C} proposed that the equivalent width of H$\beta$, the relative volume of $\rm He^{+}$ and $\rm H^{+}$ zones and
the [\ion{O}{iii}]($\lambda4959,\lambda5007$)/H$\beta$ line ratio are  good age \ion{H}{ii} region indicators.
However, the change of the value of these indicators  along the spiral disk  galaxies can be due to effects of 
variation of the hottest effective temperature of 
ionizing stars \citep{1976ApJ...203...66S} rather that evolution effects.
Since our SF sample is mostly composed of disk
\ion{H}{ii} regions, the age estimation, in principle, is not
correct and it was not carried out. Thus, a fit taken into account all the points in Fig.~\ref{icfsf} results in
\begin{equation}
\label{eqicfsf}
 \mathrm{ICF}(y^{0})=\mathrm{a\times (O^{+}/O)^{b}+c}, 
\end{equation}
where $\mathrm{a=0.385\pm0.017}$, $\mathrm{b=4.742\pm0.275}$
and $\mathrm{c=1.020\pm0.002}$.

The abundance uncertainties for this sample were
derived taken into account the error in the observational emission-line intensities published by the authors which the data were compiled.

\section{Results and Discussion}
\label{res}
\subsection{Density and temperature}

The electron density  values derived from the
$[\ion{S}{ii}]\lambda 6716/\lambda 6731$ line ratio for the Seyfert~2  sample  are in the range of 
$20 \: \la \: N_{\rm e}(\rm cm^{-3}) \: \la \: 1200$, with a median value of $\sim 500$ $\rm cm^{-3}$, which is in agreement with the estimation
obtained by \citet{2014MNRAS.443.1291D}. However, the derived mean value from our AGN sample is a factor of about two lower than the one derived by \citet{2012MNRAS.427.1266V}, who 
used optical spectra of $\sim 2\,300$ AGNs  obtained with the SDSS-DR7 \citep{2000AJ....120.1579Y}. Thus, the discrepancy between our 
mean electron density result and the one derived by \citet{2012MNRAS.427.1266V} is probably  due to the distinct sample of objects.
In any case, the $N_{\rm e}$ values are lower than the critical density for
the emission lines involved   in this work
(see, for instance, \citealt{2012MNRAS.427.1266V}), therefore, de-excitation effect has no influence on our abundance calculations (e.g. \citealt{ost06}).  

\begin{figure*}
\begin{center}
\includegraphics[scale=0.63, angle=-90]{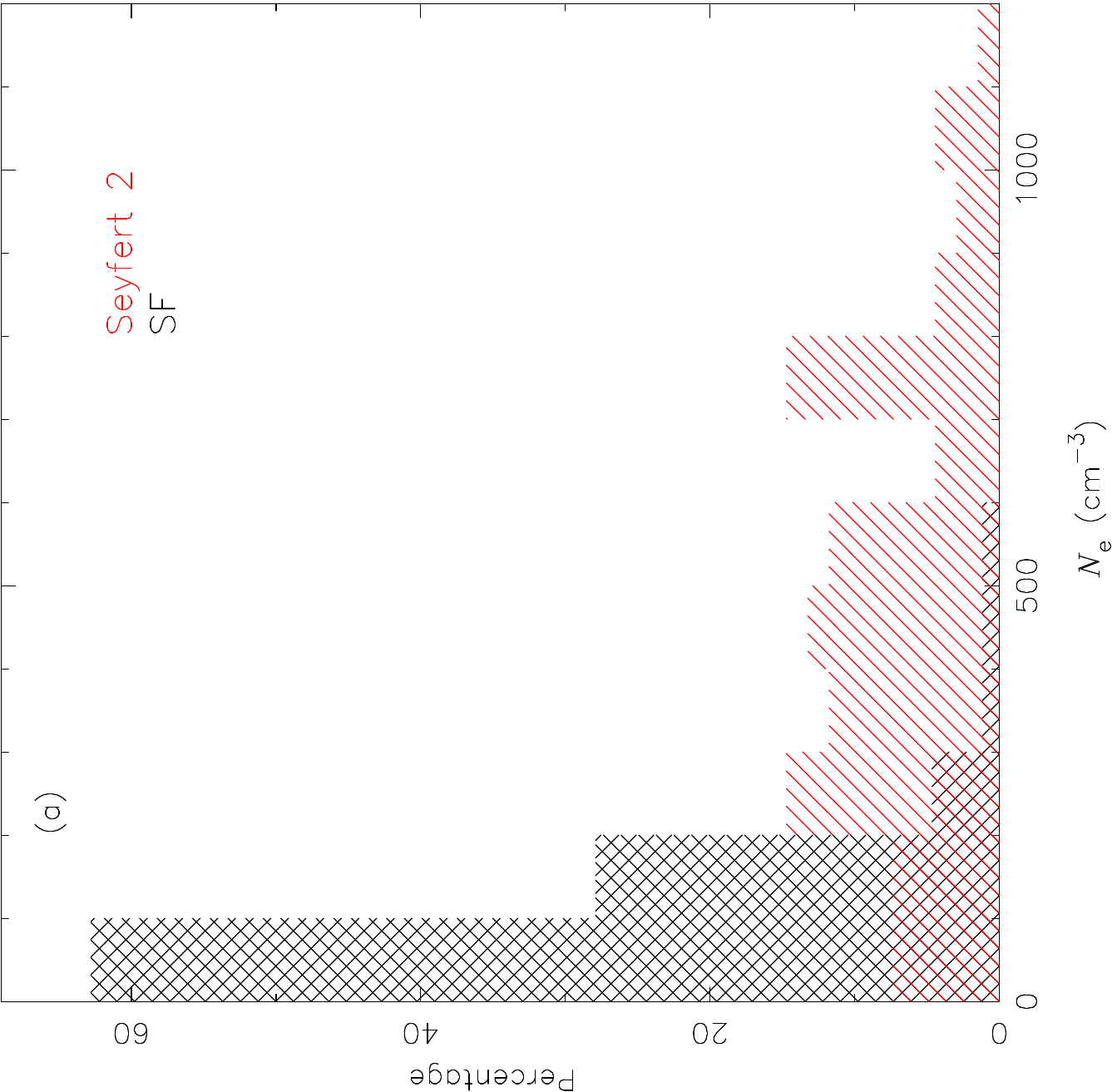}
\includegraphics[scale=0.63, angle=-90]{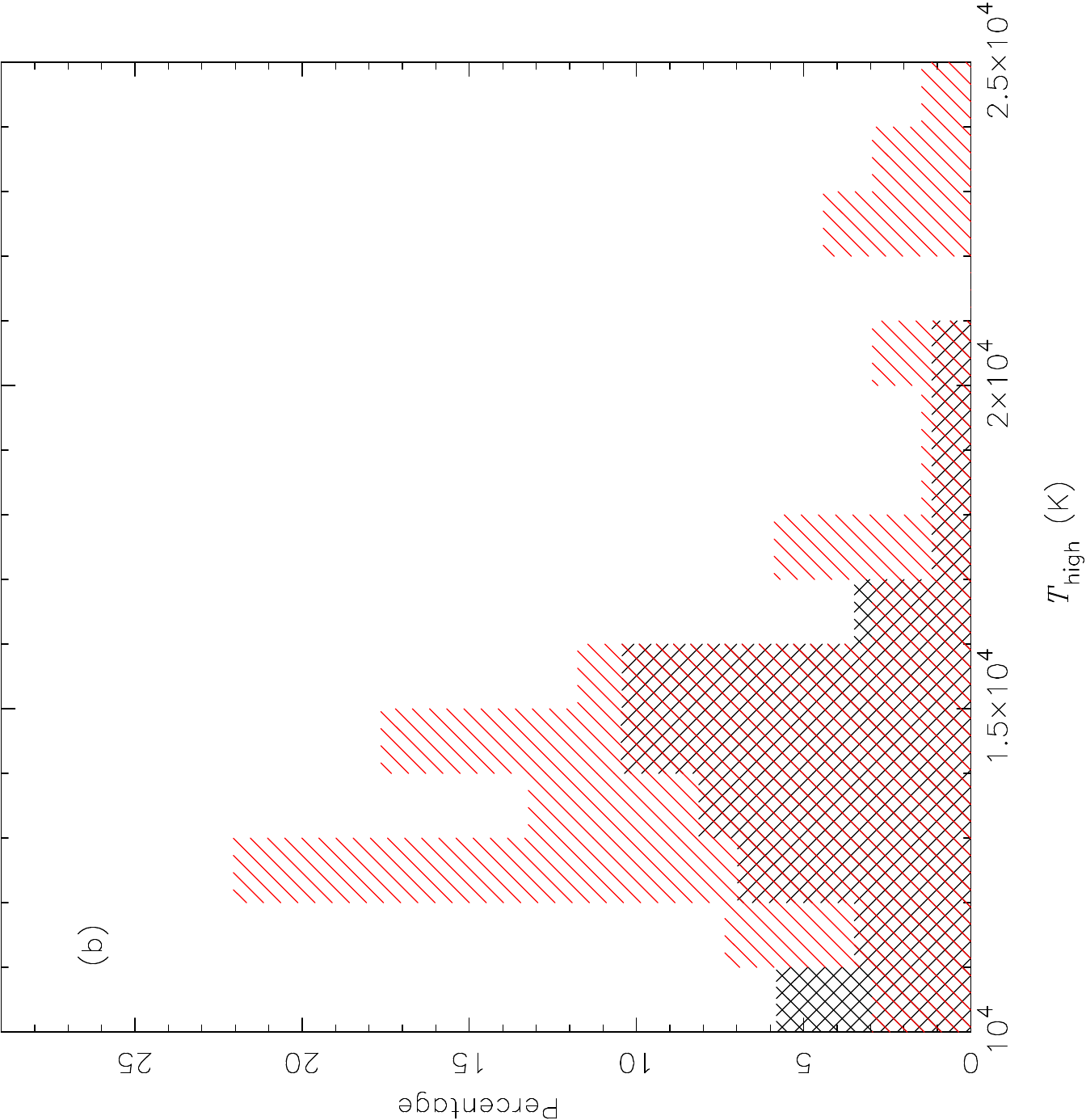}
\end{center}
\caption{Panel (a): Electron density distributions, calculated from
[\ion{S}{ii}]$\lambda6716/\lambda6731$ line ratio  (see Sect.~\ref{secTeNe})
 by using the {\sc PyNeb} \citep{Luridiana2015}, for
 distinct samples of objects (see Sect.~\ref{obs}), as indicated. Panel (b):
As left panel but for $T_{\rm high}$ calculated through the observational values of the [\ion{O}{iii}]($1.33 \times \lambda5007$)/$\lambda4363$ line ratio.}
\label{fh1}
\end{figure*}

 The values derived for the SF sample are in the range of $20 \: \la \: N_{\rm e}(\rm cm^{-3}) \: \la \: 600$, with a mean value of $\sim 100$ $\rm cm^{-3}$. This reflects the known
 discrepancy where NLRs of AGNs present higher electron density than SFs (e.g. \citealt{2013MNRAS.430.2605Z}). In Fig.~\ref{fh1},  panel (a), the distribution of values for the $N_{\rm e}$  from both Seyfert and SF samples are shown.
From this figure it is clear that AGNs have $N_{\rm e} \: \lesssim \: \rm 1200 cm^{-3}$  while most SF ($\sim$ 90 percent) have $N_{\rm e} \: < \: \rm 200\, cm^{-3}$.

Regarding the electron temperature results, we derive values for  $T_{\rm high}$ from the Seyferts ranging from $\sim9\,000$ K to $\sim25\,000$ K while  values for SFs are in the range
of $\sim7300$ K--$20\,000$ K.
In panel (b) of Fig.~\ref{fh1}, the distributions of these values  for our Seyfert and  SF samples are shown.
The Seyfert mean value is $\sim 15\:000$ K, similar to the
ones derived  by  \cite{2018ApJ...856...46R, 2018ApJ...867...88R} for the quasar type 2 Mrk\,34 and for the Seyfert~2 Mrk\,573, i.e. about
13\,000 K, respectively. \citet{1979A&A....79..350H} suggested that AGNs with 
electron temperature values higher than 20\,000K, probably, present  a secondary source of energy in addition to photoionization, possibly the presence of shocks (see also \citealt{2021MNRAS.501.1370D} and references therein). It can be seen that  majority ($\sim 92$ per cent) of the AGNs present $T_{\rm high}$ values lower than 20\,000 K. Therefore, the objects of our sample, possibly, are mainly (photo)ionized by the  radiation from the AGN accretion disk (e.g. shocks can be neglected).
For SFs, a somewhat lower mean value of
$\sim 11\:500$ K was obtained.

\citet{1967ApJ...150..825P} proposed the presence of temperature fluctuations to explain the considerable differences found between the temperature estimations relied on distinct methods in \ion{H}{ii} regions.
Recently, \citet{2021MNRAS.506L..11R}, for the first time,
by using Gemini GMOS-IFU observations of three luminous nearby Seyfert galaxies (Mrk\,79, Mrk\,348 and Mrk\,607),
quantified temperature variations in this class of object.
These authors derived a lower limit for the temperature fluctuation parameter ($t^{2}$)  to be from $\sim0.04$ to 
$\sim0.1$, in order of those derived for \ion{H}{ii} regions and Planetary Nebulae. Although  temperature fluctuations 
have  little effect on helium abundance estimates
(e.g. \citealt{2012ApJ...753...39P}),  the supposition
of $t^{2}\: > \: 0$ leads to the derivation of higher O/H abundances and, consequently, lower ($\sim 3$ per cent) values of $Y_{\rm p}$  \citep{2002ApJ...565..668P}.
Since estimations of temperature fluctuation in AGNs are barely
found in the literature, we assume the temperature values based on [\ion{O}{iii}] line ratio as  fiducial values
for the high ionization zone of Seyfert nuclei. However, we emphasise that temperature fluctuation can be present.

\begin{figure}
\begin{center}
\includegraphics[angle=-90.0,width=\columnwidth]{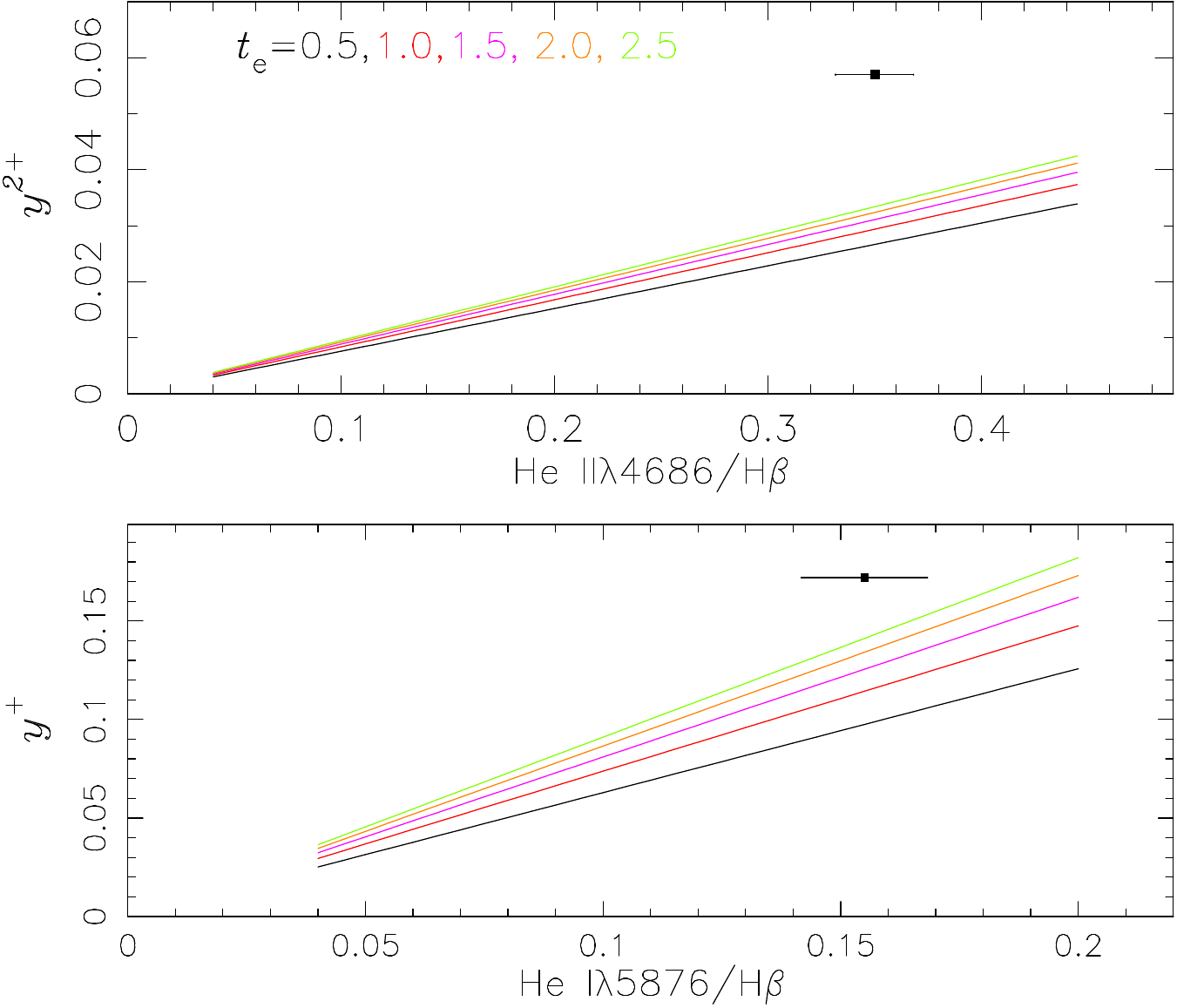}
\end{center}
\caption{Lower panel: Abundances of $y^{+}$  versus
the \ion{He}{i}$\lambda 5876$/H$\beta$ line ratio  for distinct
electron temperature values (in units of $10^{4}$ K), as indicated, assuming a
fixed value of $N_{\rm e}=100 \: \rm cm^{-3}$. Top panel: Same as
lower panel but for $y^{2+}$ versus \ion{He}{ii}$\lambda4686$/H$\beta$. In both plots, the range of emission-line ratio values (x-axes)  represents the range of
observational values of our sample of objects listed in Table~\ref{table1}. The error bars represent the mean error in the emission-line measurements of the sample of objects.}
\label{fhes}
\end{figure}

In Fig.~\ref{fhes}, we assumed the range of observational \ion{He}{i}$\lambda 5876$/H$\beta$  and \ion{He}{ii}$\lambda4686$/H$\beta$ line ratio values for our sample of objects (listed in Table~\ref{table1}) and calculated the $y^{+}$ and $y^{2+}$ abundances considering temperatures from 5\,000 K to 25\,000 K and a fixed value
of $N_{\rm e}=100 \: \rm cm^{-3}$. It can be seen that the effect of varying the temperature is in order of the uncertainty produced by the observational error for the line measurements. Therefore, we emphasize that, if temperature fluctuations exist in the objects of our sample, shocks or electron temperature fluctuations affect only the oxygen abundance estimations, while the helium estimations are reliable values.

\subsection{Ionization Correction Factor}  

\citet{2014MNRAS.440..536D} computed a large grid of photoionization models that covers a wide range of physical parameters of  planetary nebulae (PNe). They reported that the derivation of an ICF for neutral helium based on other ions is not recommended because the relative populations of helium ions depend
essentially on the effective temperature ($T_{\rm eff}$) of the central star ($T_{\rm eff}$ drives the flux of the ionizing photons), whereas those from metal ions also depend on the ionization parameter. These authors also pointed out that the correction for neutral helium is important only in models with ionizing sources at  $T_{\rm eff} \: \la \: 50\,000$ K.
In the bi-parametric expression for the Seyfert ICF($\rm He^{0})$ proposed here
(Eq.~\ref{fitbi}), both hardness of the ionizing radiation  and the gas ionization degree are
taken into account by the $y^{+}/y^{2+}$ ratio and by the ionization parameter $U$. For SFs, we notice a dependence
between ICF($\rm He^{0})$ and the age of the ionization source, while $T_{\rm eff}$ is similar to the age, since
the temperature of the hottest stars of the ionizing stellar cluster of an \ion{H}{ii} region  drives the flux of the ionizing photons \citep{2013ApJ...779...76Z} and it decreases
with the time. However, the stellar evolution  could have a minimal effect on our helium neutral fraction estimation in SFs because we considered the presence of \ion{He}{ii}$\lambda4876$ line as a selection criterion of the sample, which led to the selection of  \ion{H}{ii} regions with young cluster ages.
In fact, \citet{1999ApJ...510..104B} showed that only
photoionization models with clusters younger than 3 Myr are able to reproduce optical emission line intensities of \ion{H}{ii} regions.

The application of the ICF($\rm He^{0}$), i.e. Eq.~\ref{fitbi}, to 
our AGN data yields ICFs values ranging from 1.3 to 1.7, with a mean 
value of $\sim 1.5$. This result indicates that in NLRs of Seyfert~2 the helium is $\sim 50$ per cent in a neutral stage. As reported previously, radio and near-infrared 
(e.g. \citealt{2004A&A...425..457R, 2009MNRAS.394.1148S, 2010MNRAS.404..166R,2021MNRAS.504.3265R, 2015MNRAS.452.4128M, 2017MNRAS.464.1771S, 2022MNRAS.510..639B}, among others) 
observations have shown that AGNs harbor a neutral and warm (1000-3000 K) molecular gas  reservoir.   Therefore, a substantial proportion of neutral helium is expected in Seyfert nuclei. 

\begin{figure*}
\begin{center}
\includegraphics[scale=0.63, angle=-90]{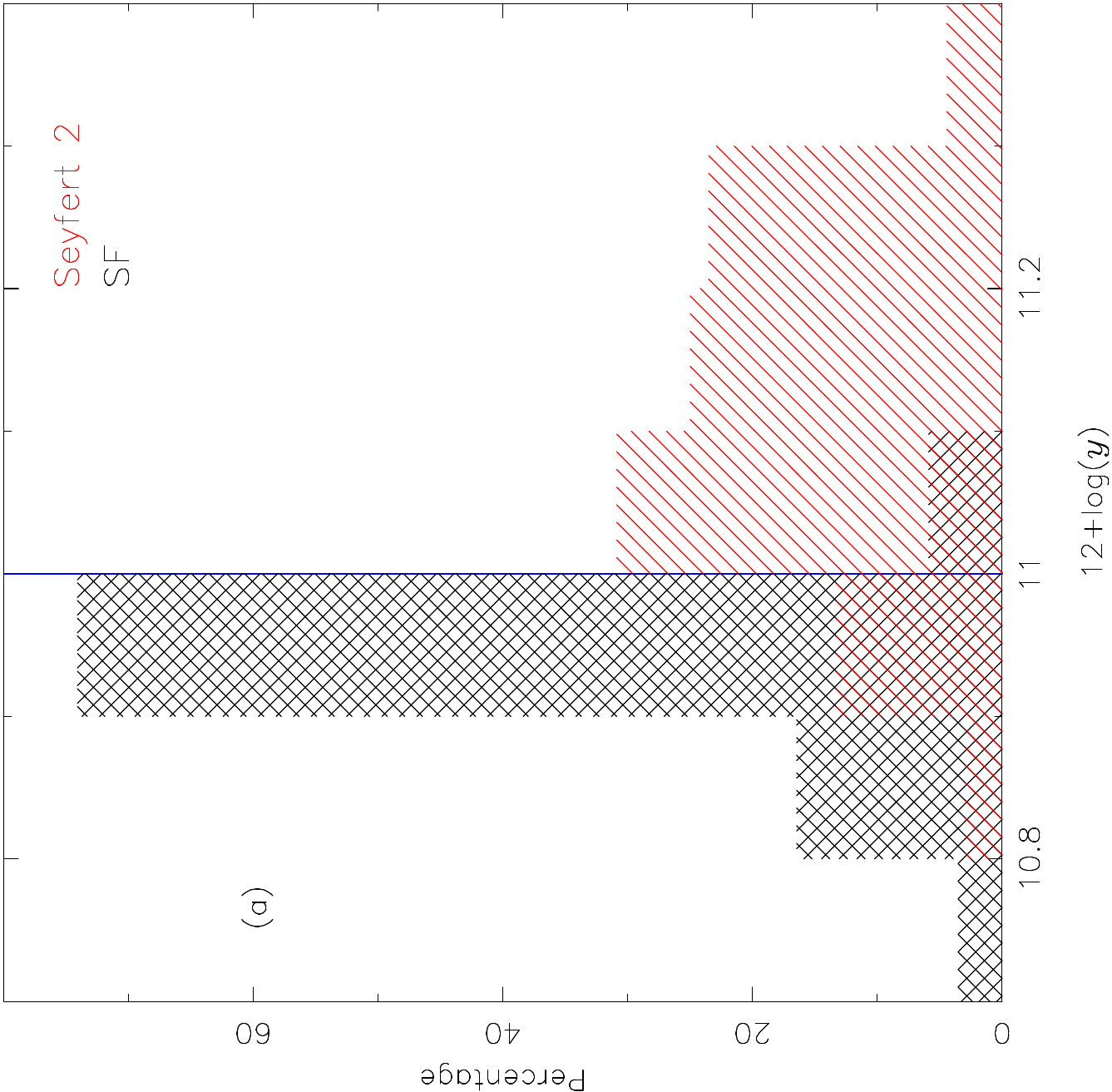}
\includegraphics[scale=0.63, angle=-90]{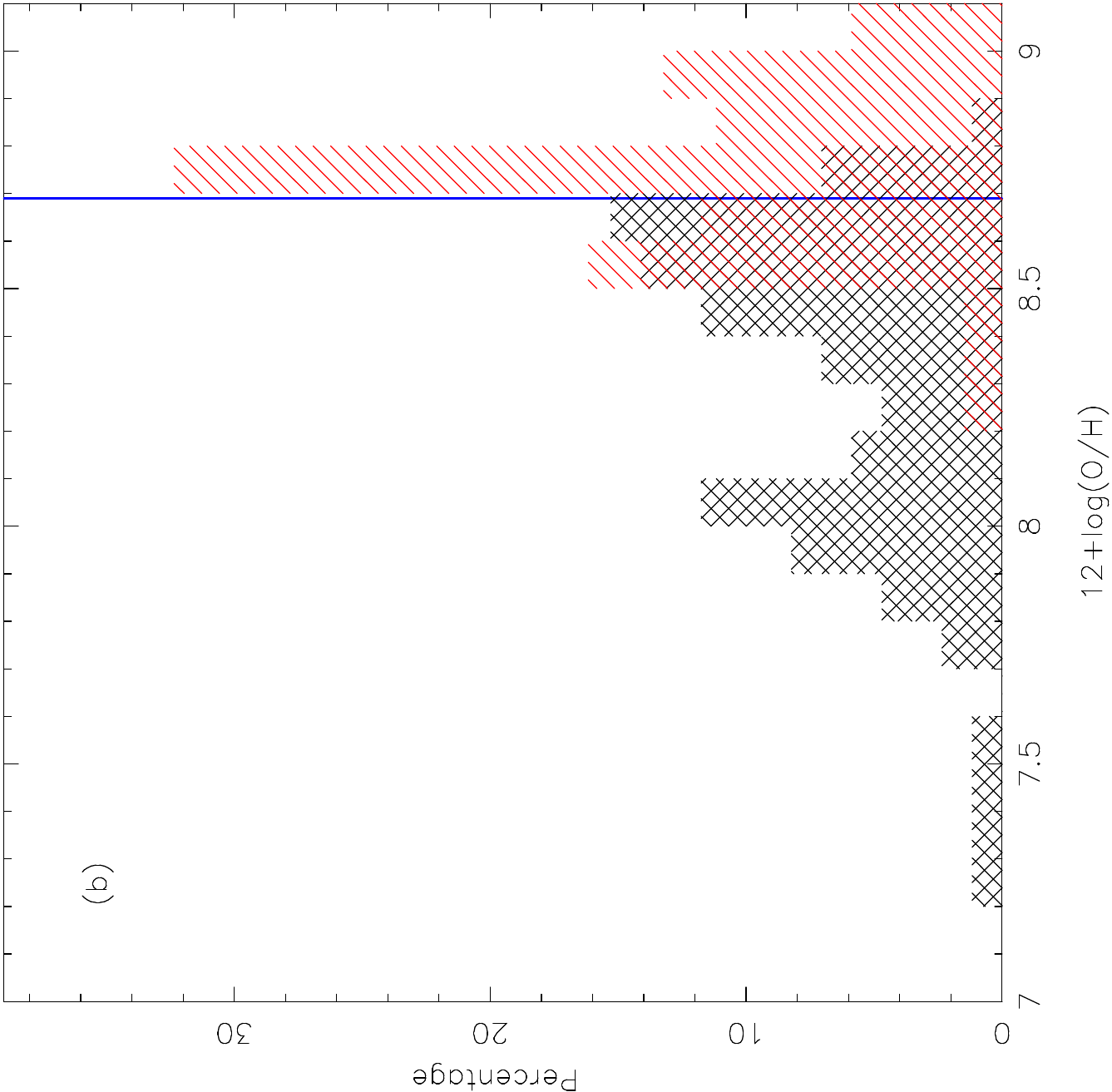}
\end{center}
\caption{As Fig.~\ref{fh1} but for 12+log($y$) [panel (a)]
and 12+log(O/H) [panel (b)]. The blue lines in panels (a)
and (b) represent the solar values derived by \citet{2001ApJ...556L..63A} and \citet{2010Ap&SS.328..179G}, respectively. These values correspond to
12+log($y$)$_{\odot}=11.0$ and 12+log(O/H)$_{\odot}=8.69$.}
\label{foxyhe}
\end{figure*} 

On the other hand, the application of  Eq.~\ref{eqicfsf}
indicates that SFs have ICF($\rm He^{0}$) ranging from
1.0 to 1.2, implying   70 per cent of the objects have
ICF($\rm He^{0}$)=1.0. The mean ICF($\rm He^{0}$) value for
SFs is 1.03, i.e. the helium is only $\sim 3$per cent in a neutral stage.
 \citet{1992MNRAS.255..325P}, following the methodology from \citet{1982ApJ...261..195M}, proposed the use of the abundance ratio
\begin{equation}
    \eta=\rm (O^{+}/S^{+})/(S^{2+}/O^{2+}),
\end{equation}
known by radiation softness parameter \citep{1988MNRAS.231..257V},
 as a means of estimating the significance of neutral
helium in SFs. \citet{1992MNRAS.255..325P} pointed out that,
for \ion{H}{ii} regions with $\log \eta \: < \: 0.9$,
the fraction of neutral helium into Str\"omgren sphere is negligible.
Recently, by using observational data from the MaNGA survey \citep{2015ApJ...798....7B}, \citet{2021MNRAS.508.1084K} determined through the $T_{\rm e}$-method values of the $\eta$ parameter for  67 star-forming galaxies. These authors found 
values in the range of $-0.4 \: \la\:  \log \eta \: \la \:0.6$.
\citet{2009ApJ...700..309B} obtained  spectrophotometric data for 
28 \ion{H}{ii} regions in the spiral galaxy NGC 300 and, also by using the $T_{\rm e}$-method, found that  the majority of the  objects present $\log \eta$ around 0.7. Otherwise, most part of Galactic \ion{H}{ii} region observed by \citet{2022MNRAS.510.4436M} present $\log \eta \: > \: 0.9$,
i.e. a non-negligible neutral helium abundance.
Our results are in consonance with those derived
by \citet{2022MNRAS.510.4436M}, who posited that, like Seyfert nuclei, SFs can have a significant neutral helium fraction.  

One of the major sources of the uncertainty in abundance estimates is
the use of ICFs, producing discrepancies by a factor of up to $\sim 4$ according to
the suppositions considered to derived them (e.g.  \citealt{2013MNRAS.432.2512D, 2020MNRAS.496.2726M}). Unfortunately, AGN ICFs for the helium are not found in the literature, therefore, it is not possible to estimate the ICF error in our estimates. In any case, detailed models built with the goal of reproducing line intensities of a large sample of AGNs (e.g. \citealt{2017MNRAS.468L.113D}) and SFs are needed to validate the neutral ICFs proposed here.

\subsection{Helium and oxygen abundances}

Initially, we compare the $y$ and O/H range of values derived from our
AGN sample with those derived from SFs. This is significant since the helium abundance at any ionization stage has hitherto been unknown for a large sample of AGNs.  

In Fig.~\ref{foxyhe}, panel (a), the distribution of 12+log($y$) for the Seyfert~2 and SF samples are shown.
Also in this plot, a line indicating the solar value of
the helium abundance, i.e. 12+log($y$)$_{\odot}=11.0$ derived by 
\citet{2010Ap&SS.328..179G} is indicated. We can see that
most Seyfert~2 present higher helium abundance in comparison
with those of SFs, with the Seyfert range  being 
$10.80 \: \la \: [12+\log(y)] \: \la \:11.40$ 
or $0.60 \: \la (y/y_{\odot}) \: \la \:2.50$, and 
a mean value of $<12+\log(y)>=11.11\pm0.11$ or $<(y/y_{\odot})>\sim1.30$.  We can also note that
the majority ($\sim 84$ per cent) of AGNs present oversolar helium abundance. Concerning the SFs, we derived the range $10.70 \: \la \: [12+\log(y)] \: \la 11.10$ or $0.50 \: \la (y/y_{\odot}) \: \la 1.20$, with a mean
value $<12+\log(y)>=10.92\pm0.05$ or $<(y/y_{\odot})>\sim0.80$.

In Fig.~\ref{foxyhe}, panel (b), the distribution of 12+log(O/H) for the Seyfert~2 and SFs samples are shown. For the Seyfert sample, we derived the range
$8.20 \: \la \: [12+\log(\rm O/H)] \: \la 9.10$, or $0.30 \: \la (Z/\rm Z_{\odot}) \: \la 2.6$, with most part ($\sim 63$ per cent) having oversolar abundance.
The mean value is $\rm <12+log(O/H)>=8.74\pm 0.16$
or $<(Z/\rm Z_{\odot})>\sim1.10$.
This is a  known result, i.e. subsolar metallicities are derived  in few  AGNs
 in the local universe (e.g., \citealt{2006MNRAS.371.1559G}), independently 
 of the method used to estimate this parameter
 \citep{2020MNRAS.492..468D}.
For SFs, the range of $7.20 \: \la \: [12+\log(\rm O/H)] \: \la 8.80$ or $0.03 \: \la (Z/\rm Z_{\odot}) \: \la 1.30$ is derived, being   $\sim8$ per cent of the objects presenting oversolar oxygen abundance. 

The lower helium and oxygen abundances in SFs
in comparison with those in AGNs is due to the fact that the latter are located in the central parts of galaxies and, according to the
inside-out scenario of galaxy formation (see \citealt{2005MNRAS.358..521M} and references therein) have  experimented a longer  time-scale of chemical enrichment of the ISM in comparison
with disk \ion{H}{ii} regions (most part of our SF sample). It is worth to mention that, our  direct helium estimates, based on the   $y^{+}$ and $y^{2}$ ionic abundances of AGNs,  seem to be the unique for the very high metallicity regime, i.e. $(Z/\rm Z_{\odot})\: > \: 1.0$. 
This fact  is very important  in the derivation of the $y$-O/H relation  and of  $Y_{\rm p}$ because our estimates,  combined those for XMPs,  produce a wide range of metallicities likely not previously considered.

\begin{figure}
\begin{center}
\includegraphics[angle=-90.0,width=\columnwidth]{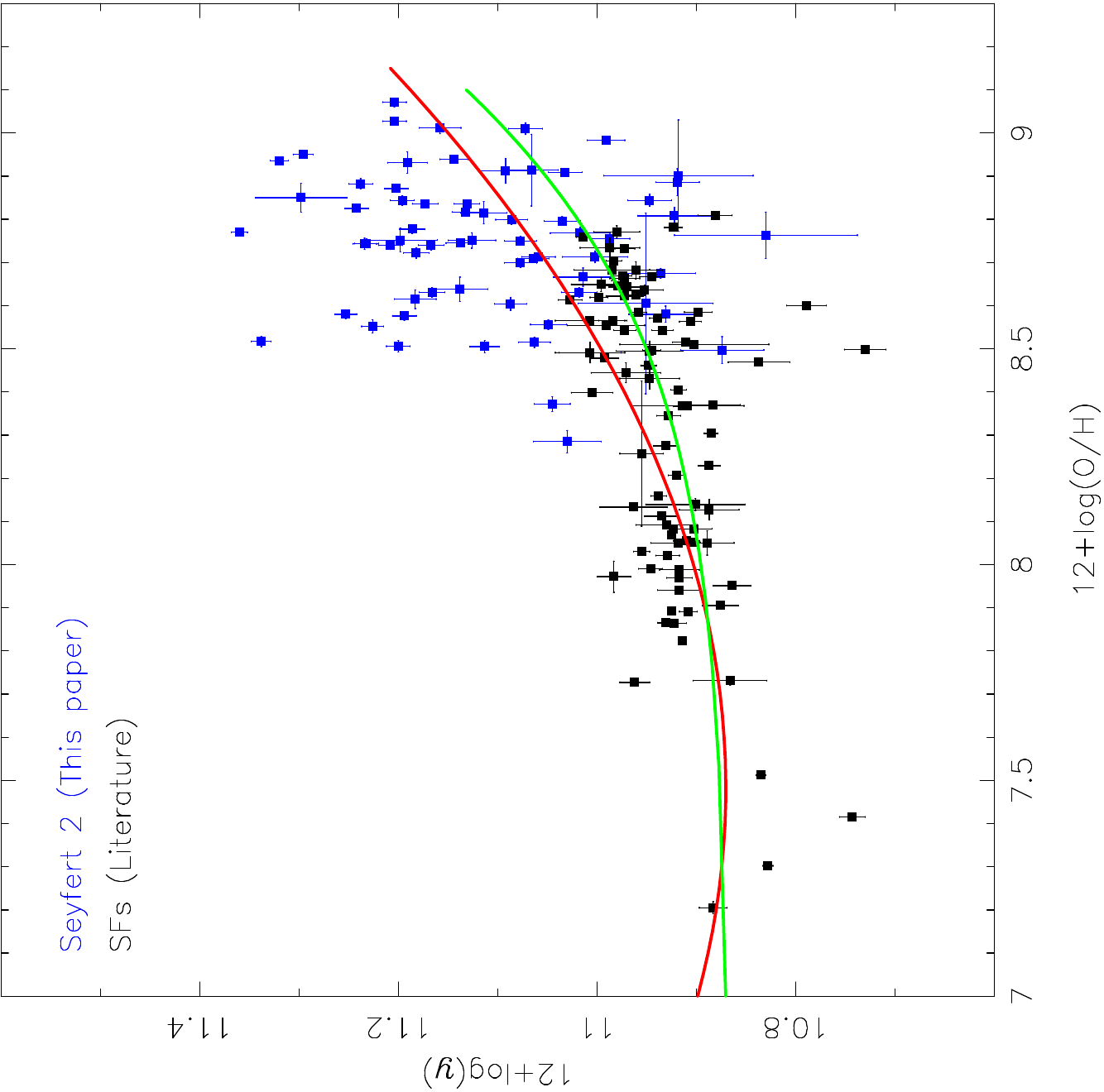}
\end{center}
\caption{Abundances of 12+log($y$) versus 12+log(O/H). Blue and black points represent the values for our sample of Seyfert~2 and SFs, respectively,   listed in Table~\ref{table2}. The red curve represents the fit to the  points given by Eq.~\ref{fit1}. The green curve represents Eq.~\ref{fdop06} derived by \citet{2006ApJS..167..177D}, but with $y$ and $Z/\mathrm{Z_{\odot}}$ values converted into 12+log($y$) and 12+log(O/H), respectively.}
\label{figheoxy}
\end{figure}

In Fig.~\ref{figheoxy}, the total helium 
abundance [in units of 12+log($y$)] versus
the oxygen abundance [in units of 12+log(O/H)]
from our  Seyfert (blue points)  and  SF (black points) samples are shown. Other than in AGNs, we note that the SFs with the highest metallicities  present  near $y$  values. Similar results were derived for nitrogen and argon by \citet{2020MNRAS.492.5675C} and
\citet{2021MNRAS.508.3023M}, respectively. 
 Moreover, we notice two clear behaviours, a smooth $y$-(O/H) relation for the low metallicity
regime [$\rm 12+\log(O/H) \: \la \: 8.5$ or ($Z/\rm Z_{\odot})\: \la \: 0.6)$] and a stepper relation for
high metallicity regime. This difference could be due to an excess of helium injected into the ISM  by winds of Wolf Rayet stars, which are more common in the high mettalicity environment (e.g. \citealt{1994A&A...287..803M, 1999ApJS..123....3L, 2002A&A...394..443P, 2004A&A...419L..17C, 2005A&A...441..981B}), containing
hydrogen burning products \citep{1992MNRAS.255..325P, 1998MNRAS.296..622C}.
 The scattering of the points observed in   both kind of objects in Fig.~\ref{figheoxy} is, probably, due to effects of different star formation rate in them (e.g. \citealt{2020ApJ...893...96B, 2022MNRAS.512.2867H}) and/or due to  radial migration of stars   \citep{2021arXiv210603912V, 2022arXiv220204666J}
 rather than evolutionary processes, by virtue of the fact that majority of the AGNs and SFs of our sample are located in the local universe. 
 
In order to derive an expression for the $y$-O/H relation, we performed 
 1000 bootstrap realisations \citep{davison1997bootstrap} 
 with Huber Regressor model \citep{owen2007robust}. The resulting fitting  
 to the points obtained  is given by
 \begin{equation}
\label{fit1}
\begin{split}
{\rm w} =&(0.1215\pm0.0422) \times {\rm x^{2} } -
(1.8183\pm0.6977) \times {\rm x}  \\
&+ \mathrm{ (17.6732\pm2.8798)}
\end{split}
\end{equation}
 where w=12+log($y$) and x=12+log(O/H).
\citet{2006ApJS..167..177D}, by using  abundance estimates of SFs obtained by \citet{1992MNRAS.255..325P} and \citet{1992ApJ...384..508R}, derived the relation
\begin{equation}
\label{fdop06}
    y=0.0737+0.024 \times (Z/\mathrm{Z_{\odot}}).
\end{equation}
We converted the $y$ and $Z/\mathrm{Z_{\odot}}$  values of this expression into the same units assumed previously as well as extrapolated it for
the high metallicity regime. In Fig.~\ref{figheoxy}, we compare the \citet{2006ApJS..167..177D}  relation with our relation (Eq.~\ref{fit1}), which it can be seen a good agreement between both.

\subsection{Primordial helium abundance}

\begin{figure*}
\begin{center}
\includegraphics[angle=-90.0,width=\columnwidth]{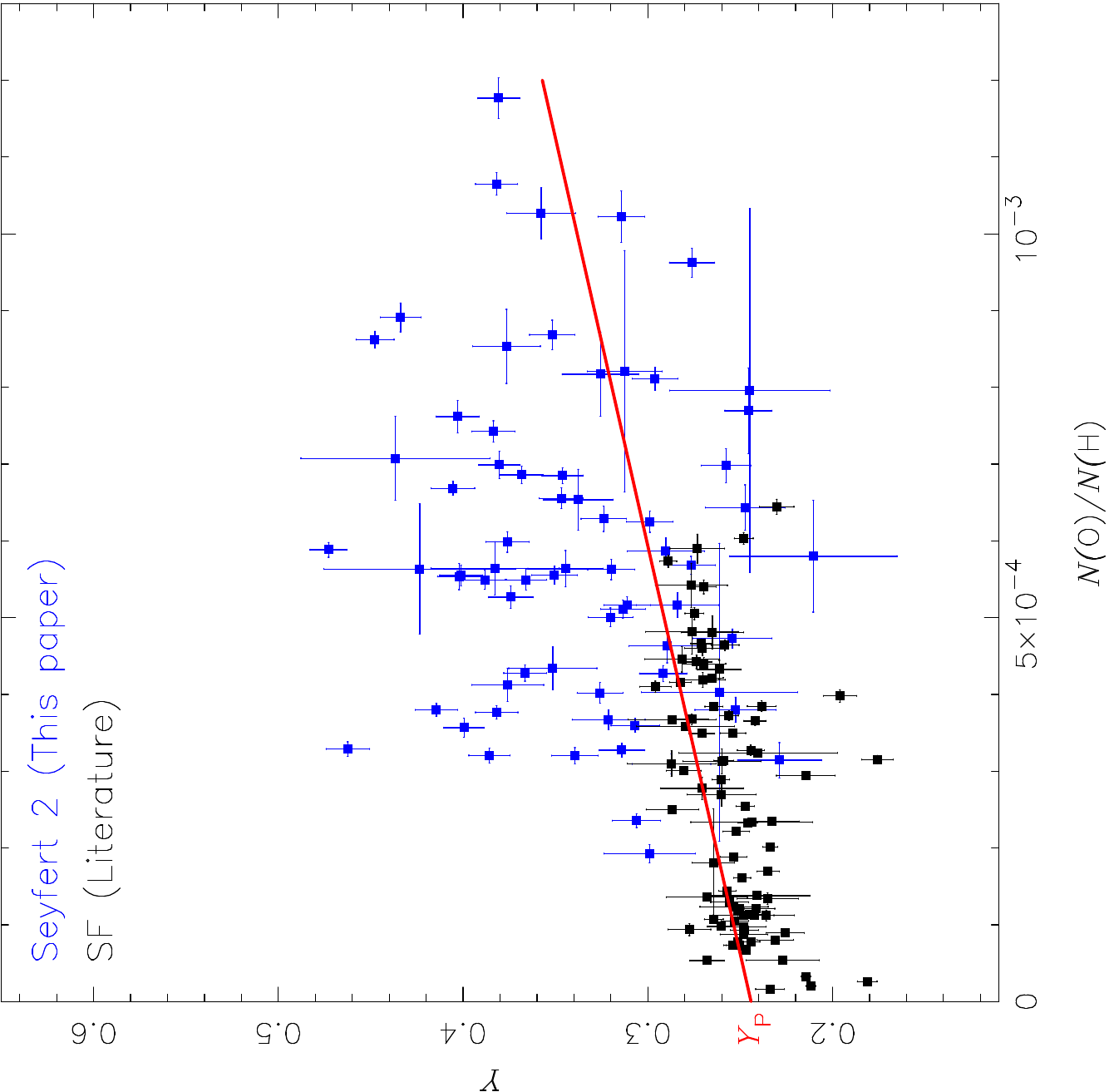}
\includegraphics[angle=-90.0,width=\columnwidth]{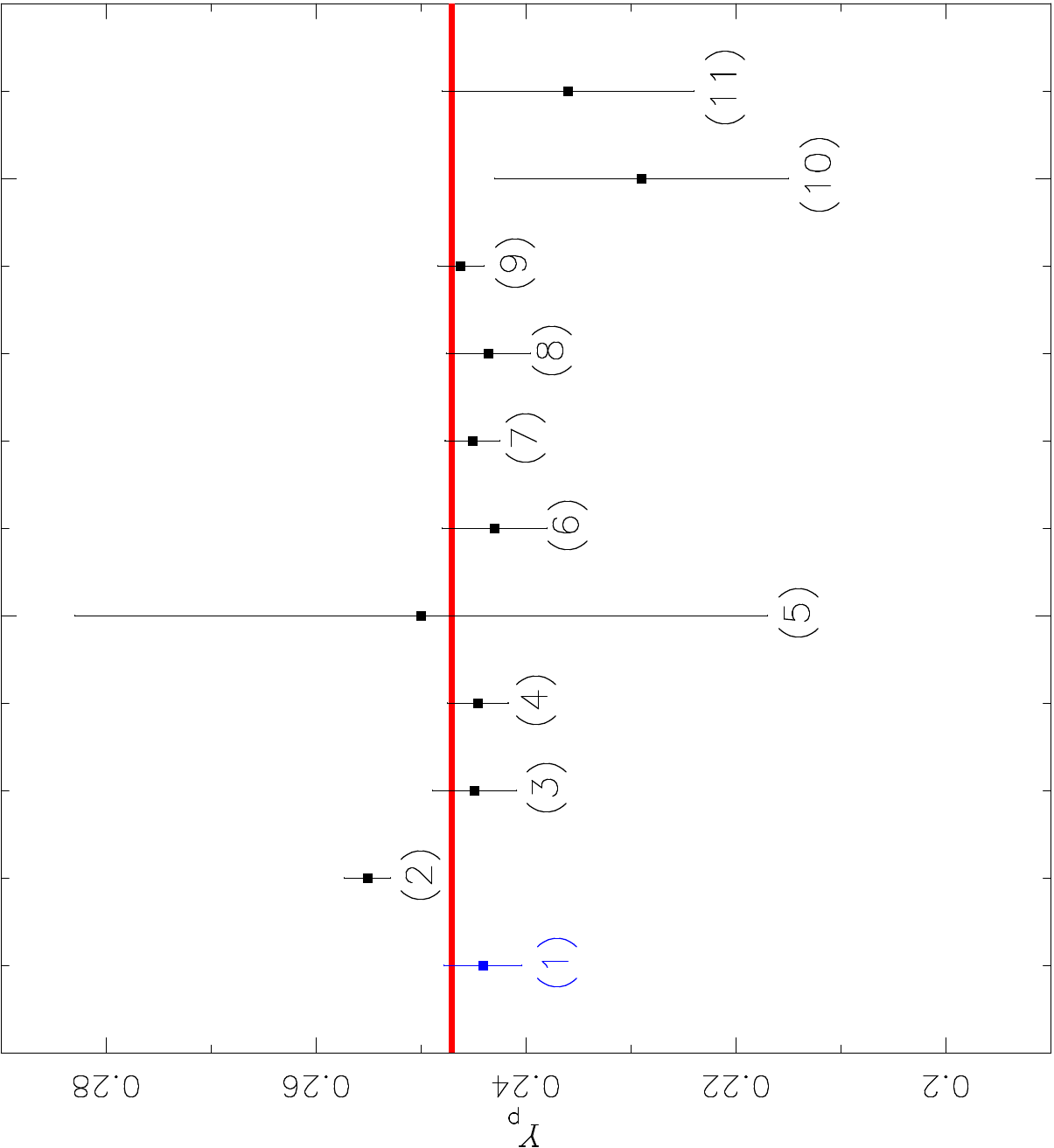}
\end{center}
\caption{Left panel: $Y$ versus the abundance ratio $N(\mathrm{O})/N(\mathrm{H})$ derived by using 
the $T_{\rm e}$-method. $Y$ values are calculated as in Eq.~\ref{eq_1}. Blue and black points represent 
estimates for our Seyfert~2 and SF samples, respectively. Red line represents the linear regression to
the points  given by Eq.~\ref{fineqa}. The primordial helium abundance derived from this linear regression, 
$Y_{\mathrm{p}}=0.2441 \pm 0.0037$, is indicated. Right panel: Estimates of the primordial helium abundance 
($Y_{\rm p}$) derived by different authors. The red line represents the estimation by \citet{planck2018}, i.e. 
$Y_{\mathrm{ p}}=0.2471\pm0.0003$. Other estimates correspond to:
(1) This work ($Y_{\mathrm{p}}=0.2441 \pm 0.0037$),
(2) \citet{2014MNRAS.445..778I}, (3) \citet{2015JCAP...07..011A}, (4) \citet{2016RMxAA..52..419P},
(5) \citet{2018NatAs...2..957C}, (6) \citet{2019MNRAS.487.3221F}, (7) \citet{2019ApJ...876...98V}, (8) \citet{2020ApJ...896...77H}, 
(9) \citet{2021MNRAS.502.3045K},  (10) and (11) estimates
by  \citet{2022MNRAS.510.4436M} for $t^{2}=0$
and $t^{2}\:> \:0$, respectively. The  $Y_{\rm p}$ value obtained by  each author is listed by \citet{2021MNRAS.502.3045K}, 
with exception of the one derived by \citet{2022MNRAS.510.4436M}.}
\label{fig10}
\end{figure*}

\citet{1972ApJ...173...25S} presented  abundance analysis of the two compact dwarf galaxies I\,Zw18 and II\,Zw40, and
for the first time, suggested that metal-poor
objects, such as these,  would be  used to estimate the
primordial helium abundance. After this  pioneering work,
several studies have been undertaken in an effort to address the observation of 
XMPs and objects in a wide range of metallicities in order
to derive $Y_{\rm p}$ (for a review see \citealt{2008arXiv0811.2980P, 2010IAUS..268..163F,   2012MSAIS..22..164S, 2017PASP..129h2001P}).

\citet{2007ApJ...666..636P} listed the main error sources   in the $Y_{\rm p}$  determination, which
 produce uncertainties from $\pm0.0005$ to $\pm0.0015$ 
in the final value of $Y_{\rm p}$, where the collisional excitation of the \ion{H}{i} lines and temperature structure were considered as the most important physical conditions  in the gaseous nebulae. Also, \citet{2020MNRAS.496.2726M} pointed out that an important source of error in the $Y_{\rm p}$ estimates is related to the ionization structure of the nebulae, i.e. the use of
ICFs in the  estimation of the total helium abundance, since  spectra of typical SF only show recombination lines of $\rm He^{+}$.

 In our case, probably, the main source of uncertainty in the $y$ estimation and, consequently of $Y_{\rm p}$, is the fact that observational data of our  AGN sample, in opposite to the SF sample,  do not have 
 errors in the line measurements for most of the objects and  typical values  for these
 were considered (see Sect.~\ref{obs}). This procedure, combined 
 with the uncertainty produced by the use of ICF($\rm He^{0})$, certainly 
 produced larger errors than those discussed by \citet{2007ApJ...666..636P} and \citet{2020MNRAS.496.2726M}.  Despite this drawback, our AGN estimates combined with estimations from SFs are useful in the  derivation of $Y_{\rm p}$ taking objects with  very high metallicity into
 account.

We calculated $Y$ values using our Seyfert~2 and SFs results, listed in Table~\ref{table2} together with  Eq.~\ref{eq_1} for the purpose of estimating $Y_{\rm p}$ which rely on a wide range of metallicities.  These values are shown in panel (a) of Fig.~\ref{fig10}. It can be seen that Seyfert~2s show higher $Y$ and
O/H values in comparison with those in SFs.  As previously, we performed 1000 bootstrap realisations to obtain a linear fit that yielded the following relation
\begin{equation}
\label{fineqa}
    Y=(94.0946\pm9.9534)\times N(\mathrm{O})/N(\mathrm{H}) + (0.2441 \pm 0.0037)
\end{equation}
with a Pearson correlation coefficient of $R=0.57$.
 Thus, we derived the value of $Y_{\mathrm{p}}=0.2441 \pm 0.0037$.

In Fig.~\ref{fig10}, panel (b), estimates of $Y_{\mathrm{p}}$ obtained by different authors and methods are compared to our estimate.
Also in this figure the line representing the value derived by the \cite{planck2018} is shown. It is worthwhile to note  that our $Y_{\mathrm{p}}$ value is in a
good agreement than the others, with exception of the estimation by \citet{2018NatAs...2..957C} which presents the highest error bars. We performed a linear fit to the points  considering only our SF estimates and found the expression:
\begin{equation}
\label{fineq1}
    Y=(36.5264\pm7.1944)\times N(\mathrm{O})/N(\mathrm{H}) + (0.2459 \pm 0.0024),
\end{equation}
with $Y_{\mathrm{p}}=0.2459 \pm 0.0024$, i.e.  a value
somewhat higher than the one obtained by assuming
AGN and SF estimates. However, this value is in
consonance with  the one inferred by the \cite{planck2018}
and by other authors.  

\section{Conclusion}
\label{conc}

We used optical emission line intensities [$3000 \: < \lambda($\AA$) \: < \: 7000$] taken from SDSS DR15 and  additional compilations from the literature to derive, through $T_{\rm e}$-method, the helium and oxygen abundances relative to hydrogen  in the  narrow line regions (NLRs)  of a local ($z\: < \: 0.2$) sample of 65 Seyfert 2 nuclei and  85 star-forming regions (SFs; i.e. \ion{H}{ii} region, star-forming galaxies). 
Photoionization model grids, built with the \textsc{Cloudy} code and
simulating NLRs and SFs, were used to obtain  expressions for the Ionization Correction Factor (ICF) of the neutral helium.  The application of these ICFs indicates that the NLRs of Seyfert~2 present a neutral helium fraction 
in relation to the total helium abundance ranging from 
$\sim20$ to $\sim70$ per cent (mean value $50$ per cent) and SFs from 
$\sim0$ to $\sim20$ per cent (mean value 3 per cent). The high neutral helium
abundance in AGNs it is due
to these objects harbour a neutral and warm (1000-3000 K) molecular gas  reservoir. 
We found that Seyfert~2 nuclei present helium abundance
ranging from 0.60 to 2.50 times the solar value, which implies
that for $\sim 85$ per cent of the sample an oversolar helium abundance was derived. 
Our results indicate that NLR of Seyfert~2 nuclei have a stepper (He/H)-(O/H) relation than SFs. This difference could be due to excess of helium injected into the ISM  by winds of Wolf Rayet stars, which are more common in the high mettalicity environment, i.e. $\rm 12+\log(O/H)\: \ga \: 8.7$ [$(Z/Z_{\odot}) \: \ga \: 1.0$]. From a regression to zero metallicity, by using  Seyfert~2  and SFs estimates, we derived  a primordial helium  mass fraction  $Y_{\rm p}=0.2441 \pm 0.0037$, a value in good agreement with the one  inferred from the temperature fluctuations of the cosmic microwave background by the Planck Collaboration, i.e.   $Y_{\rm p}^{Planck}=0.2471\pm0.0003$.  

\section*{Data availability}
The data underlying this article will be shared on reasonable request to the corresponding author.

\section*{Acknowledgements}
We are grateful to the referee for his/her dedicated work in reviewing our paper.
OLD and ACK are grateful to the Fundação de Amparo à Pesquisa do Estado de São Paulo (FAPESP) and to Conselho Nacional de Desenvolvimento Científico e Tecnológico (CNPq)
 for the financial support.
CBO is grateful to the FAPESP for the support under grant 2019/11934-0 and to the Coordenação de Aperfeiçoamento de Pessoal de
Nível Superior (CAPES).
Funding for the Sloan Digital Sky Survey  has been provided by the Alfred P. Sloan Foundation, the U.S. Department of Energy Office of Science, and the Participating Institutions. SDSS acknowledges support and resources from the Center for High-Performance Computing at the University of Utah. The SDSS web site is www.sdss.org.

\bibliographystyle{mn2e}
\bibliography{publicacao}

\appendix

\label{app}

\begin{table*}
\addtolength{\tabcolsep}{-3pt}
\caption{Reddening corrected emission line intensities (in relation to
$\rm H\beta=1.0$) of the  Seyfert~2 sample  described in Sect.~\ref{sampleagn}. }
\label{table1}
\begin{tabular}{lccccccccccc}
\hline
ObjID	           &  [\ion{O}{ii}]            &    [\ion{O}{iii}]   &     \ion{He}{ii}   &   [\ion{O}{iii}]   &     \ion{He}{i}       &  $\rm H\alpha$  &	   [\ion{N}{ii}]         &   [\ion{S}{ii}]           &	[\ion{S}{ii}]	 &  Ref. \\
\noalign{\smallskip}

           	   &	    $\lambda$3727      &    $\lambda$4363    &    $\lambda$4686   &    $\lambda$5007   &      $\lambda$5876    &   $\lambda$6563 &	   $\lambda$6583	 &     $\lambda$6716	    &	 $\lambda$6731   &	 \\
\hline
J013957.81-004504.2  & 2.31$\pm$0.22 & 0.06$\pm$0.01 & 0.16$\pm$0.01 & \; 7.45$\pm$0.11 & 0.06$\pm$0.02 & 2.86$\pm$0.04 & 2.51$\pm$0.04 & 0.79$\pm$0.02 & 0.67$\pm$0.02 & 1\\
J033923.14-054841.5  & 2.63$\pm$0.20 & 0.09$\pm$0.01 & 0.26$\pm$0.01 & \; 7.90$\pm$0.02 & 0.09$\pm$0.01 & 2.86$\pm$0.01 & 1.52$\pm$0.01 & 0.53$\pm$0.01 & 0.45$\pm$0.01 & 1\\
J074257.23+333217.9  & 1.93$\pm$0.09 & 0.04$\pm$0.01 & 0.08$\pm$0.01 & 3.03$\pm$0.01 & 0.10$\pm$0.01 & 2.86$\pm$0.01 & 1.63$\pm$0.01 & 0.49$\pm$0.01 & 0.43$\pm$0.01 & 1\\
J093509.12+002557.4 & 2.58$\pm$0.10 & 0.07$\pm$0.01 & 0.11$\pm$0.01 & 3.86$\pm$0.02 & 0.07$\pm$0.01 & 2.86$\pm$0.01 & 2.12$\pm$0.01 & 0.59$\pm$0.01 & 0.54$\pm$0.01 & 1\\
J095759.45+022810.5 & 3.17$\pm$0.23 & 0.08$\pm$0.01 & 0.26$\pm$0.01 & \; 7.59$\pm$0.07 & 0.08$\pm$0.01 & 2.86$\pm$0.03 & 2.12$\pm$0.02 & 0.84$\pm$0.01 & 0.69$\pm$0.01 & 1\\
J100602.50+071131.8 & 6.21$\pm$0.04 & 0.15$\pm$0.01 & 0.27$\pm$0.01 & \; 7.61$\pm$0.01 & 0.09$\pm$0.01 & 2.86$\pm$0.01 & 1.85$\pm$0.01 & 0.62$\pm$0.01 & 0.51$\pm$0.01 & 1\\
J100921.26+013334.5 & 3.31$\pm$0.28 & 0.14$\pm$0.01 & 0.31$\pm$0.01 & 12.29$\pm$0.07& 0.11$\pm$0.01 & 2.86$\pm$0.02 & 2.89$\pm$0.02 & 0.72$\pm$0.01 & 0.74$\pm$0.01 & 1\\
J101754.72-002811.9 & 2.34$\pm$0.14 & 0.12$\pm$0.01 & 0.27$\pm$0.01 & 6.05$\pm$0.03 & 0.11$\pm$0.01 & 2.86$\pm$0.01 & 1.75$\pm$0.01 & 0.53$\pm$0.01 & 0.43$\pm$0.01 & 1\\
J102039.81+642435.8 & 4.38$\pm$1.00 & 0.17$\pm$0.01 & 0.25$\pm$0.01 & 8.82$\pm$0.06 & 0.07$\pm$0.01 & 2.86$\pm$0.02 & 2.72$\pm$0.02 & 0.76$\pm$0.01 & 0.58$\pm$0.01 & 1\\
J112850.39+021016.2 & 2.97$\pm$0.20 & 0.28$\pm$0.02 & 0.28$\pm$0.02 & 6.75$\pm$0.15 & 0.16$\pm$0.03 & 2.86$\pm$0.07 & 1.79$\pm$0.05 & 0.64$\pm$0.03 & 0.51$\pm$0.03 & 1\\
Mrk\,176 & 3.54 & 0.32 & 0.43 & 14.36 &  0.10 & 2.81 & 2.99 & 0.56 & 0.54 &  2 \\ 
3C\,33 & 4.93  & 0.32 & 0.26 & 12.68 &  0.096 & 2.63 & 1.76 & 0.87 & 0.73 &  2 \\ 
Mrk\,3 & 3.52  & 0.24 & 0.18 & 12.67 &  0.084 & 3.10 & 3.18 & 0.73 & 0.82 &  2 \\ 
Mrk\,573 & 2.92  & 0.18 & 0.36 & 12.12 &  0.10 & 2.95 & 2.57 & 0.75 & 0.80 &  2 \\ 
Mrk\,78  & 4.96  & 0.14 & 0.35 & 11.94 &  0.10 & 2.46 & 2.32 & 0.68 & 0.61 &  2 \\ 
Mrk\,348 & 4.45  & 0.26 & 0.22 & 11.74 &  0.12 & 2.76 & 2.28 & 1.09 & 1.25 &  2 \\ 
Mrk\,34  & 3.43  & 0.15 & 0.28 & 11.46 &  0.12 & 2.99 & 2.18 & 0.82 & 0.80 &  2 \\ 
Mrk\,1 & 2.79  & 0.21 & 0.30 & 10.95 &  0.10 & 2.66 & 2.21 & 0.49 & 0.52 &  2 \\ 
Mrk\,270 & 5.64 & 0.28 & 0.22 & 8.71 &  0.19 & 3.14 & 2.93 & 1.21 & 1.39 &  2 \\
III\,Zw\,55  & 3.19  & 0.25 & 0.20 & 6.92 &  0.090 & 2.86 & 3.87 & 0.66 & 0.74 &  2  \\
3C\,452  & 4.81  & 0.18 & 0.059  & 6.85 &  0.17 & 2.98 & 3.58 & 1.10 & 0.77 &  2  \\
Mrk\,198 & 2.51  & 0.12 & 0.075  & 5.56 &  0.11 & 3.02 & 2.26 & 0.89 & 0.68 &  2  \\
Mrk\,268 & 3.75  & 0.25 & 0.078  & 4.82 &  0.095 & 3.38 & 4.94 & 1.28 & 1.08 &  2  \\
NGC\,2110 & 4.38 & 0.17 & 0.06 & 4.76 &  0.14 & 2.66 & 3.76 & 1.52 & 1.42 &  3 \\
ESO\,138\,G1 & $2.35\pm0.08$ & $0.34\pm0.02$  & $0.30\pm0.02$  & $8.71\pm0.36$  &  $0.11\pm0.02$  & $3.00\pm0.13$ & $0.68\pm0.03$ & $0.47\pm0.03$ & $0.48\pm0.03$ &  4 \\ 
NGC\,3081  & 2.19 & 0.24  & 0.44  & 12.62 &  0.10 & 2.84  & 2.47 & 0.60  & 0.66  &  5 \\
NGC\,4388  & 2.73 & 0.16  & 0.21 & 10.55 &  0.11 & 2.84  & 1.50 & 0.72  & 0.62  &  5 \\
NGC\,5135  & 2.06 & 0.11  & 0.20 & 4.42 &  0.10 & 2.83  & 2.50 & 0.40  & 0.38  &  5 \\
NGC\,5728  & 3.50 & 0.47  & 0.32  & 10.85 &  0.15 & 2.83  & 3.93 & 0.44  & 0.435 &  5 \\
Akn\,347$^{a}$ & 2.72 & 0.42 & 0.33  & 15.15 & 0.12  & 3.18 & 3.93 & 0.93 & 0.89  & 6 \\
UM\,16$^{a}$ & 3.05 & 0.25 & 0.37 & 13.8 & 0.18 & 2.77 & 1.66 & 0.47  & 0.48 & 6 \\
Mrk\,612$^{a}$  & 1.90 & 0.20 & 0.16  &	9.09 & 0.20 & 2.81 & 3.81 & 0.78 & 0.57 & 6 \\
Mrk\,573  & $2.13\pm0.05$ & $0.14\pm0.011$  & $0.29\pm0.02$  & $10.26\pm0.05$ & $0.09\pm0.01$ & $2.86\pm0.05$  &  $2.34\pm0.04$ & $0.79\pm0.01$ &  $0.73\pm0.02$ &  7 \\
NGC\,2992 & $3.38\pm0.11$ & $0.21\pm0.027$  & $0.07\pm0.02$  & $ 5.25\pm0.09$ & $0.08\pm0.01$ & $2.86\pm0.02$  &  $2.70\pm0.03$ & $0.52\pm0.01$ &  $0.47\pm0.01$ &  7 \\
IC\,2560  & $1.65\pm0.02$ & $0.16\pm0.017$  & $0.26\pm0.06$  & $10.69\pm0.10$ & $0.10\pm0.01$ & $2.86\pm0.08$  &  $2.88\pm0.07$ & $0.74\pm0.02$ &  $0.79\pm0.01$ &  7 \\
NGC\,5664 & $1.83\pm0.05$ & $0.02\pm0.006$  & $0.12\pm0.01$  & $ 2.95\pm0.07$ & $0.09\pm0.01$ & $2.86\pm0.12$  &  $1.70\pm0.11$ & $0.58\pm0.03$ &  $0.43\pm0.07$ &  7 \\
NGC\,5728 & $2.32\pm 0.03$ & $ 0.11\pm 0.016$  & $ 0.17\pm0.01$  & $9.10\pm0.11$ & $0.08\pm0.01$ & $2.86\pm0.13$  &  $3.53\pm0.12$ & $0.97\pm0.06$ &  $0.66\pm0.03$ &  7 \\
ESO\,339-G11 & $2.39\pm0.12$ & $0.14\pm0.074$  & $0.13\pm0.01$  & $ 7.86\pm0.05$ & $0.09\pm0.01$ & $2.86\pm0.12$  &  $3.97\pm0.09$ 	& $0.69\pm0.02$ &  $0.64\pm0.02$ &  7 \\
NGC\,6890 & $1.66\pm0.04$ & $0.23\pm0.037$  & $0.19\pm0.01$  & $10.90\pm0.03$ & $0.06\pm0.02$ & $2.86\pm0.10$  &  $2.85\pm0.09$ 	& $0.43\pm0.01$ &  $0.39\pm0.01$ &  7 \\
IC\,5063 & $2.88\pm0.06$ & $0.12\pm0.008$  & $0.09\pm0.01$  & $ 7.83\pm0.07$ & $0.08\pm0.01$ & $2.86\pm0.06$  &  $1.81\pm0.06$ 	& $0.77\pm0.06$ &  $0.77\pm0.02$ &  7 \\
NGC\,7130 & $1.75\pm0.03$ & $0.09\pm0.005$  & $0.15\pm0.01$  & $ 4.62\pm0.05$ & $0.05\pm0.01$ & $2.86\pm0.01$  &  $3.64\pm0.00$ 	& $0.50\pm0.05$ &  $0.50\pm0.01$ &  7 \\
NGC\,7582 & $1.24\pm0.02$ & $0.03\pm0.002$  & $0.11\pm0.05$  & $ 2.14\pm0.10$ & $0.10\pm0.01$ & $2.86\pm0.02$  &  $1.86\pm0.02$ 	& $0.40\pm0.01$ &  $0.38\pm0.01$ &  7 \\
NGC\,7590 & $3.32\pm0.19$ & $0.05\pm0.027$  & $0.07\pm0.03$  & $ 3.47\pm0.16$ & $0.06\pm0.02$ & $2.86\pm0.11$  &  $2.39\pm0.18$ 	& $0.98\pm0.08$ &  $0.78\pm0.04$ &  7 \\
Cygnus\,A & 5.01 & 0.20  & 0.27 & 12.30  & 0.07 & 3.09 & 6.16 & 1.65 & 1.51 &  8 \\
Mrk\,1157$^{a}$ &  7.32 & 0.25  & 0.24 & 9.81  & 0.12 & 2.82 &  3.07  &  0.65  & 0.69 &  9 \\
ESO\,428-G14  & 2.22 & 0.24  & 0.19 & 11.2  & 0.16 & 2.90 &  3.13  &  0.83  & 0.88 &  10 \\ 
ESO\,137-G34  & 3.07 & 0.11  & 0.21 & 9.35  & 0.11 & 3.10 & 3.35  &  0.98  & 1.20 &  11 \\
IC\,3639 &  1.24 & 0.09  & 0.12 & 5.68 & 0.10 & 3.10 & 2.52 &	0.68 & 0.73  &  11 \\
IC\,4777 &  2.32 & 0.07  & 0.17 & 6.34 & 0.10 & 3.10 & 3.82 &	0.92 & 0.87  &  11 \\
IC\,4995 &  1.32 & 0.20  & 0.27 & 11.87 & 0.08 & 3.10 & 2.49 &	0.68 & 0.76  &  11 \\
IRAS\,11215-2806 &  2.48 & 0.14  & 0.25 & \; 7.25 & 0.11 & 3.10 & 1.74 &	0.84 & 0.79  &  11 \\

\hline												\end{tabular}
\end{table*} 
  
 \begin{table*}
\addtolength{\tabcolsep}{-3pt}
\centering
\contcaption{}
\begin{tabular}{lccccccccccc}
\hline
ObjID	           &  [\ion{O}{ii}]            &    [\ion{O}{iii}]   &     \ion{He}{ii}   &   [\ion{O}{iii}]   &     \ion{He}{i}       &  $\rm H\alpha$     &	   [\ion{N}{ii}         &   [\ion{S}{ii}]           &	[\ion{S}{ii}]	 &  Ref. \\
\noalign{\smallskip}

           	   &	    $\lambda$3727      &    $\lambda$4363    &    $\lambda$4686   &    $\lambda$5007   &      $\lambda$5876    &   $\lambda$6563    &	   $\lambda$6583	&     $\lambda$6716	    &	 $\lambda$6731   &	 \\
\hline

MCG-01-24-012 &  3.31 & 0.16  & 0.20 & 8.51 & 0.09 & 3.10 & 1.77 &	0.97 & 0.86  &  11 \\
MCG-02-51-008 &  2.35 & 0.05  & 0.09 & 4.23 & 0.10 & 3.10 & 2.10 &	0.84 & 0.63  &  11 \\
NGC\,1125 &  3.13 & 0.08  & 0.14 & 5.86 & 0.10 & 3.10 & 2.30 &	0.79 & 0.78  &  11 \\
NGC\,1194 &  1.64 & 0.08  & 0.06 & 6.21 & 0.11 & 3.10 & 1.61 &	1.01 & 0.77  &  11 \\
NGC\,1320 &  0.98 & 0.13  & 0.37 & 9.34 & 0.11 & 3.10 & 2.21 &	0.65 & 0.62  &  11 \\
NGC\,3281 &  2.06 & 0.09  & 0.29 & \; 7.71 & 0.11 & 3.10 & 2.85 &	0.88 & 0.76  &  11 \\
NGC\,3393 &  2.23 & 0.10  & 0.26 & 9.428 & 0.12 & 3.10 & 3.13 &	0.92 & 0.93  &  11 \\
NGC\,4939 &  2.05 & 0.12  & 0.30 & 10.03 & 0.12 & 3.10 & 4.16 &	0.91 & 0.99  &  11 \\
NGC\,4968 &  1.64 & 0.22  & 0.27 & 9.45 & 0.12 & 3.10 & 3.28 &	0.62 & 0.65  &  11 \\
NGC\,5427 &  2.15 & 0.13  & 0.18 & \; 7.96 & 0.14 & 3.10 & 3.95 &	0.80 & 0.88  &  11 \\
NGC\,5643 &  2.74 & 0.12  & 0.24 & 8.60 & 0.04 & 3.10 & 2.98 &	0.97 & 0.91  &  11 \\
NGC\,5990 &  1.88 & 0.01  & 0.07 & 2.16 & 0.09 & 3.10 & 2.06 &	0.48 & 0.42  &  11 \\
NGC\,7682 &  2.85 & 0.16  & 0.19 & 9.34 & 0.09 & 3.10 & 3.03 &	1.09 & 1.17  &  11 \\
PKS\,1306-241 &  3.47 & 0.04  & 0.05 & 3.82 & 0.10 & 3.10 & 1.25 &	0.69 & 0.61  &  11 \\
\hline
\end{tabular}
\begin{minipage}[l]{16.5cm}
 References: 
 (1) SDSS-DR7 \citep{2000AJ....120.1579Y}, 
 (2) \citet{1978ApJ...223...56K}, 
 (3) \citet{1980ApJ...240...32S}, 
 (4) \citet{1992A&A...266..117A}, 
 (5) \citet{1983ApJ...266..485P},
 (6) \citet{1981ApJ...250...55S}, 
 (7) \citet{2015ApJS..217...12D}, 
 (8) \citet{1975ApJ...197..535O}, 
 (9) \citet{1994A&AS..105...57D}, 
 (10) \citet{1986A&A...166...92B}.
 (11) \citet{2017ApJS..232...11T}.
\end{minipage}
\end{table*} 

\begin{table*}
\addtolength{\tabcolsep}{-5pt}
\centering
\caption{Reddening corrected emission line intensities (in relation to
$\rm H\beta=1.0$) of the  SF sample  described in Sect.~\ref{samplesf}. }
\label{table1a}
\begin{tabular}{lccccccccccc}
\hline
ObjID	           &  [\ion{O}{ii}]         &    [\ion{O}{iii}]   &     \ion{He}{ii}   &   [\ion{O}{iii}]   &     \ion{He}{i}       &  $\rm H\alpha$      &  [\ion{N}{ii}]       &   [\ion{S}{ii}]    &	[\ion{S}{ii}]	 &  Ref. \\
\noalign{\smallskip}

           	       & $\lambda$3727      &    $\lambda$4363    &    $\lambda$4686   &    $\lambda$5007   &      $\lambda$5876    &   $\lambda$6563     &  $\lambda$6583	 &   $\lambda$6716     &   $\lambda$6731    &	 \\
\hline 
NGC3184-72.8+91.3      & $1.82\pm0.01$      &  $0.002\pm0.0010$   & $0.011\pm0.0010$   &  $0.384\pm0.002$   &	$0.083\pm0.001$     &	$2.801\pm0.027$   &   $0.923\pm0.008$	 &   $0.352\pm0.003$   &   $0.253\pm0.002$  &  1  \\		   
NGC3184+14.9-139.6     & $2.47\pm0.04$      &  $0.008\pm0.0010$   & $0.024\pm0.0030$   &  $1.445\pm0.016$   &	$0.081\pm0.006$     & 	$2.883\pm0.046$   &   $0.789\pm0.012$	 &   $0.424\pm0.006$   &   $0.303\pm0.005$  &  1  \\
NGC3184+80.0-148.2     & $2.88\pm0.03$      &  $0.004\pm0.0010$   & $0.007\pm0.0010$   &  $0.860\pm0.018$   &	$0.110\pm0.010$     & 	$2.937\pm0.056$   &   $0.828\pm0.016$	 &   $0.439\pm0.009$   &   $0.321\pm0.007$  &  1  \\
NGC3184-93.3-142.3     & $2.20\pm0.03$      &  $0.008\pm0.0020$   & $0.001\pm0.0010$   &  $1.902\pm0.031$   &	$0.129\pm0.010$     & 	$2.840\pm0.067$   &   $0.543\pm0.013$	 &   $0.385\pm0.009$   &   $0.263\pm0.006$  &  1  \\
NGC3184-172.5-30.2     & $2.32\pm0.01$      &  $0.014\pm0.0010$   & $0.003\pm0.0010$   &  $2.434\pm0.017$   &	$0.124\pm0.004$     & 	$2.801\pm0.023$   &   $0.485\pm0.005$	 &   $0.251\pm0.002$   &   $0.183\pm0.001$  &  1  \\
+164.6+9.9             & $1.94\pm0.04$      &  $0.002\pm0.0002$   & $0.009\pm0.0010$   &  $0.980\pm0.020$   &	$0.116\pm0.002$     & 	$3.015\pm0.060$   &   $0.650\pm0.027$	 &   $0.161\pm0.006$   &   $0.116\pm0.005$  &  2  \\
+17.3-235.4  	       & $2.43\pm0.03$      &  $0.004\pm0.0005$   & $0.001\pm0.0004$   &  $1.170\pm0.023$   &	$0.110\pm0.002$     & 	$2.865\pm0.057$   &   $0.650\pm0.020$	 &   $0.240\pm0.007$   &   $0.170\pm0.005$  &  2  \\
+189.2-136.3 	       & $1.60\pm0.03$      &  $0.006\pm0.0002$   & $0.010\pm0.0017$   &  $2.603\pm0.052$   &	$0.140\pm0.002$     & 	$2.815\pm0.056$   &   $0.400\pm0.014$	 &   $0.113\pm0.004$   &   $0.090\pm0.003$  &  2  \\
-183.9-179.0 	       & $1.82\pm0.03$      &  $0.003\pm0.0007$   & $0.001\pm0.0004$   &  $1.534\pm0.031$   &	$0.110\pm0.002$     & 	$2.955\pm0.059$   &   $0.510\pm0.020$	 &   $0.223\pm0.009$   &   $0.155\pm0.007$  &  2  \\
+225.6-124.1 	       & $2.49\pm0.04$      &  $0.006\pm0.0004$   & $0.002\pm0.0004$   &  $1.747\pm0.035$   &	$0.118\pm0.002$     & 	$2.874\pm0.057$   &   $0.530\pm0.011$	 &   $0.252\pm0.005$   &   $0.179\pm0.003$  &  2  \\
+117.9-235.0 	       & $2.57\pm0.06$      &  $0.006\pm0.0008$   & $0.018\pm0.0017$   &  $1.381\pm0.028$   &	$0.115\pm0.003$     & 	$2.850\pm0.057$   &   $0.420\pm0.015$	 &   $0.202\pm0.007$   &   $0.146\pm0.005$  &  2  \\  
-200.3-193.6 	       & $2.00\pm0.04$      &  $0.006\pm0.0006$   & $0.010\pm0.0017$   &  $1.966\pm0.039$   &	$0.114\pm0.003$     & 	$2.697\pm0.054$   &   $0.380\pm0.015$	 &   $0.230\pm0.009$   &   $0.161\pm0.007$  &  2  \\ 
+96.7+266.9  	       & $2.50\pm0.05$      &  $0.003\pm0.0006$   & $0.003\pm0.0004$   &  $0.820\pm0.016$   &	$0.091\pm0.002$     & 	$2.976\pm0.060$   &   $0.710\pm0.026$	 &   $0.430\pm0.016$   &   $0.300\pm0.012$  &  2  \\ 
+252.2-109.8 	       & $2.44\pm0.05$      &  $0.007\pm0.0005$   & $0.004\pm0.0003$   &  $1.623\pm0.032$   &	$0.099\pm0.002$     & 	$2.794\pm0.056$   &   $0.600\pm0.023$	 &   $0.370\pm0.014$   &   $0.260\pm0.010$  &  2  \\ 
+254.6-107.2 	       & $1.60\pm0.05$      &  $0.013\pm0.0003$   & $0.008\pm0.0010$   &  $3.455\pm0.069$   &	$0.144\pm0.004$     & 	$2.421\pm0.048$   &   $0.370\pm0.020$	 &   $0.144\pm0.008$   &   $0.120\pm0.006$  &  2  \\ 
+281.4-71.8  	       & $2.84\pm0.04$      &  $0.007\pm0.0005$   & $0.011\pm0.0013$   &  $1.621\pm0.032$   &	$0.120\pm0.002$     & 	$2.885\pm0.058$   &   $0.430\pm0.015$	 &   $0.191\pm0.007$   &   $0.130\pm0.005$  &  2  \\ 
-243.0+159.6 	       & $2.81\pm0.05$      &  $0.002\pm0.0004$   & $0.003\pm0.0005$   &  $0.860\pm0.017$   &	$0.089\pm0.003$     & 	$2.798\pm0.056$   &   $0.670\pm0.027$	 &   $0.400\pm0.016$   &   $0.280\pm0.012$  &  2  \\ 
-297.7+87.1  	       & $3.01\pm0.08$      &  $0.005\pm0.0005$   & $0.010\pm0.0013$   &  $1.197\pm0.024$   &	$0.109\pm0.004$     & 	$2.811\pm0.056$   &   $0.630\pm0.029$	 &   $0.370\pm0.018$   &   $0.270\pm0.015$  &  2  \\ 
-309.4+56.9  	       & $2.45\pm0.06$      &  $0.004\pm0.0005$   & $0.008\pm0.0013$   &  $1.094\pm0.022$   &	$0.073\pm0.003$     & 	$2.785\pm0.056$   &   $0.640\pm0.028$	 &   $0.400\pm0.017$   &   $0.280\pm0.013$  &  2  \\ 
+354.1+71.2  	       & $2.43\pm0.06$      &  $0.019\pm0.0005$   & $0.001\pm0.0002$   &  $3.520\pm0.070$   &	$0.113\pm0.002$     & 	$2.887\pm0.058$   &   $0.330\pm0.014$	 &   $0.218\pm0.009$   &   $0.159\pm0.007$  &  2  \\ 
-164.9-333.9 	       & $2.28\pm0.05$      &  $0.008\pm0.0002$   & $0.007\pm0.0011$   &  $2.441\pm0.049$   &	$0.125\pm0.002$     & 	$2.796\pm0.056$   &   $0.290\pm0.013$	 &   $0.152\pm0.006$   &   $0.108\pm0.004$  &  2  \\ 
+360.9+75.3  	       & $1.75\pm0.04$      &  $0.017\pm0.0005$   & $0.002\pm0.0002$   &  $3.537\pm0.071$   &	$0.114\pm0.003$     & 	$2.884\pm0.058$   &   $0.250\pm0.012$	 &   $0.168\pm0.008$   &   $0.120\pm0.005$  &  2  \\ 
-377.9-64.9  	       & $2.87\pm0.07$      &  $0.007\pm0.0003$   & $0.004\pm0.0003$   &  $1.324\pm0.026$   &	$0.095\pm0.002$     & 	$2.797\pm0.056$   &   $0.570\pm0.026$	 &   $0.430\pm0.020$   &   $0.300\pm0.017$  &  2  \\ 
-99.6-388.0  	       & $1.87\pm0.04$      &  $0.020\pm0.0004$   & $0.005\pm0.0008$   &  $3.882\pm0.078$   &	$0.125\pm0.003$     & 	$2.663\pm0.053$   &   $0.260\pm0.012$	 &   $0.150\pm0.007$   &   $0.122\pm0.005$  &  2  \\ 
-397.4-71.7  	       & $2.41\pm0.05$      &  $0.010\pm0.0016$   & $0.014\pm0.0031$   &  $1.847\pm0.037$   &	$0.065\pm0.003$     & 	$2.691\pm0.054$   &   $0.440\pm0.016$	 &   $0.310\pm0.012$   &   $0.212\pm0.008$  &  2  \\ 
-226.9-366.4 	       & $3.94\pm0.09$      &  $0.008\pm0.0006$   & $0.006\pm0.0007$   &  $1.514\pm0.030$   &	$0.109\pm0.002$     & 	$2.752\pm0.055$   &   $0.400\pm0.016$	 &   $0.290\pm0.012$   &   $0.204\pm0.008$  &  2  \\ 
-405.5-157.7 	       & $3.14\pm0.06$      &  $0.010\pm0.0009$   & $0.011\pm0.0020$   &  $1.299\pm0.026$   &	$0.114\pm0.005$     & 	$2.801\pm0.056$   &   $0.510\pm0.021$	 &   $0.360\pm0.015$   &   $0.250\pm0.011$  &  2  \\ 
-345.5+273.8 	       & $3.52\pm0.07$      &  $0.039\pm0.0014$   & $0.008\pm0.0016$   &  $4.450\pm0.100$   &	$0.100\pm0.017$     & 	$2.810\pm0.056$   &   $0.380\pm0.021$	 &   $0.550\pm0.030$   &   $0.400\pm0.022$  &  2  \\ 
-410.3-206.3 	       & $3.18\pm0.08$      &  $0.006\pm0.0007$   & $0.004\pm0.0010$   &  $1.462\pm0.029$   &	$0.111\pm0.003$     & 	$2.736\pm0.055$   &   $0.460\pm0.021$	 &   $0.250\pm0.012$   &   $0.173\pm0.008$  &  2  \\ 
-371.1-280.0 	       & $2.52\pm0.04$      &  $0.016\pm0.0003$   & $0.004\pm0.0005$   &  $2.665\pm0.053$   &	$0.111\pm0.002$     & 	$2.865\pm0.057$   &   $0.360\pm0.014$	 &   $0.270\pm0.010$   &   $0.195\pm0.008$  &  2  \\ 
-368.3-285.6 	       & $2.36\pm0.04$      &  $0.016\pm0.0003$   & $0.003\pm0.0004$   &  $3.231\pm0.065$   &	$0.118\pm0.002$     & 	$2.932\pm0.059$   &   $0.350\pm0.013$	 &   $0.222\pm0.008$   &   $0.166\pm0.006$  &  2  \\ 
-392.0-270.1 	       & $1.55\pm0.03$      &  $0.021\pm0.0004$   & $0.008\pm0.0012$   &  $3.721\pm0.074$   &	$0.117\pm0.002$     & 	$2.879\pm0.058$   &   $0.223\pm0.009$	 &   $0.138\pm0.005$   &   $0.105\pm0.004$  &  2  \\ 
-481.4-0.5   	       & $3.17\pm0.07$      &  $0.039\pm0.0007$   & $0.009\pm0.0013$   &  $2.843\pm0.057$   &	$0.091\pm0.002$     & 	$2.443\pm0.049$   &   $0.190\pm0.011$	 &   $0.250\pm0.015$   &   $0.180\pm0.011$  &  2  \\ 
-453.8-191.8 	       & $2.65\pm0.06$      &  $0.036\pm0.0009$   & $0.003\pm0.0004$   &  $3.747\pm0.075$   &	$0.103\pm0.002$     & 	$2.620\pm0.052$   &   $0.185\pm0.007$	 &   $0.204\pm0.008$   &   $0.142\pm0.005$  &  2  \\ 
+331.9+401.0 	       & $2.05\pm0.04$      &  $0.029\pm0.0015$   & $0.003\pm0.0009$   &  $3.610\pm0.072$   &	$0.098\pm0.006$     & 	$2.812\pm0.056$   &   $0.157\pm0.007$	 &   $0.128\pm0.005$   &   $0.089\pm0.004$  &  2  \\ 
+509.5+264.1 	       & $2.04\pm0.04$      &  $0.033\pm0.0006$   & $0.004\pm0.0001$   &  $4.173\pm0.083$   &	$0.107\pm0.002$     & 	$2.832\pm0.057$   &   $0.153\pm0.006$	 &   $0.157\pm0.005$   &   $0.110\pm0.004$  &  2  \\ 
+266.0+534.1 	       & $2.85\pm0.07$      &  $0.053\pm0.0024$   & $0.030\pm0.0054$   &  $4.707\pm0.094$   &	$0.099\pm0.014$     & 	$2.759\pm0.055$   &   $0.190\pm0.012$	 &   $0.230\pm0.014$   &   $0.160\pm0.010$  &  2  \\ 
+667.9+174.1 	       & $1.21\pm0.03$      &  $0.088\pm0.0027$   & $0.006\pm0.0002$   &  $6.740\pm0.100$   &	$0.110\pm0.002$     & 	$2.714\pm0.054$   &   $0.082\pm0.004$	 &   $0.105\pm0.004$   &   $0.086\pm0.003$  &  2  \\ 
+1.0+885.8   	       & $2.56\pm0.06$      &  $0.027\pm0.0007$   & $0.017\pm0.0008$   &  $1.839\pm0.037$   &	$0.097\pm0.007$     & 	$2.654\pm0.053$   &   $0.118\pm0.006$	 &   $0.167\pm0.007$   &   $0.115\pm0.005$  &  2  \\ 
+6.6+886.3   	       & $2.37\pm0.05$      &  $0.039\pm0.0010$   & $0.009\pm0.0008$   &  $2.688\pm0.054$   &	$0.103\pm0.004$     & 	$2.682\pm0.054$   &   $0.111\pm0.005$	 &   $0.163\pm0.006$   &   $0.112\pm0.004$  &  2  \\ 
NGC5194+30.2+2.2       & $1.51\pm0.03$      &  $0.013\pm0.0035$   &  $0.020\pm0.0023$  &  $0.341\pm0.006$   &	$0.062\pm0.001$     &  $2.960\pm0.059$    &  $1.394\pm0.028$	&   $0.266\pm0.005$   &  $0.246\pm0.004$   &   3     \\
NGC 2403-38+51         & $2.03\pm0.13$      &  $0.006\pm0.0010$   &  $0.018\pm0.0110$  &  $1.446\pm0.114$   &	$0.110\pm0.008$     &  $2.949\pm0.182$    &  $0.478\pm0.028$	&   $0.173\pm0.013$   &  $0.135\pm0.011$   &   4     \\
NGC 2403+7+37          & $2.12\pm0.14$      &  $0.007\pm0.0010$   &  $0.006\pm0.0040$  &  $1.429\pm0.121$   &	$0.114\pm0.008$     &  $2.952\pm0.187$    &  $0.455\pm0.034$	&   $0.178\pm0.013$   &  $0.131\pm0.010$   &   4     \\
NGC 2403+119-28        & $2.47\pm0.10$      &  $0.006\pm0.0020$   &  $0.008\pm0.0050$  &  $2.098\pm0.094$   &	$0.123\pm0.006$     &  $3.017\pm0.126$    &  $0.437\pm0.027$	&   $0.221\pm0.014$   &  $0.167\pm0.014$   &   4     \\
NGC 2403-59+118        & $1.47\pm0.10$      &  $0.005\pm0.0010$   &  $0.020\pm0.0130$  &  $2.230\pm0.184$   &	$0.130\pm0.009$     &  $2.939\pm0.182$    &  $0.266\pm0.182$	&   $0.147\pm0.012$   &  $0.104\pm0.009$   &   4     \\
NGC 2403+96+30         & $2.38\pm0.05$      &  $0.008\pm0.0010$   &  $0.008\pm0.0040$  &  $2.159\pm0.070$   &	$0.120\pm0.005$     &  $2.909\pm0.092$    &  $0.361\pm0.020$	&   $0.190\pm0.011$   &  $0.151\pm0.009$   &   4     \\
NGC 2403+44+82         & $2.99\pm0.21$      &  $0.005\pm0.0010$   &  $0.014\pm0.0090$  &  $1.383\pm0.117$   &	$0.114\pm0.007$     &  $3.067\pm0.189$    &  $0.453\pm0.033$	&   $0.204\pm0.014$   &  $0.144\pm0.014$   &   4     \\
NGC 2403+166-140       & $1.99\pm0.14$      &  $0.010\pm0.0010$   &  $0.014\pm0.0090$  &  $2.132\pm0.180$   &	$0.129\pm0.010$     &  $2.923\pm0.183$    &  $0.260\pm0.021$	&   $0.182\pm0.015$   &  $0.133\pm0.011$   &   4     \\
NGC 2403-99-59         & $3.31\pm0.18$      &  $0.007\pm0.0010$   &  $0.003\pm0.0020$  &  $1.699\pm0.108$   &	$0.109\pm0.005$     &  $2.992\pm0.145$    &  $0.482\pm0.034$	&   $0.345\pm0.017$   &  $0.247\pm0.013$   &   4     \\
NGC 2403-196+58        & $1.90\pm0.07$      &  $0.007\pm0.0010$   &  $0.006\pm0.0040$  &  $2.160\pm0.084$   &	$0.117\pm0.005$     &  $2.902\pm0.102$    &  $0.315\pm0.023$	&   $0.150\pm0.008$   &  $0.110\pm0.006$   &   4     \\
NGC 2403-22-162	       & $3.63\pm0.15$      &  $0.009\pm0.0020$   &  $0.020\pm0.0160$  &  $1.792\pm0.083$   &	$0.116\pm0.007$     &  $2.881\pm0.108$    &  $0.394\pm0.024$	&   $0.331\pm0.019$   &  $0.241\pm0.016$   &   4     \\
NGC 2403+160-251       & $2.83\pm0.20$      &  $0.026\pm0.0030$   &  $0.015\pm0.0030$  &  $3.160\pm0.261$   &	$0.109\pm0.007$     &  $2.976\pm0.178$    &  $0.204\pm0.017$	&   $0.201\pm0.014$   &  $0.143\pm0.011$   &   4     \\
SDSS J1455             & $1.11\pm0.01$      &  $0.102\pm0.0030$   &  $0.007\pm0.0009$  &  $6.135\pm0.033$   &	$0.114\pm0.003$     &  $2.775\pm0.018$    &  $0.079\pm0.002$	&   $0.100\pm0.002$   &  $0.008\pm0.002$   &   5     \\
SDSS J1657	       & $0.84\pm0.02$      &  $0.085\pm0.2690$   &  $0.032\pm0.0043$  &  $6.151\pm0.037$   &	$0.106\pm0.007$     &  $2.789\pm0.014$    &  $0.043\pm0.002$	&   $0.077\pm0.002$   &  $0.005\pm0.002$   &   5     \\
Reg-1	               & $0.27\pm0.01$      &  $0.057\pm0.0020$   &  $0.023\pm0.0070$  &  $1.810\pm0.080$   &	$0.093\pm0.003$     &  $2.810\pm0.100$    &  $0.010\pm0.002$	&   $0.021\pm0.001$   &  $0.015\pm0.001$   &   6     \\
\hline		              
\end{tabular}	             
\end{table*}

\begin{table*}
\addtolength{\tabcolsep}{-5pt}
\centering
\contcaption{}
\begin{tabular}{lccccccccccc}
\hline
ObjID	           &  [\ion{O}{ii}]         &    [\ion{O}{iii}]   &     \ion{He}{ii}   &   [\ion{O}{iii}]   &     \ion{He}{i}       &  $\rm H\alpha$      &  [\ion{N}{ii}]      &   [\ion{S}{ii}]    &	[\ion{S}{ii}]	   &  Ref.   \\
\noalign{\smallskip}

           	       & $\lambda$3727      &    $\lambda$4363    &    $\lambda$4686   &    $\lambda$5007   &      $\lambda$5876    &   $\lambda$6563     &  $\lambda$6583	&   $\lambda$6716     &   $\lambda$6731    &	     \\
\hline 
J0118+3512             & $0.84\pm0.01$      &  $0.064\pm0.0016$   &  $0.032\pm0.0020$  &  $3.062\pm0.005$   &	$0.114\pm0.004$     &  $3.349\pm0.005$    &  $0.039\pm0.002$	&   $0.110\pm0.003$   &  $0.085\pm0.002$   &   7     \\
J1322+5425             & $0.43\pm0.01$      &  $0.075\pm0.0013$   &  $0.010\pm0.0008$  &  $2.938\pm0.005$   &	$0.084\pm0.001$     &  $2.670\pm0.008$    &  $0.015\pm0.001$	&   $0.044\pm0.001$   &  $0.031\pm0.001$   &   7     \\
0723+692A   	       & $0.61\pm0.01$      &  $0.143\pm0.0010$   &  $0.009\pm0.0010$  &  $6.591\pm0.007$   &	$0.106\pm0.001$     &  $2.788\pm0.004$    &  $0.024\pm0.001$	&   $0.049\pm0.001$   &  $0.037\pm0.001$   &   8     \\
0723+692B   	       & $1.57\pm0.01$      &  $0.082\pm0.0040$   &  $0.014\pm0.0040$  &  $4.303\pm0.031$   &	$0.102\pm0.003$     &  $2.794\pm0.023$    &  $0.055\pm0.003$	&   $0.114\pm0.003$   &  $0.082\pm0.003$   &   8     \\
0749+568    	       & $1.66\pm0.04$      &  $0.098\pm0.0110$   &  $0.018\pm0.0080$  &  $4.880\pm0.099$   &	$0.111\pm0.009$     &  $2.797\pm0.064$    &  $0.076\pm0.007$	&   $0.178\pm0.011$   &  $0.114\pm0.008$   &   8     \\
0907+543    	       & $0.94\pm0.02$      &  $0.121\pm0.0080$   &  $0.029\pm0.0010$  &  $6.838\pm0.109$   &	$0.106\pm0.007$     &  $2.803\pm0.053$    &  $0.033\pm0.005$	&   $0.065\pm0.006$   &  $0.050\pm0.005$   &   8     \\
0917+527    	       & $1.88\pm0.01$      &  $0.092\pm0.0040$   &  $0.023\pm0.0030$  &  $4.680\pm0.003$   &	$0.103\pm0.003$     &  $2.797\pm0.019$    &  $0.058\pm0.002$	&   $0.164\pm0.003$   &  $0.114\pm0.003$   &   8     \\
0926+606               & $1.78\pm0.01$      &  $0.083\pm0.0030$   &  $0.016\pm0.0020$  &  $4.772\pm0.026$   &	$0.104\pm0.003$     &  $2.804\pm0.017$    &  $0.083\pm0.002$	&   $0.182\pm0.003$   &  $0.146\pm0.003$   &   8     \\
0930+554    	       & $0.40\pm0.00$      &  $0.060\pm0.0020$   &  $0.027\pm0.0020$  &  $1.961\pm0.008$   &	$0.081\pm0.001$     &  $2.930\pm0.012$    &  $0.013\pm0.001$	&   $0.031\pm0.001$   &  $0.023\pm0.001$   &   8     \\
1030+583    	       & $0.96\pm0.00$      &  $0.104\pm0.0020$   &  $0.024\pm0.0020$  &  $5.028\pm0.021$   &	$0.099\pm0.002$     &  $2.786\pm0.013$    &  $0.031\pm0.001$	&   $0.096\pm0.002$   &  $0.067\pm0.001$   &   8     \\
1116+583B   	       & $0.58\pm0.02$      &  $0.117\pm0.0120$   &  $0.025\pm0.0100$  &  $4.842\pm0.111$   &	$0.101\pm0.009$     &  $2.777\pm0.072$    &  $0.027\pm0.007$	&   $0.068\pm0.007$   &  $0.065\pm0.008$   &   8     \\
1205+557    	       & $2.13\pm0.02$      &  $0.083\pm0.0070$   &  $0.018\pm0.0070$  &  $3.714\pm0.043$   &	$0.099\pm0.005$     &  $2.714\pm0.035$    &  $0.083\pm0.005$	&   $0.183\pm0.007$   &  $0.130\pm0.007$   &   8     \\
1222+614    	       & $1.16\pm0.00$      &  $0.102\pm0.0020$   &  $0.017\pm0.0020$  &  $5.955\pm0.022$   &	$0.100\pm0.002$     &  $2.761\pm0.012$    &  $0.038\pm0.001$	&   $0.090\pm0.001$   &  $0.065\pm0.001$   &   8     \\
1223+487    	       & $0.71\pm0.00$      &  $0.127\pm0.0010$   &  $0.012\pm0.0010$  &  $5.543\pm0.008$   &	$0.102\pm0.001$     &  $2.777\pm0.005$    &  $0.029\pm0.001$	&   $0.061\pm0.001$   &  $0.045\pm0.001$   &   8     \\
1256+351     	       & $1.10\pm0.00$      &  $0.089\pm0.0010$   &  $0.011\pm0.0010$  &  $5.802\pm0.009$   &	$0.106\pm0.001$     &  $2.819\pm0.005$    &  $0.046\pm0.001$	&   $0.095\pm0.001$   &  $0.071\pm0.001$   &   8     \\
1319+579A   	       & $1.02\pm0.00$      &  $0.093\pm0.0020$   &  $0.008\pm0.0010$  &  $6.700\pm0.028$   &	$0.113\pm0.002$     &  $2.827\pm0.014$    &  $0.050\pm0.001$	&   $0.101\pm0.001$   &  $0.079\pm0.001$   &   8     \\
1319+579B4  	       & $2.44\pm0.06$      &  $0.050\pm0.0150$   &  $0.038\pm0.0230$  &  $3.282\pm0.073$   &	$0.094\pm0.011$     &  $2.815\pm0.069$    &  $0.161\pm0.013$	&   $0.374\pm0.017$   &  $0.273\pm0.015$   &   8     \\
1319+579C   	       & $2.38\pm0.01$      &  $0.033\pm0.0030$   &  $0.014\pm0.0040$  &  $3.520\pm0.025$   &	$0.106\pm0.003$     &  $2.852\pm0.022$    &  $0.150\pm0.003$	&   $0.269\pm0.004$   &  $0.192\pm0.004$   &   8     \\
1358+576      	       & $1.67\pm0.01$      &  $0.089\pm0.0030$   &  $0.012\pm0.0020$  &  $4.874\pm0.026$   &	$0.112\pm0.002$     &  $2.803\pm0.017$    &  $0.108\pm0.003$	&   $0.153\pm0.003$   &  $0.111\pm0.002$   &   8     \\
1441+294    	       & $1.50\pm0.03$      &  $0.069\pm0.0100$   &  $0.019\pm0.0130$  &  $4.927\pm0.099$   &	$0.116\pm0.009$     &  $2.824\pm0.064$    &  $0.083\pm0.008$	&   $0.204\pm0.011$   &  $0.138\pm0.010$   &   8     \\
1533+574B   	       & $2.03\pm0.01$      &  $0.065\pm0.0040$   &  $0.011\pm0.0030$  &  $5.329\pm0.035$   &	$0.108\pm0.003$     &  $2.849\pm0.021$    &  $0.087\pm0.002$	&   $0.167\pm0.003$   &  $0.117\pm0.002$   &   8     \\
Pox 105 	       & $1.25\pm0.00$      &  $0.113\pm0.0000$   &  $0.015\pm0.0000$  &  $5.540\pm0.000$   &	$0.096\pm0.000$     &  $2.800\pm0.000$    &  $0.064\pm0.000$	&   $0.084\pm0.000$   &  $0.058\pm0.000$   &   9     \\
Pox 120 	       & $0.79\pm0.00$      &  $0.135\pm0.0000$   &  $0.016\pm0.0000$  &  $6.290\pm0.000$   &	$0.101\pm0.000$     &  $2.780\pm0.000$    &  $0.048\pm0.000$	&   $0.074\pm0.000$   &  $0.047\pm0.000$   &   9     \\
Pox 139 	       & $1.47\pm0.00$      &  $0.096\pm0.0000$   &  $0.023\pm0.0000$  &  $5.710\pm0.000$   &	$0.101\pm0.000$     &  $2.800\pm0.000$    &  $0.019\pm0.000$	&   $0.117\pm0.000$   &  $0.077\pm0.000$   &   9     \\
UM 160 A               & $1.33\pm0.03$      &  $0.085\pm0.0068$   &  $0.004\pm0.0012$  &  $5.240\pm0.524$   &	$0.100\pm0.006$     &  $2.800\pm0.028$    &  $0.048\pm0.003$	&   $0.111\pm0.006$   &  $0.085\pm0.005$   &  10     \\
UM 160 B     	       & $1.48\pm0.03$      &  $0.066\pm0.0066$   &  $0.023\pm0.0035$  &  $4.740\pm0.474$   &	$0.096\pm0.007$     &  $2.780\pm0.055$    &  $0.071\pm0.005$	&   $0.152\pm0.009$   &  $0.102\pm0.006$   &  10     \\
UM 420 B     	       & $2.57\pm0.05$      &  $0.061\pm0.0073$   &  $0.023\pm0.0058$  &  $4.260\pm4.260$   &	$0.113\pm0.006$     &  $2.810\pm0.056$    &  $0.245\pm0.009$	&   $0.250\pm0.010$   &  $0.196\pm0.009$   &  10     \\
TOL 0513-393 	       & $0.53\pm0.01$      &  $0.155\pm0.0061$   &  $0.010\pm0.0015$  &  $7.760\pm0.776$   &	$0.128\pm0.005$     &  $2.820\pm0.028$    &  $0.045\pm0.002$	&   $0.062\pm0.003$   &  $0.053\pm0.003$   &  10     \\
TOL 2146-391-C         & $0.61\pm0.00$      &  $0.127\pm0.0026$   &  $0.017\pm0.0009$  &  $5.941\pm0.059$   &	$0.109\pm0.002$     &  $2.820\pm0.028$    &  $0.027\pm0.001$	&   $0.066\pm0.001$   &  $0.052\pm0.001$   &  11     \\
TOL 2146-391-E         & $0.64\pm0.00$      &  $0.127\pm0.0038$   &  $0.018\pm0.0013$  &  $5.897\pm0.058$   &	$0.106\pm0.003$     &  $2.770\pm0.027$    &  $0.028\pm0.001$	&   $0.067\pm0.002$   &  $0.052\pm0.002$   &  11     \\
TOL 0357-3915-C        & $0.81\pm0.01$      &  $0.123\pm0.0043$   &  $0.019\pm0.0016$  &  $6.549\pm0.065$   &	$0.111\pm0.003$     &  $2.800\pm0.028$    &  $0.055\pm0.002$	&   $0.065\pm0.002$   &  $0.049\pm0.002$   &  11     \\
TOL 0357-3915-E        & $0.85\pm0.01$      &  $0.116\pm0.0058$   &  $0.013\pm0.0019$  &  $6.300\pm0.063$   &	$0.107\pm0.005$     &  $2.810\pm0.042$    &  $0.058\pm0.004$	&   $0.070\pm0.004$   &  $0.054\pm0.003$   &  11     \\
NGC 346                & $0.80\pm0.01$      &  $0.070\pm0.0014$   &  $0.002\pm0.0002$  &  $5.220\pm0.044$   &	$0.108\pm0.001$     &  $2.820\pm0.023$    &  $0.038\pm0.001$	&   $0.074\pm0.001$   &  $0.053\pm0.001$   &  12     \\
\hline		           											\end{tabular}
\begin{minipage}[l]{16.5cm}
 References: (1) \citet{2020ApJ...893...96B}, (2) \citet{2016ApJ...830....4C}, (3) \citet{2015ApJ...808...42C}, (4) \citet{2021ApJ...915...21R}, (5) \citet{2008MNRAS.383..209H}, (6) \citet{2019MNRAS.482.3892A},
(7) \citet{2020ApJ...896...77H}, (8) \citet{1997ApJS..108....1I}, (9) \citet{1983ApJ...273...81K}, (10) \citet{2021MNRAS.505.3624V}, (11) \citet{2012ApJ...753...39P}, (12) \citet{2019ApJ...876...98V}.
\end{minipage}
\end{table*}

\begin{table*}
\addtolength{\tabcolsep}{-5pt}
\caption{Chemical abundances for the Seyfert~2 and Star-forming samples.}
\label{table2}
\begin{tabular}{lcccccccc}
\hline
ObjID & $^{\rm a}$ O$^{+}$/H$^{+}$  & $^{\rm a}$ O$^{2+}$/H$^{+}$ & ICF(O)  &  $^{\rm b}$ O/H  & $^{\rm a}$ He$^{+}$/H$^{+}$ & $^{\rm a}$ He$^{2+}$/H$^{+}$ &  ICF(He) & $^{\rm b}$ He/H  \\
\hline
\multicolumn{9}{c}{Seyfert~2} \\
\hline
J013957.81-004504.2 & \; 8.564 $\pm$ 0.041 & \; 8.324 $\pm$ 0.006 & 1.3 & \; 8.901 $\pm$ 0.129 & \; 10.626 $\pm$ 0.144 & \; 10.120 $\pm$ 0.027 & 1.5 & \; 10.918 $\pm$ 0.075 \\
J033923.14-054841.5 & \; 8.522 $\pm$ 0.033 & \; 8.185 $\pm$ 0.001 & 1.3 & \; 8.815 $\pm$ 0.026 & \; 10.805 $\pm$ 0.048 & \; 10.339 $\pm$ 0.017 & 1.5 & \; 11.114 $\pm$ 0.023 \\
J074257.23+333217.9 & \; 8.366 $\pm$ 0.021 & \; 7.700 $\pm$ 0.001 & 1.1 & \; 8.491 $\pm$ 0.018 & \; 10.852 $\pm$ 0.044 & \; 9.833 $\pm$ 0.055 & 1.3 & \; 11.007 $\pm$ 0.031 \\
J093509.12+002557.4 & \; 8.438 $\pm$ 0.017 & \; 7.647 $\pm$ 0.002 & 1.2 & \; 8.580 $\pm$ 0.019 & \; 10.700 $\pm$ 0.061 & \; 9.981 $\pm$ 0.039 & 1.4 & \; 10.931 $\pm$ 0.037 \\
J095759.45+022810.5 & \; 8.631 $\pm$ 0.033 & \; 8.208 $\pm$ 0.004 & 1.4 & \; 8.912 $\pm$ 0.029 & \; 10.757 $\pm$ 0.056 & \; 10.338 $\pm$ 0.017 & 1.6 & \; 11.092 $\pm$ 0.026 \\
J100602.50+071131.8 & \; 8.825 $\pm$ 0.003 & \; 7.904 $\pm$ 0.001 & 1.4 & \; 9.012 $\pm$ 0.014 & \; 10.807 $\pm$ 0.046 & \; 10.370 $\pm$ 0.015 & 1.6 & \; 11.158 $\pm$ 0.021 \\
J100921.26+013334.5 & \; 8.599 $\pm$ 0.036 & \; 8.377 $\pm$ 0.002 & 1.3 & \; 8.931 $\pm$ 0.025 & \; 10.888 $\pm$ 0.038 & \; 10.416 $\pm$ 0.014 & 1.5 & \; 11.191 $\pm$ 0.020 \\
J101754.72-002811.9  & \; 8.404 $\pm$ 0.026 & \; 7.801 $\pm$ 0.002 & 1.3 & \; 8.615 $\pm$ 0.022 & \; 10.895 $\pm$ 0.040 & \; 10.370 $\pm$ 0.016 & 1.5 & \; 11.183 $\pm$ 0.021 \\
J102039.81+642435.8 & \; 8.681 $\pm$ 0.099 & \; 7.979 $\pm$ 0.003 & 1.4 & \; 8.914 $\pm$ 0.083 & \; 10.699 $\pm$ 0.061 & \; 10.334 $\pm$ 0.017 & 1.6 & \; 11.066 $\pm$ 0.027 \\
J112850.39+021016.2 & \; 8.734 $\pm$ 0.029 & \; 7.456 $\pm$ 0.010 & 1.2 & \; 8.850 $\pm$ 0.033 & \; 11.049 $\pm$ 0.082 & \; 10.409 $\pm$ 0.031 & 1.4 & \; 11.298 $\pm$ 0.046 \\
Mrk176 & \; 8.564 $\pm$ 0.008 & \; 8.115 $\pm$ 0.004 & 1.5 & \; 8.882 $\pm$ 0.012 & \; 10.849 $\pm$ 0.028 & \; 10.576 $\pm$ 0.007 & 1.6 & \; 11.238 $\pm$ 0.012 \\
3C33 & \; 8.740 $\pm$ 0.006 & \; 7.998 $\pm$ 0.005 & 1.3 & \; 8.939 $\pm$ 0.010 & \; 10.835 $\pm$ 0.028 & \; 10.361 $\pm$ 0.011 & 1.5 & \; 11.144 $\pm$ 0.014 \\
Mrk3 & \; 8.549 $\pm$ 0.008 & \; 8.144 $\pm$ 0.005 & 1.3 & \; 8.796 $\pm$ 0.010 & \; 10.771 $\pm$ 0.032 & \; 10.194 $\pm$ 0.015 & 1.5 & \; 11.035 $\pm$ 0.018 \\
Mrk573 & \; 8.492 $\pm$ 0.010 & \; 8.244 $\pm$ 0.005 & 1.4 & \; 8.844 $\pm$ 0.011 & \; 10.848 $\pm$ 0.028 & \; 10.488 $\pm$ 0.008 & 1.6 & \; 11.196 $\pm$ 0.012 \\
Mrk78 & \; 8.782 $\pm$ 0.006 & \; 8.351 $\pm$ 0.005 & 1.4 & \; 9.071 $\pm$ 0.010 & \; 10.851 $\pm$ 0.027 & \; 10.469 $\pm$ 0.008 & 1.6 & \; 11.204 $\pm$ 0.012 \\
Mrk348 & \; 8.654 $\pm$ 0.007 & \; 8.032 $\pm$ 0.005 & 1.2 & \; 8.836 $\pm$ 0.007 & \; 10.927 $\pm$ 0.023 & \; 10.286 $\pm$ 0.013 & 1.4 & \; 11.173 $\pm$ 0.013 \\
Mrk34 & \; 8.590 $\pm$ 0.008 & \; 8.280 $\pm$ 0.005 & 1.3 & \; 8.871 $\pm$ 0.008 & \; 10.929 $\pm$ 0.022 & \; 10.376 $\pm$ 0.010 & 1.5 & \; 11.202 $\pm$ 0.012 \\
Mrk1 & \; 8.450 $\pm$ 0.010 & \; 8.073 $\pm$ 0.005 & 1.4 & \; 8.740 $\pm$ 0.011 & \; 10.849 $\pm$ 0.027 & \; 10.416 $\pm$ 0.009 & 1.5 & \; 11.167 $\pm$ 0.013 \\
Mrk270 & \; 8.844 $\pm$ 0.005 & \; 7.707 $\pm$ 0.007 & 1.2 & \; 8.936 $\pm$ 0.006 & \; 11.123 $\pm$ 0.014 & \; 10.298 $\pm$ 0.013 & 1.4 & \; 11.320 $\pm$ 0.009 \\
IIIZw55 & \; 8.655 $\pm$ 0.009 & \; 7.543 $\pm$ 0.009 & 1.3 & \; 8.799 $\pm$ 0.011 & \; 10.799 $\pm$ 0.031 & \; 10.260 $\pm$ 0.014 & 1.5 & \; 11.086 $\pm$ 0.016 \\
3C452 & \; 8.771 $\pm$ 0.006 & \; 7.710 $\pm$ 0.009 & 1.0 & \; 8.825 $\pm$ 0.006 & \; 11.088 $\pm$ 0.017 & \; 9.719 $\pm$ 0.048 & 1.4 & \; 11.242 $\pm$ 0.012 \\
Mrk198 & \; 8.446 $\pm$ 0.012 & \; 7.722 $\pm$ 0.011 & 1.1 & \; 8.556 $\pm$ 0.011 & \; 10.898 $\pm$ 0.025 & \; 9.815 $\pm$ 0.037 & 1.3 & \; 11.049 $\pm$ 0.018 \\
Mrk268 & \; 8.930 $\pm$ 0.007 & \; 7.241 $\pm$ 0.012 & 1.1 & \; 8.983 $\pm$ 0.008 & \; 10.824 $\pm$ 0.027 & \; 9.861 $\pm$ 0.033 & 1.3 & \; 10.991 $\pm$ 0.019 \\
NGC2110 & \; 8.796 $\pm$ 0.006 & \; 7.387 $\pm$ 0.012 & 1.1 & \; 8.836 $\pm$ 0.007 & \; 10.992 $\pm$ 0.018 & \; 9.736 $\pm$ 0.043 & 1.3 & \; 11.131 $\pm$ 0.013 \\
ESO138G1 & \; 8.572 $\pm$ 0.015 & \; 7.596 $\pm$ 0.018 & 1.4 & \; 8.752 $\pm$ 0.027 & \; 10.885 $\pm$ 0.079 & \; 10.439 $\pm$ 0.029 & 1.5 & \; 11.198 $\pm$ 0.037 \\
NGC3081 & \; 8.344 $\pm$ 0.013 & \; 8.139 $\pm$ 0.005 & 1.5 & \; 8.743 $\pm$ 0.013 & \; 10.848 $\pm$ 0.027 & \; 10.582 $\pm$ 0.006 & 1.6 & \; 11.234 $\pm$ 0.012 \\
NGC4388 & \; 8.481 $\pm$ 0.011 & \; 8.173 $\pm$ 0.006 & 1.2 & \; 8.745 $\pm$ 0.009 & \; 10.894 $\pm$ 0.025 & \; 10.254 $\pm$ 0.013 & 1.4 & \; 11.137 $\pm$ 0.015 \\
NGC5135 & \; 8.344 $\pm$ 0.014 & \; 7.546 $\pm$ 0.014 & 1.3 & \; 8.505 $\pm$ 0.014 & \; 10.850 $\pm$ 0.028 & \; 10.247 $\pm$ 0.014 & 1.5 & \; 11.113 $\pm$ 0.015 \\
NGC5728 & \; 8.813 $\pm$ 0.009 & \; 7.634 $\pm$ 0.006 & 1.3 & \; 8.950 $\pm$ 0.009 & \; 11.020 $\pm$ 0.019 & \; 10.471 $\pm$ 0.009 & 1.5 & \; 11.296 $\pm$ 0.010 \\
Akn347a & \; 8.485 $\pm$ 0.011 & \; 8.025 $\pm$ 0.004 & 1.3 & \; 8.744 $\pm$ 0.010 & \; 10.927 $\pm$ 0.023 & \; 10.467 $\pm$ 0.008 & 1.5 & \; 11.232 $\pm$ 0.012 \\
UM16a & \; 8.493 $\pm$ 0.010 & \; 8.202 $\pm$ 0.004 & 1.3 & \; 8.770 $\pm$ 0.007 & \; 11.103 $\pm$ 0.015 & \; 10.505 $\pm$ 0.008 & 1.4 & \; 11.360 $\pm$ 0.008 \\
Mrk612a & \; 8.334 $\pm$ 0.016 & \; 7.925 $\pm$ 0.007 & 1.1 & \; 8.517 $\pm$ 0.012 & \; 11.159 $\pm$ 0.014 & \; 10.145 $\pm$ 0.018 & 1.4 & \; 11.338 $\pm$ 0.010 \\
Mrk573 & \; 8.382 $\pm$ 0.011 & \; 8.213 $\pm$ 0.002 & 1.4 & \; 8.751 $\pm$ 0.018 & \; 10.802 $\pm$ 0.050 & \; 10.390 $\pm$ 0.031 & 1.5 & \; 11.126 $\pm$ 0.024 \\
NGC2992 & \; 8.743 $\pm$ 0.014 & \; 7.375 $\pm$ 0.008 & 1.1 & \; 8.808 $\pm$ 0.020 & \; 10.752 $\pm$ 0.055 & \; 9.813 $\pm$ 0.125 & 1.3 & \; 10.922 $\pm$ 0.037 \\
IC2560 & \; 8.245 $\pm$ 0.005 & \; 8.185 $\pm$ 0.004 & 1.3 & \; 8.638 $\pm$ 0.028 & \; 10.847 $\pm$ 0.045 & \; 10.344 $\pm$ 0.105 & 1.5 & \; 11.138 $\pm$ 0.028 \\
NGC5664 & \; 8.538 $\pm$ 0.012 & \; 8.006 $\pm$ 0.010 & 1.2 & \; 8.713 $\pm$ 0.014 & \; 10.801 $\pm$ 0.049 & \; 9.991 $\pm$ 0.037 & 1.4 & \; 11.002 $\pm$ 0.033 \\
NGC5728 & \; 8.482 $\pm$ 0.006 & \; 8.217 $\pm$ 0.005 & 1.2 & \; 8.768 $\pm$ 0.013 & \; 10.762 $\pm$ 0.054 & \; 10.157 $\pm$ 0.025 & 1.4 & \; 11.017 $\pm$ 0.030 \\
ESO339-G11 & \; 8.480 $\pm$ 0.023 & \; 7.957 $\pm$ 0.003 & 1.2 & \; 8.666 $\pm$ 0.021 & \; 10.802 $\pm$ 0.050 & \; 10.050 $\pm$ 0.035 & 1.4 & \; 11.014 $\pm$ 0.030 \\
NGC6890 & \; 8.246 $\pm$ 0.010 & \; 8.025 $\pm$ 0.001 & 1.4 & \; 8.605 $\pm$ 0.209 & \; 10.626 $\pm$ 0.146 & \; 10.219 $\pm$ 0.023 & 1.5 & \; 10.951 $\pm$ 0.068 \\
IC5063 & \; 8.486 $\pm$ 0.009 & \; 8.038 $\pm$ 0.004 & 1.1 & \; 8.675 $\pm$ 0.011 & \; 10.751 $\pm$ 0.055 & \; 9.886 $\pm$ 0.049 & 1.3 & \; 10.936 $\pm$ 0.035 \\
NGC7130 & \; 8.250 $\pm$ 0.008 & \; 7.690 $\pm$ 0.005 & 1.4 & \; 8.497 $\pm$ 0.031 & \; 10.546 $\pm$ 0.090 & \; 10.115 $\pm$ 0.030 & 1.6 & \; 10.874 $\pm$ 0.042 \\
NGC7582 & \; 8.139 $\pm$ 0.007 & \; 7.517 $\pm$ 0.021 & 1.1 & \; 8.285 $\pm$ 0.026 & \; 10.848 $\pm$ 0.045 & \; 9.967 $\pm$ 0.206 & 1.3 & \; 11.030 $\pm$ 0.034 \\
NGC7590 & \; 8.650 $\pm$ 0.024 & \; 7.717 $\pm$ 0.020 & 1.1 & \; 8.763 $\pm$ 0.054 & \; 10.633 $\pm$ 0.141 & \; 9.780 $\pm$ 0.180 & 1.4 & \; 10.830 $\pm$ 0.092 \\
CygnusA & \; 8.728 $\pm$ 0.006 & \; 8.206 $\pm$ 0.005 & 1.5 & \; 9.010 $\pm$ 0.014 & \; 10.696 $\pm$ 0.040 & \; 10.365 $\pm$ 0.011 & 1.6 & \; 11.072 $\pm$ 0.017 \\
Mrk1157a & \; 8.889 $\pm$ 0.004 & \; 7.882 $\pm$ 0.006 & 1.3 & \; 9.027 $\pm$ 0.006 & \; 10.927 $\pm$ 0.023 & \; 10.327 $\pm$ 0.011 & 1.5 & \; 11.204 $\pm$ 0.012 \\
ESO428-G14 & \; 8.351 $\pm$ 0.014 & \; 8.028 $\pm$ 0.006 & 1.1 & \; 8.580 $\pm$ 0.010 & \; 11.051 $\pm$ 0.018 & \; 10.220 $\pm$ 0.015 & 1.4 & \; 11.253 $\pm$ 0.011 \\
ESO137-G34 & \; 8.554 $\pm$ 0.010 & \; 8.242 $\pm$ 0.007 & 1.2 & \; 8.817 $\pm$ 0.009 & \; 10.886 $\pm$ 0.025 & \; 10.249 $\pm$ 0.013 & 1.4 & \; 11.132 $\pm$ 0.014 \\
IC3639 & \; 8.111 $\pm$ 0.024 & \; 7.882 $\pm$ 0.011 & 1.1 & \; 8.372 $\pm$ 0.017 & \; 10.847 $\pm$ 0.028 & \; 10.011 $\pm$ 0.024 & 1.4 & \; 11.045 $\pm$ 0.018 \\
IC4777 & \; 8.461 $\pm$ 0.012 & \; 8.103 $\pm$ 0.009 & 1.2 & \; 8.699 $\pm$ 0.010 & \; 10.849 $\pm$ 0.026 & \; 10.154 $\pm$ 0.016 & 1.4 & \; 11.077 $\pm$ 0.016 \\
IC4995 & \; 8.130 $\pm$ 0.022 & \; 8.173 $\pm$ 0.005 & 1.4 & \; 8.604 $\pm$ 0.015 & \; 10.750 $\pm$ 0.033 & \; 10.366 $\pm$ 0.010 & 1.5 & \; 11.087 $\pm$ 0.016 \\
IRAS11215-2806 & \; 8.409 $\pm$ 0.011 & \; 7.892 $\pm$ 0.008 & 1.3 & \; 8.631 $\pm$ 0.011 & \; 10.891 $\pm$ 0.024 & \; 10.336 $\pm$ 0.011 & 1.5 & \; 11.166 $\pm$ 0.013 \\
MCG-01-24-012 & \; 8.542 $\pm$ 0.009 & \; 7.975 $\pm$ 0.007 & 1.3 & \; 8.750 $\pm$ 0.010 & \; 10.804 $\pm$ 0.031 & \; 10.238 $\pm$ 0.014 & 1.5 & \; 11.077 $\pm$ 0.016 \\
MCG-02-51-008 & \; 8.486 $\pm$ 0.012 & \; 7.900 $\pm$ 0.014 & 1.1 & \; 8.631 $\pm$ 0.011 & \; 10.856 $\pm$ 0.026 & \; 9.879 $\pm$ 0.029 & 1.3 & \; 11.018 $\pm$ 0.018 \\
NGC1125 & \; 8.542 $\pm$ 0.009 & \; 7.968 $\pm$ 0.010 & 1.2 & \; 8.713 $\pm$ 0.009 & \; 10.850 $\pm$ 0.027 & \; 10.076 $\pm$ 0.020 & 1.4 & \; 11.059 $\pm$ 0.017 \\
NGC1194 & \; 8.306 $\pm$ 0.018 & \; 8.023 $\pm$ 0.010 & 1.1 & \; 8.515 $\pm$ 0.012 & \; 10.896 $\pm$ 0.024 & \; 9.703 $\pm$ 0.045 & 1.4 & \; 11.063 $\pm$ 0.016 \\
\hline
\end{tabular}
\end{table*}

\begin{table*}
\addtolength{\tabcolsep}{-5pt}
\contcaption{}
\begin{tabular}{lcccccccc}
\hline
ObjID & $^{\rm a}$ O$^{+}$/H$^{+}$  & $^{\rm a}$ O$^{2+}$/H$^{+}$ & ICF(O)  &  $^{\rm b}$ O/H  & $^{\rm a}$ He$^{+}$/H$^{+}$ & $^{\rm a}$ He$^{2+}$/H$^{+}$ &  ICF(He) & $^{\rm b}$ He/H  \\
\hline
NGC1320 & \; 8.036 $\pm$ 0.029 & \; 8.160 $\pm$ 0.006 & 1.4 & \; 8.552 $\pm$ 0.015 & \; 10.890 $\pm$ 0.024 & \; 10.498 $\pm$ 0.007 & 1.5 & \; 11.226 $\pm$ 0.011 \\
NGC3281 & \; 8.407 $\pm$ 0.014 & \; 8.163 $\pm$ 0.008 & 1.3 & \; 8.722 $\pm$ 0.012 & \; 10.892 $\pm$ 0.025 & \; 10.387 $\pm$ 0.009 & 1.5 & \; 11.182 $\pm$ 0.013 \\
NGC3393 & \; 8.446 $\pm$ 0.013 & \; 8.294 $\pm$ 0.006 & 1.3 & \; 8.777 $\pm$ 0.010 & \; 10.926 $\pm$ 0.023 & \; 10.338 $\pm$ 0.011 & 1.4 & \; 11.186 $\pm$ 0.013 \\
NGC4939 & \; 8.377 $\pm$ 0.014 & \; 8.265 $\pm$ 0.006 & 1.3 & \; 8.740 $\pm$ 0.010 & \; 10.926 $\pm$ 0.023 & \; 10.403 $\pm$ 0.009 & 1.5 & \; 11.208 $\pm$ 0.012 \\
NGC4968 & \; 8.227 $\pm$ 0.017 & \; 7.911 $\pm$ 0.006 & 1.3 & \; 8.506 $\pm$ 0.013 & \; 10.927 $\pm$ 0.022 & \; 10.375 $\pm$ 0.010 & 1.5 & \; 11.200 $\pm$ 0.012 \\
NGC5427 & \; 8.346 $\pm$ 0.014 & \; 8.014 $\pm$ 0.008 & 1.2 & \; 8.576 $\pm$ 0.010 & \; 10.993 $\pm$ 0.019 & \; 10.189 $\pm$ 0.015 & 1.4 & \; 11.194 $\pm$ 0.012 \\
NGC5643 & \; 8.485 $\pm$ 0.011 & \; 8.124 $\pm$ 0.007 & 1.7 & \; 8.886 $\pm$ 0.031 & \; 10.449 $\pm$ 0.067 & \; 10.309 $\pm$ 0.011 & 1.7 & \; 10.919 $\pm$ 0.022 \\
NGC5990 & \; 8.721 $\pm$ 0.015 & \; 8.059 $\pm$ 0.027 & 1.1 & \; 8.844 $\pm$ 0.014 & \; 10.794 $\pm$ 0.029 & \; 9.749 $\pm$ 0.038 & 1.3 & \; 10.947 $\pm$ 0.022 \\
NGC7682 & \; 8.465 $\pm$ 0.010 & \; 8.059 $\pm$ 0.006 & 1.3 & \; 8.709 $\pm$ 0.010 & \; 10.802 $\pm$ 0.030 & \; 10.214 $\pm$ 0.014 & 1.5 & \; 11.064 $\pm$ 0.016 \\
PKS1306-241 & \; 8.659 $\pm$ 0.008 & \; 7.908 $\pm$ 0.016 & 1.1 & \; 8.755 $\pm$ 0.009 & \; 10.849 $\pm$ 0.027 & \; 9.620 $\pm$ 0.055 & 1.3 & \; 10.987 $\pm$ 0.020 \\
\hline
\multicolumn{9}{c}{Star-forming regions} \\
\hline
NGC3184-72.8-91.3   & \; 8.258$\pm$0.002 & \; 7.241$\pm$0.002 & \; 1.015 & \; 8.304$\pm$0.002 & \; 10.780$\pm$0.005 & \; 8.947$\pm$0.039 & \; 1.253 & \; 10.885$\pm$0.007 \\
NGC3184-14.9-139.6  & \; 8.350$\pm$0.007 & \; 7.784$\pm$0.005 & \; 1.033 & \; 8.469$\pm$0.006 & \; 10.771$\pm$0.033 & \; 9.286$\pm$0.056 & \; 1.127 & \; 10.837$\pm$0.031 \\
NGC3184-80.0-148.2  & \; 8.494$\pm$0.005 & \; 7.639$\pm$0.009 & \; 1.007 & \; 8.554$\pm$0.004 & \; 10.901$\pm$0.040 & \; 8.749$\pm$0.063 & \; 1.221 & \; 10.991$\pm$0.039 \\
NGC3184-93.3-142.3  & \; 8.414$\pm$0.006 & \; 8.032$\pm$0.007 & \; 1.001 & \; 8.565$\pm$0.005 & \; 10.968$\pm$0.034 & \; 7.892$\pm$0.452 & \; 1.094 & \; 11.007$\pm$0.035 \\
NGC3184-172.5-30.2  & \; 8.304$\pm$0.002 & \; 7.992$\pm$0.003 & \; 1.003 & \; 8.478$\pm$0.002 & \; 10.958$\pm$0.014 & \; 8.387$\pm$0.146 & \; 1.078 & \; 10.992$\pm$0.014 \\
164.6-9.9           & \; 8.626$\pm$0.009 & \; 8.052$\pm$0.009 & \; 1.009 & \; 8.732$\pm$0.007 & \; 10.910$\pm$0.008 & \; 8.845$\pm$0.049 & \; 1.142 & \; 10.972$\pm$0.010 \\
17.3-235.4          & \; 8.532$\pm$0.005 & \; 7.907$\pm$0.009 & \; 1.001 & \; 8.625$\pm$0.005 & \; 10.896$\pm$0.008 & \; 7.899$\pm$0.175 & \; 1.161 & \; 10.961$\pm$0.010 \\
189.2-136.3         & \; 8.483$\pm$0.008 & \; 8.423$\pm$0.009 & \; 1.008 & \; 8.759$\pm$0.006 & \; 10.993$\pm$0.006 & \; 8.894$\pm$0.074 & \; 1.039 & \; 11.014$\pm$0.007 \\
183.9-179.0         & \; 8.621$\pm$0.007 & \; 8.268$\pm$0.009 & \; 1.001 & \; 8.781$\pm$0.006 & \; 10.886$\pm$0.008 & \; 7.891$\pm$0.173 & \; 1.088 & \; 10.923$\pm$0.009 \\
225.6-124.1         & \; 8.539$\pm$0.007 & \; 8.077$\pm$0.009 & \; 1.002 & \; 8.669$\pm$0.006 & \; 10.926$\pm$0.007 & \; 8.199$\pm$0.087 & \; 1.115 & \; 10.974$\pm$0.009 \\
117.9-235.0         & \; 8.457$\pm$0.010 & \; 7.873$\pm$0.009 & \; 1.017 & \; 8.565$\pm$0.008 & \; 10.920$\pm$0.011 & \; 9.158$\pm$0.041 & \; 1.140 & \; 10.984$\pm$0.014 \\
200.3-193.6         & \; 8.489$\pm$0.009 & \; 8.182$\pm$0.009 & \; 1.010 & \; 8.667$\pm$0.006 & \; 10.909$\pm$0.012 & \; 8.897$\pm$0.074 & \; 1.075 & \; 10.945$\pm$0.012 \\
96.7-266.9          & \; 8.518$\pm$0.009 & \; 7.726$\pm$0.009 & \; 1.004 & \; 8.585$\pm$0.008 & \; 10.814$\pm$0.010 & \; 8.374$\pm$0.060 & \; 1.209 & \; 10.898$\pm$0.014 \\
252.2-109.8         & \; 8.440$\pm$0.009 & \; 7.946$\pm$0.009 & \; 1.004 & \; 8.563$\pm$0.007 & \; 10.854$\pm$0.009 & \; 8.503$\pm$0.033 & \; 1.122 & \; 10.906$\pm$0.011 \\
254.6-107.2         & \; 8.288$\pm$0.013 & \; 8.330$\pm$0.009 & \; 1.006 & \; 8.613$\pm$0.008 & \; 11.010$\pm$0.012 & \; 8.804$\pm$0.054 & \; 1.031 & \; 11.027$\pm$0.012 \\
281.4-71.8          & \; 8.510$\pm$0.006 & \; 7.946$\pm$0.009 & \; 1.010 & \; 8.619$\pm$0.005 & \; 10.938$\pm$0.007 & \; 8.944$\pm$0.051 & \; 1.137 & \; 10.998$\pm$0.009 \\
243.0-159.6         & \; 8.744$\pm$0.008 & \; 7.945$\pm$0.009 & \; 1.004 & \; 8.809$\pm$0.007 & \; 10.797$\pm$0.015 & \; 8.370$\pm$0.073 & \; 1.209 & \; 10.881$\pm$0.017 \\
297.7-87.1          & \; 8.541$\pm$0.012 & \; 7.827$\pm$0.009 & \; 1.010 & \; 8.622$\pm$0.010 & \; 10.896$\pm$0.016 & \; 8.903$\pm$0.057 & \; 1.180 & \; 10.972$\pm$0.019 \\
309.4-56.9          & \; 8.509$\pm$0.011 & \; 7.849$\pm$0.009 & \; 1.012 & \; 8.600$\pm$0.009 & \; 10.718$\pm$0.018 & \; 8.801$\pm$0.071 & \; 1.163 & \; 10.789$\pm$0.020 \\
354.1-71.2          & \; 8.345$\pm$0.011 & \; 8.178$\pm$0.009 & \; 1.001 & \; 8.571$\pm$0.007 & \; 10.916$\pm$0.008 & \; 7.905$\pm$0.088 & \; 1.053 & \; 10.939$\pm$0.008 \\
164.9-333.9         & \; 8.515$\pm$0.009 & \; 8.243$\pm$0.009 & \; 1.006 & \; 8.704$\pm$0.007 & \; 10.951$\pm$0.007 & \; 8.743$\pm$0.068 & \; 1.070 & \; 10.983$\pm$0.008 \\
360.9-75.3          & \; 8.250$\pm$0.010 & \; 8.233$\pm$0.009 & \; 1.002 & \; 8.543$\pm$0.007 & \; 10.918$\pm$0.011 & \; 8.205$\pm$0.043 & \; 1.036 & \; 10.934$\pm$0.011 \\
377.9-64.9          & \; 8.428$\pm$0.010 & \; 7.764$\pm$0.008 & \; 1.005 & \; 8.515$\pm$0.009 & \; 10.840$\pm$0.009 & \; 8.508$\pm$0.032 & \; 1.171 & \; 10.911$\pm$0.013 \\
99.6-388.0          & \; 8.238$\pm$0.009 & \; 8.242$\pm$0.009 & \; 1.004 & \; 8.543$\pm$0.006 & \; 10.955$\pm$0.010 & \; 8.605$\pm$0.070 & \; 1.034 & \; 10.972$\pm$0.011 \\
397.4-71.7          & \; 8.356$\pm$0.009 & \; 7.906$\pm$0.009 & \; 1.024 & \; 8.498$\pm$0.008 & \; 10.677$\pm$0.021 & \; 9.056$\pm$0.098 & \; 1.102 & \; 10.730$\pm$0.021 \\
226.9-366.4         & \; 8.571$\pm$0.010 & \; 7.826$\pm$0.009 & \; 1.006 & \; 8.646$\pm$0.009 & \; 10.900$\pm$0.008 & \; 8.685$\pm$0.051 & \; 1.193 & \; 10.979$\pm$0.013 \\
405.5-157.7         & \; 8.320$\pm$0.008 & \; 7.585$\pm$0.009 & \; 1.011 & \; 8.398$\pm$0.007 & \; 10.926$\pm$0.019 & \; 8.955$\pm$0.079 & \; 1.186 & \; 11.005$\pm$0.021 \\
345.5-273.8         & \; 8.314$\pm$0.009 & \; 8.060$\pm$0.010 & \; 1.009 & \; 8.510$\pm$0.007 & \; 10.872$\pm$0.075 & \; 8.821$\pm$0.089 & \; 1.066 & \; 10.902$\pm$0.075 \\
410.3-206.3         & \; 8.574$\pm$0.011 & \; 7.921$\pm$0.009 & \; 1.004 & \; 8.663$\pm$0.009 & \; 10.903$\pm$0.012 & \; 8.503$\pm$0.109 & \; 1.167 & \; 10.972$\pm$0.015 \\
371.1-280.0         & \; 8.322$\pm$0.007 & \; 8.011$\pm$0.009 & \; 1.004 & \; 8.497$\pm$0.005 & \; 10.911$\pm$0.008 & \; 8.512$\pm$0.054 & \; 1.078 & \; 10.945$\pm$0.009 \\
368.3-285.6         & \; 8.364$\pm$0.007 & \; 8.181$\pm$0.009 & \; 1.003 & \; 8.584$\pm$0.006 & \; 10.933$\pm$0.007 & \; 8.382$\pm$0.058 & \; 1.055 & \; 10.958$\pm$0.008 \\
392.0-270.1         & \; 8.129$\pm$0.009 & \; 8.183$\pm$0.009 & \; 1.008 & \; 8.461$\pm$0.006 & \; 10.932$\pm$0.008 & \; 8.811$\pm$0.066 & \; 1.030 & \; 10.948$\pm$0.008 \\
481.4-0.5           & \; 8.088$\pm$0.009 & \; 7.654$\pm$0.009 & \; 1.011 & \; 8.229$\pm$0.007 & \; 10.840$\pm$0.009 & \; 8.884$\pm$0.061 & \; 1.103 & \; 10.887$\pm$0.011 \\
453.8-191.8         & \; 8.160$\pm$0.010 & \; 7.943$\pm$0.008 & \; 1.003 & \; 8.367$\pm$0.007 & \; 10.887$\pm$0.008 & \; 8.396$\pm$0.056 & \; 1.061 & \; 10.914$\pm$0.009 \\
331.9-401.0         & \; 8.119$\pm$0.008 & \; 8.009$\pm$0.009 & \; 1.003 & \; 8.370$\pm$0.006 & \; 10.862$\pm$0.026 & \; 8.393$\pm$0.129 & \; 1.045 & \; 10.883$\pm$0.027 \\
509.5-264.1         & \; 8.124$\pm$0.009 & \; 8.080$\pm$0.009 & \; 1.004 & \; 8.405$\pm$0.006 & \; 10.900$\pm$0.008 & \; 8.517$\pm$0.011 & \; 1.038 & \; 10.918$\pm$0.008 \\
266.0-534.1         & \; 8.124$\pm$0.010 & \; 7.967$\pm$0.008 & \; 1.034 & \; 8.368$\pm$0.008 & \; 10.873$\pm$0.059 & \; 9.401$\pm$0.076 & \; 1.047 & \; 10.909$\pm$0.057 \\
667.9-174.1         & \; 7.677$\pm$0.011 & \; 8.051$\pm$0.006 & \; 1.006 & \; 8.207$\pm$0.005 & \; 10.908$\pm$0.008 & \; 8.705$\pm$0.014 & \; 1.021 & \; 10.920$\pm$0.008 \\
1.0-885.8           & \; 7.973$\pm$0.010 & \; 7.432$\pm$0.009 & \; 1.020 & \; 8.092$\pm$0.008 & \; 10.870$\pm$0.031 & \; 9.160$\pm$0.020 & \; 1.127 & \; 10.930$\pm$0.031 \\
6.6-886.3           & \; 7.945$\pm$0.009 & \; 7.602$\pm$0.009 & \; 1.010 & \; 8.112$\pm$0.007 & \; 10.897$\pm$0.017 & \; 8.884$\pm$0.038 & \; 1.083 & \; 10.935$\pm$0.017 \\
NGC5194-30.2-2.2    & \; 7.364$\pm$0.009 & \; 6.264$\pm$0.008 & \; 1.042 & \; 7.415$\pm$0.008 & \; 10.632$\pm$0.007 & \; 9.258$\pm$0.050 & \; 1.242 & \; 10.743$\pm$0.013 \\
NGC2403-38-51       & \; 8.356$\pm$0.028 & \; 7.909$\pm$0.034 & \; 1.018 & \; 8.495$\pm$0.023 & \; 10.895$\pm$0.032 & \; 9.153$\pm$0.267 & \; 1.110 & \; 10.947$\pm$0.034 \\
NGC2403-7-37        & \; 8.319$\pm$0.028 & \; 7.831$\pm$0.036 & \; 1.006 & \; 8.444$\pm$0.023 & \; 10.917$\pm$0.030 & \; 8.683$\pm$0.287 & \; 1.126 & \; 10.971$\pm$0.035 \\
NGC2403-119-28      & \; 8.610$\pm$0.018 & \; 8.252$\pm$0.019 & \; 1.007 & \; 8.771$\pm$0.014 & \; 10.940$\pm$0.021 & \; 8.800$\pm$0.270 & \; 1.088 & \; 10.980$\pm$0.023 \\
NGC2403-59-118      & \; 8.470$\pm$0.029 & \; 8.378$\pm$0.035 & \; 1.016 & \; 8.734$\pm$0.023 & \; 10.961$\pm$0.029 & \; 9.189$\pm$0.275 & \; 1.044 & \; 10.987$\pm$0.030 \\
NGC2403-96-30       & \; 8.476$\pm$0.009 & \; 8.139$\pm$0.014 & \; 1.007 & \; 8.643$\pm$0.008 & \; 10.933$\pm$0.018 & \; 8.806$\pm$0.214 & \; 1.082 & \; 10.970$\pm$0.018 \\
NGC2403-44-82       & \; 8.587$\pm$0.031 & \; 7.948$\pm$0.037 & \; 1.014 & \; 8.683$\pm$0.026 & \; 10.913$\pm$0.027 & \; 9.048$\pm$0.279 & \; 1.169 & \; 10.985$\pm$0.038 \\
\hline
\end{tabular}
\end{table*}

\begin{table*}
\addtolength{\tabcolsep}{-5pt}
\contcaption{}
\begin{tabular}{lcccccccc}
\hline
ObjID & $^{\rm a}$ O$^{+}$/H$^{+}$  & $^{\rm a}$ O$^{2+}$/H$^{+}$ & ICF(O)  &  $^{\rm b}$ O/H  & $^{\rm a}$ He$^{+}$/H$^{+}$ & $^{\rm a}$ He$^{2+}$/H$^{+}$ &  ICF(He) & $^{\rm b}$ He/H  \\
\hline
NGC2403-166-140     & \; 8.304$\pm$0.030 & \; 8.020$\pm$0.036 & \; 1.012 & \; 8.491$\pm$0.024 & \; 10.970$\pm$0.033 & \; 9.057$\pm$0.269 & \; 1.075 & \; 11.007$\pm$0.035 \\
NGC2403-99-59       & \; 8.583$\pm$0.024 & \; 7.982$\pm$0.028 & \; 1.003 & \; 8.682$\pm$0.020 & \; 10.895$\pm$0.020 & \; 8.376$\pm$0.291 & \; 1.159 & \; 10.961$\pm$0.027 \\
NGC2403-196-58      & \; 8.435$\pm$0.016 & \; 8.197$\pm$0.017 & \; 1.006 & \; 8.636$\pm$0.012 & \; 10.922$\pm$0.019 & \; 8.687$\pm$0.284 & \; 1.064 & \; 10.952$\pm$0.019 \\
NGC2403-22-162      & \; 8.549$\pm$0.018 & \; 7.926$\pm$0.020 & \; 1.019 & \; 8.649$\pm$0.016 & \; 10.926$\pm$0.026 & \; 9.211$\pm$0.344 & \; 1.154 & \; 10.996$\pm$0.030 \\
NGC2403-160-251     & \; 8.248$\pm$0.031 & \; 7.943$\pm$0.036 & \; 1.015 & \; 8.431$\pm$0.024 & \; 10.908$\pm$0.028 & \; 9.093$\pm$0.088 & \; 1.078 & \; 10.947$\pm$0.030 \\
Reg-1               & \; 6.672$\pm$0.016 & \; 7.037$\pm$0.019 & \; 1.029 & \; 7.205$\pm$0.014 & \; 10.861$\pm$0.014 & \; 9.316$\pm$0.131 & \; 1.021 & \; 10.883$\pm$0.014 \\
J0118-3512          & \; 7.334$\pm$0.005 & \; 7.479$\pm$0.001 & \; 1.032 & \; 7.727$\pm$0.002 & \; 10.937$\pm$0.015 & \; 9.443$\pm$0.027 & \; 1.025 & \; 10.962$\pm$0.015 \\
J1322-5425          & \; 6.969$\pm$0.010 & \; 7.358$\pm$0.001 & \; 1.013 & \; 7.512$\pm$0.003 & \; 10.820$\pm$0.005 & \; 8.946$\pm$0.035 & \; 1.021 & \; 10.835$\pm$0.005 \\
0723-692A           & \; 7.182$\pm$0.007 & \; 7.792$\pm$0.001 & \; 1.010 & \; 7.892$\pm$0.002 & \; 10.912$\pm$0.004 & \; 8.895$\pm$0.048 & \; 1.020 & \; 10.925$\pm$0.004 \\
0723-692B           & \; 7.650$\pm$0.003 & \; 7.672$\pm$0.003 & \; 1.015 & \; 7.969$\pm$0.003 & \; 10.896$\pm$0.013 & \; 9.084$\pm$0.123 & \; 1.032 & \; 10.917$\pm$0.013 \\
0907-543            & \; 7.441$\pm$0.009 & \; 7.909$\pm$0.007 & \; 1.032 & \; 8.050$\pm$0.006 & \; 10.895$\pm$0.029 & \; 9.398$\pm$0.015 & \; 1.021 & \; 10.918$\pm$0.028 \\
0917-527            & \; 7.722$\pm$0.002 & \; 7.695$\pm$0.001 & \; 1.025 & \; 8.021$\pm$0.002 & \; 10.903$\pm$0.012 & \; 9.300$\pm$0.056 & \; 1.035 & \; 10.929$\pm$0.012 \\
0926-606            & \; 7.730$\pm$0.002 & \; 7.762$\pm$0.002 & \; 1.018 & \; 8.055$\pm$0.002 & \; 10.889$\pm$0.013 & \; 9.139$\pm$0.055 & \; 1.031 & \; 10.910$\pm$0.013 \\
0930-554            & \; 6.853$\pm$0.001 & \; 7.086$\pm$0.002 & \; 1.038 & \; 7.302$\pm$0.002 & \; 10.801$\pm$0.005 & \; 9.382$\pm$0.032 & \; 1.023 & \; 10.828$\pm$0.005 \\
1030-583            & \; 7.409$\pm$0.001 & \; 7.700$\pm$0.002 & \; 1.027 & \; 7.891$\pm$0.001 & \; 10.887$\pm$0.009 & \; 9.320$\pm$0.036 & \; 1.022 & \; 10.908$\pm$0.009 \\
1116-583B           & \; 7.086$\pm$0.015 & \; 7.602$\pm$0.010 & \; 1.032 & \; 7.731$\pm$0.010 & \; 10.845$\pm$0.038 & \; 9.340$\pm$0.172 & \; 1.020 & \; 10.866$\pm$0.037 \\
1205-557            & \; 7.716$\pm$0.004 & \; 7.526$\pm$0.005 & \; 1.020 & \; 7.941$\pm$0.005 & \; 10.886$\pm$0.022 & \; 9.194$\pm$0.170 & \; 1.053 & \; 10.917$\pm$0.022 \\
1222-614            & \; 7.566$\pm$0.001 & \; 7.867$\pm$0.002 & \; 1.019 & \; 8.051$\pm$0.001 & \; 10.886$\pm$0.009 & \; 9.163$\pm$0.051 & \; 1.022 & \; 10.904$\pm$0.008 \\
1223-487            & \; 7.229$\pm$0.001 & \; 7.689$\pm$0.001 & \; 1.013 & \; 7.824$\pm$0.001 & \; 10.900$\pm$0.004 & \; 9.021$\pm$0.036 & \; 1.021 & \; 10.914$\pm$0.004 \\
1256-351            & \; 7.583$\pm$0.001 & \; 7.909$\pm$0.001 & \; 1.012 & \; 8.082$\pm$0.001 & \; 10.908$\pm$0.004 & \; 8.972$\pm$0.039 & \; 1.022 & \; 10.923$\pm$0.004 \\
1319-579A           & \; 7.584$\pm$0.001 & \; 8.020$\pm$0.002 & \; 1.008 & \; 8.159$\pm$0.001 & \; 10.926$\pm$0.008 & \; 8.831$\pm$0.055 & \; 1.021 & \; 10.938$\pm$0.008 \\
1319-579B4          & \; 7.933$\pm$0.011 & \; 7.663$\pm$0.010 & \; 1.047 & \; 8.139$\pm$0.014 & \; 10.856$\pm$0.050 & \; 9.512$\pm$0.259 & \; 1.062 & \; 10.901$\pm$0.050 \\
1319-579C           & \; 8.125$\pm$0.002 & \; 7.929$\pm$0.003 & \; 1.015 & \; 8.345$\pm$0.002 & \; 10.899$\pm$0.012 & \; 9.065$\pm$0.122 & \; 1.055 & \; 10.928$\pm$0.012 \\
1358-576            & \; 7.697$\pm$0.003 & \; 7.748$\pm$0.002 & \; 1.012 & \; 8.030$\pm$0.002 & \; 10.936$\pm$0.008 & \; 9.015$\pm$0.074 & \; 1.030 & \; 10.955$\pm$0.008 \\
1441-294            & \; 7.758$\pm$0.009 & \; 7.882$\pm$0.009 & \; 1.018 & \; 8.133$\pm$0.009 & \; 10.945$\pm$0.035 & \; 9.194$\pm$0.318 & \; 1.027 & \; 10.963$\pm$0.034 \\
1533-574B           & \; 7.953$\pm$0.002 & \; 7.985$\pm$0.003 & \; 1.011 & \; 8.275$\pm$0.002 & \; 10.913$\pm$0.012 & \; 8.966$\pm$0.120 & \; 1.032 & \; 10.931$\pm$0.012 \\
Pox105              & \; 7.502$\pm$0.001 & \; 7.747$\pm$0.010 & \; 1.019 & \; 7.951$\pm$0.006 & \; 10.846$\pm$0.019 & \; 9.120$\pm$0.083 & \; 1.023 & \; 10.864$\pm$0.019 \\
Pox120              & \; 7.276$\pm$0.001 & \; 7.777$\pm$0.009 & \; 1.019 & \; 7.905$\pm$0.007 & \; 10.859$\pm$0.018 & \; 9.144$\pm$0.079 & \; 1.020 & \; 10.876$\pm$0.018 \\
Pox139              & \; 7.663$\pm$0.001 & \; 7.857$\pm$0.010 & \; 1.026 & \; 8.083$\pm$0.006 & \; 10.880$\pm$0.018 & \; 9.297$\pm$0.054 & \; 1.024 & \; 10.902$\pm$0.018 \\
UM160-A             & \; 7.638$\pm$0.010 & \; 7.833$\pm$0.045 & \; 1.005 & \; 8.050$\pm$0.028 & \; 10.875$\pm$0.027 & \; 8.533$\pm$0.135 & \; 1.025 & \; 10.889$\pm$0.027 \\
UM160-B             & \; 7.753$\pm$0.009 & \; 7.868$\pm$0.042 & \; 1.027 & \; 8.127$\pm$0.024 & \; 10.865$\pm$0.031 & \; 9.291$\pm$0.064 & \; 1.027 & \; 10.887$\pm$0.030 \\
UM420-B             & \; 8.019$\pm$0.008 & \; 7.845$\pm$0.426 & \; 1.023 & \; 8.257$\pm$0.168 & \; 10.926$\pm$0.023 & \; 9.292$\pm$0.109 & \; 1.052 & \; 10.955$\pm$0.022 \\
TOL0513-393         & \; 7.134$\pm$0.008 & \; 7.900$\pm$0.043 & \; 1.009 & \; 7.972$\pm$0.036 & \; 10.971$\pm$0.017 & \; 8.941$\pm$0.065 & \; 1.020 & \; 10.983$\pm$0.017 \\
TOL2146-391-C       & \; 7.182$\pm$0.001 & \; 7.754$\pm$0.004 & \; 1.018 & \; 7.865$\pm$0.003 & \; 10.914$\pm$0.008 & \; 9.171$\pm$0.023 & \; 1.020 & \; 10.931$\pm$0.008 \\
TOL2146-391-E       & \; 7.202$\pm$0.001 & \; 7.747$\pm$0.004 & \; 1.020 & \; 7.864$\pm$0.003 & \; 10.906$\pm$0.012 & \; 9.196$\pm$0.032 & \; 1.020 & \; 10.922$\pm$0.012 \\
TOL0357-3915-C      & \; 7.364$\pm$0.005 & \; 7.861$\pm$0.004 & \; 1.019 & \; 7.990$\pm$0.004 & \; 10.929$\pm$0.012 & \; 9.215$\pm$0.037 & \; 1.020 & \; 10.946$\pm$0.012 \\
TOL0357-3915-E      & \; 7.387$\pm$0.005 & \; 7.855$\pm$0.004 & \; 1.014 & \; 7.988$\pm$0.004 & \; 10.902$\pm$0.021 & \; 9.050$\pm$0.065 & \; 1.021 & \; 10.917$\pm$0.020 \\
NGC346              & \; 7.506$\pm$0.005 & \; 7.930$\pm$0.004 & \; 1.002 & \; 8.069$\pm$0.003 & \; 10.915$\pm$0.004 & \; 8.229$\pm$0.042 & \; 1.021 & \; 10.925$\pm$0.004 \\
\hline
\end{tabular}
\begin{minipage}[l]{15.cm}
$^{\rm a}$ Ionic abundance in units of $12+{\rm log}_{10}({\rm X}^{\rm n+}/{\rm H}{^+})$ \\
$^{\rm b}$ Total abundance in units of $12+{\rm log}_{10}({\rm X}/{\rm H})$ \\
\end{minipage}
\end{table*}

\label{lastpage}
\end{document}